\documentstyle[12pt]{article}
\makeatletter

\newdimen\hoogte    \hoogte=12pt    
\newdimen\breedte   \breedte=14pt   
\newdimen\dikte     \dikte=0.5pt    
\def\beginYoung{
       \begingroup
       \def\vr{\vrule height0.8\hoogte width\dikte depth 0.2\hoogte}
       \def\fbox##1{\vbox{\offinterlineskip
                    \hrule height\dikte
                    \hbox to \breedte{\vr\hfill##1\hfill\vr}
                    \hrule height\dikte}}
       \vbox\bgroup \offinterlineskip \tabskip=-\dikte \lineskip=-\dikte
            \halign\bgroup &\fbox{##\unskip}\unskip  \crcr }
\def\End@Young{\egroup\egroup\endgroup}
\newenvironment{Young}{\beginYoung}{\End@Young}
\makeatother

\newcommand{\Pslash}{\mbox{$\not{\hspace{-1.1mm}P}$}}

\newcommand{\vslash}{\mbox{$\not{\hspace{-0.8mm}v}$}}

\newcommand{\epsslash}{\mbox{$\not{\hspace{-0.8mm}\epsilon}$}}

\newcommand{\nc}{\newcommand}
\newcommand{\rnc}{\renewcommand}
\makeatletter
\@addtoreset{equation}{section}                        
\rnc{\theequation}{\thesection.\arabic{equation}} 
\makeatother
\nc{\be}{\begin{equation}}
\nc{\ee}{\end{equation}}
\nc{\bea}{\begin{eqnarray}}
\nc{\eea}{\end{eqnarray}}
\nc{\g}{\gamma}
\nc{\m}{\mu}
\nc{\n}{\nu}
\nc{\r}{\rho}
\nc{\s}{\sigma}
\nc{\e}{\epsilon}
\nc{\k}{\kappa}
\nc{\G}{\Gamma}
\nc{\f}{\phi}

\begin{document}

\begin{titlepage}
\begin{flushright}
 IC/93/314 \\ MZ-TH/93-23
 \end{flushright}
 \begin{center}
  {\Huge Hadrons of Arbitrary Spin and Heavy Quark Symmetry }

 \vspace*{1cm}

 {\large
  F. Hussain, G. Thompson}

 \vspace*{0.5cm}

 {\large
  International Centre for Theoretical Physics, Trieste, Italy\\}

\vspace*{1cm}

  {\large
   and\\}

\vspace*{0.5cm}

  {\large
         J.G. K\"orner}
  \footnote{Supported in part by the
  BMFT, FRG, under contract 06MZ730}

 \vspace*{0.5cm}

 {\large
  Institut f\"ur Physik \\
  Johannes Gutenberg-Universit\"at \\
  Staudinger Weg 7 \\
  D-55099 Mainz, Germany \\

  \vspace*{0.5cm}

  November 1993}

  \vspace*{1cm}

 \begin{abstract}
We present a general construction of the spin content of the Bethe-Salpeter
amplitudes (covariant wave functions) for heavy hadrons with arbitrary orbital
excitations, using representations of ${\cal L}\otimes O(3,1)$. These
wave functions incorporate the symmetries manifest in the heavy quark
limit. In the baryonic sector we clearly differentiate between the
$\Lambda$ and $\Sigma$-type excited baryons.
We then use the trace formalism to evaluate the weak transitions of
ground state heavy hadrons to arbitrary excited heavy hadrons. The
contributions of excited states to the Bjorken sum rule are also worked
out in detail.

 \end{abstract}

 \end{center}
\end{titlepage}

\section{Introduction}

  Recent advances in heavy quark physics have not only led to the
development of the heavy quark effective theory
[HQET] \cite{hqet}-\cite{kt} but also have
contributed to a better understanding of relativistic wave functions
(Bethe-Salpeter amplitudes) of both heavy and light
hadrons \cite{hkkt}-\cite{bhkt}. In our
previous work, in the mesonic sector, we have used the Bethe-Salpeter
formulation to derive the consequences of the heavy quark symmetry for
weak transitions involving both s-wave and p-wave mesons \cite{bhkt}. In
\cite{hkkt},\cite{hlkkt}, we developed the B-S approach to heavy
quark symmetry for s-wave baryons.

 In this paper we extend the B-S approach to present covariant
wavefunctions for heavy hadrons of arbitrary spin. We then derive explicit
expressions for the current induced
transitions of s-wave heavy hadrons to arbitrary spin heavy hadrons, in both
the
mesonic and baryonic sectors. In \cite{f} wavefunctions for arbitrary
spin states have also been proposed. The approach followed in \cite{f} is
different from the present work as the wave functions are developed there
explicitly from the heavy quark spinor with subsequent projections to
particular spin states, whereas we construct the wave functions directly
in terms of the polarisation tensors (generalised Rarita-Schwinger spinors
in the case of baryons) of the
hadrons using representations of the Lorentz group. Although in the
mesonic sector the wave functions look different one can recast the
forms proposed in \cite{f} into the simpler forms given in the present
work. During the course of this work we became aware that heavy
meson Bethe-Salpeter wave functions have been also constructed in \cite{dhj}
albeit following a different approach. Their wave functions are
equivalent to the ones developed here from considering representations
of the Lorentz group.

As far as the heavy baryon wave functions are concerned, in \cite{f} only a
certain subset of the possible baryonic states are constructed. This is an
artifact of the manner of construction
which involves taking the product of a totally symmetric tensor of rank
$j$ with the heavy quark spinor. Such states do not exhaust the possible
baryonic excited states. As we shall see in section {\bf 3} on
baryons other symmetry types are also possible. These are discussed in
detail in the present work. We present a systematic approach
towards constructing any $\Lambda$-type or $\Sigma$-type excited resonance.

In this work we consider hadrons as bound states of quarks and antiquarks
and we concentrate on mesons and baryons containing one heavy quark $Q$.
The natural object to describe a bound state is the Bethe-Salpeter
amplitude \cite{hw} which we will define for mesons as
\be
M_{\alpha}\,^{\beta}(x_{1},x_{2}) = \langle 0 \vert T
\psi_{Q\alpha}(x_{1})\bar{\psi}_{q_{2}}^{\beta}
(x_{2})\vert M,P \rangle \, , \label{1}
\ee
where $\vert M,P \rangle$ is a mesonic state with a certain
momentum P, spin and other quantum numbers.

Similarly, for heavy baryons the corresponding
BS amplitude is defined as
\be
B_{\alpha \beta \g}(x_{1},x_{2},x_{3}) = \langle 0 \vert T
\psi_{q_{1}\alpha}(x_{1})
\psi_{q_{2}\beta}(x_{2}) \psi_{Q\g}(x_{3})\vert B,P \rangle \, . \label{2}
\ee
Here $\vert B \rangle$ represents a particular baryon state. In eqs.
(\ref{1}) and (\ref{2}), $\psi_{Q}$ represents the heavy quark field
while the $\psi_{q}$'s represent the light quark fields.

As is by now very well known \cite{hkstw}-\cite{bhkt}, \cite{bj}, in the
heavy quark mass limit,
$m_{Q}\rightarrow\infty$, in the leading order of the HQET, the heavy
quark spin is decoupled from the light degrees of freedom and as a
consequence
the B-S amplitudes satisfy the Bargmann-Wigner(Dirac) equation on the
heavy label (in momentum space):
\be
(\vslash-1)_{\alpha}\,^{\delta}M_{\delta}\,^{\beta}(p_{1},p_{2})=0 \label{bw1}
\ee
and
\be
(\vslash-1)_{\g}\,^{\delta}B_{\alpha \beta \delta}(p_{1},p_{2},p_{3})=0\,,
\label{bw2}
\ee
where $v$ is the four velocity of the hadron.
These equations imply that the heavy meson and heavy baryon B-S
amplitudes can be written in momentum space as
\be
M_{\alpha}\,^{\beta}(p_{1},p_{2}) =
\chi_{\alpha}\,^{\delta}(p_{1},p_{2})A_{\delta}\,^{\beta}(p_{1},p_{2}) \,
\label{3}
\ee
and
\be
B_{\alpha \beta \g}(p_{1},p_{2},p_{3}) = \chi _{\r \delta
\g}(p_{1},p_{2},p_{3})A^{\r \delta}_{\alpha \beta}(p_{1},p_{2},p_{3})\, ,
\label{4}
\ee
where the $\chi$'s are projection operators satisfying
\be
(\vslash-1)_{\alpha}\,^{\alpha^{\prime}}\chi_{\alpha^{\prime}}\,^{\delta}
=(\vslash-1)_{\g}\,^{\g^{\prime}}\chi_{\r \delta \g^{\prime}}=0\,.
\label{po}
\ee
They project out particular spin and parity states from the orbital wave
functions $A$. The Dirac indices on these operators indicate that these
transform as product representations of the Lorentz group, whereas the
momentum arguments indicate that they also have specific transformation
properties under the 4-dimensional rotation group. A particular spin
parity projection operator corresponds to reducing such a product
representation down to its required spin content. This can be
done in an elegant manner through the use of the Bargmann-Wigner wave
functions \cite{bw} and by a careful consideration of the Lorentz group.
We will do this in detail in the next two sections for mesons and
baryons. Although the projection operators developed here are used in
the context of heavy hadrons, they have nothing to do, per se, with
heavy quark physics. In fact, one can use them also in studying light
hadrons \cite{hlkkt}, \cite{bhkt}, \cite{bhkt2}. After all, we are just
constructing representations of the Lorentz group, albeit
representations which are particularly suited to the heavy quark limit.

The consequences of eqs. (\ref{3}) and (\ref{4}) have been worked out for
s-wave to s-wave and s-wave to p-wave heavy meson semileptonic decays in
\cite{bhkt}
and for heavy baryon s-wave to s-wave semileptonic decays in \cite{hkkt} and
\cite{hlkkt}. In sections {\bf 4} and {\bf 5} we will consider semileptonic
decays
of ground state heavy hadrons to arbitrary heavy orbital excitations.
We also calculate the contributions of these excitations to the Bjorken
sum rules. We mention that our covariant description is also well suited
to calculate one-pion and photon transitions between heavy hadrons of
the same flavour \cite{ikkl}.

\section{Wave Functions of Heavy Mesons of Arbitrary Spin}

First let us consider the mesonic case. Essentially we would like to
discover how $\chi_{\alpha}\,^{\beta}$ reduces under the Lorentz group, $\cal
L$.
Now the infinitesimal generators of Lorentz transformations acting on
$\chi_{\alpha}\,^{\beta}(p_{1},p_{2})$ are
\be
J_{\m \n} = S_{\m \n} + i\sum_{j=1}^{2}(p_{j \m}\frac{\partial}{\partial
p_{j}^{\n}}-p_{j \n}\frac{\partial}{\partial p_{j}^{\m}}) \, , \label{5}
\ee
where $S_{\m \n}$ are the generators of the Lorentz group acting on the
product space \newline
$[(\frac{1}{2},0)\oplus(0,\frac{1}{2})]\otimes
[(\frac{1}{2},0)\oplus(0,\frac{1}{2})]$
or
\be
S_{\m \n} = \frac{1}{2}\s_{\m \n}\otimes{\bf 1}+{\bf1}\otimes
\frac{1}{2}\s_{\m \n}\, , \label{6}
\ee
with the $S_{\m \n}$ acting on the Dirac labels $\alpha$ and $\beta$ of
$\chi_{\alpha}\,^{\beta}$.
Therefore, $\chi_{\alpha}\,^{\beta}(p_{1},p_{2})$ has to be reduced in terms of
representations of ${\cal L}\otimes O(3,1)$. It is easy to show that the
$O(3,1)$ generators can be written as
\be
\sum_{j=1}^{2}(p_{j \m}\frac{\partial}{\partial p_{j}^{\n}}-p_{j \n}
\frac{\partial}{\partial p_{j}^{\m}}) = (P_{\m}\frac{\partial}{\partial
P^{\n}} - P_{\n}\frac{\partial}{\partial P^{\m}})+
(k_{\m}\frac{\partial}{\partial k^{\n}} - k_{\n}\frac{\partial}{\partial
k^{\m}})\, , \label{7}
\ee
where $P = p_{1} + p_{2}$ and $k= p_{1} - p_{2}$.

Hence
\be
J_{\m \n} = S_{\m \n} + i(k_{\m}\frac{\partial}{\partial k^{\n}} -
k_{\n}\frac{\partial}{\partial k^{\m}}) +
i(P_{\m}\frac{\partial}{\partial P^{\m}} -
P_{\n}\frac{\partial}{\partial P^{\m}}). \,  \label{8}
\ee
The last term in eq. (\ref{8}) describes the overall orbital angular
momentum of the centre of mass of the system with respect to an external
origin. The first two terms give the genuine total internal angular
momentum operator of the $Q \bar{q}$ system,
\be
M_{\m \n} = S_{\m \n} + L_{\m \n}\, , \label{9}
\ee
with
\be
L_{\m \n} = i(k_{\m}\frac{\partial}{\partial k^{\n}}-
k_{\n}\frac{\partial}{\partial k^{\m}}).\, \label{10}
\ee
The ``spin", i.e. the total angular momentum of the $Q\bar{q}$ pair is
described by the Pauli-Lubanski operator
\be
W_{\m} = \frac{1}{2}\e_{\m \n \k \lambda}P^{\n}J^{\k \lambda} =
 \frac{1}{2}\e_{\m \n \k \lambda}P^{\n}M^{\k \lambda}. \,  \label{11}
\ee
Because of the $\e_{\m \n \k \lambda}$ the overall orbital angular momentum
operator drops out, as expected, and we are left with the relative
internal orbital momentum $L_{\m \n}$ plus $S_{\m \n}$. The square of
the Pauli-Lubansky operator gives the magnitude of the ``total spin" of
the system
\be
W^{2} = -M^{2}J(J+1) \, , \label{12}
\ee
where $M^{2} = P^{2}$ is the invariant mass of the system.

 It is easy now to look at the $S^{\m \n}$ and orbital parts separately.
In the rest frame, $P=(M,\underline{0})$,
\be
W^{\m} = M(0,M_{23},M_{31},M_{12}) = M(0,J_{1},J_{2},J_{3}) \, ,
\label{13}
\ee
with
\be
[J_{i},J_{j}] = i\e_{ijk}J_{k} \,  \label{14}
\ee
and
\bea
J_{i} = \e_{ijk}(S_{jk}+L_{jk}) \, , \label{15}
      = S_{i}+L_{i} \, , \label{16}
\eea
where the $S_{i}$ are the usual spin matrices and the $L_{i}$ are the orbital
angular momenta. We have thus reduced the group from ${\cal L}\otimes O(3,1)$
to $SU(2)\otimes O(3)$. As expected, the total angular momentum is given by
the usual vector addition of the $Q\bar{q}$ spin and the relative
orbital angular momentum. The operators $S_{\m \n}$ and $L_{\m \n}$(or
$S_{i},L_{i}$) act on different spaces. So we have to separately find
the appropriate irreducible representations of $\cal L$ and $O(3,1)$.

The reduction of multispinor representations of the Lorentz group to
their irreducible $SU(2)$ components was first done by Bargmann and
Wigner \cite{bw} and was later worked out in detail in the 60's by Delbourgo,
Salam and Strathdee \cite{sds},\cite{sdrs}, and other workers \cite{u12}
in the context of relativistic $SU(6)$. Of course such
reductions have nothing to do, per se, with relativistic $SU(6)$. This is a
point
emphasized in a modern review of the procedure as given in \cite{hkt}.

 The basic approach is very simple. The four component Dirac indices of
a multispinor
$\chi_{\alpha_{1}\ldots\alpha_{n}}^{\beta_{1}\ldots\beta_{m}}(p_{i})$
are reduced to two components by imposing the Bargmann-Wigner equations on
each of the upper and lower indices
\bea
(\vslash - 1)_{\alpha}\,^{\alpha_{1}}\chi_{\alpha_{1}\ldots
\alpha_{n}}^{\beta_{1}\ldots\beta_{m}} & = & 0  \nonumber \\
\vdots & & \nonumber\\
(\vslash - 1)_{\alpha}\,^{\alpha_{n}}\chi_{\alpha_{1}
\ldots\alpha_{n}}^{\beta_{1}\ldots\beta_{m}} & = & 0
\label{17}
\eea
and
\bea
\chi_{\alpha_{1}\ldots\alpha_{n}}^{\beta_{1}\ldots\beta_{m}}
(\vslash+1)_{\beta_{1}}\,^{\beta} & = & 0
\nonumber \\
\vdots &&\nonumber \\
\chi_{\alpha_{1}\ldots\alpha_{n}}^{\beta_{1}\ldots\beta_{m}}
(\vslash+1)_{\beta_{m}}\,^{\beta} & = & 0.
\label{18}
\eea

 The $\chi$'s are spinor functions of the total momentum,
$P=\sum_{i}p_{i}$, or in modern usage the total velocity, $v=P/M$, and
the relative momenta $k_{i}$. Also following current usage, we have
written the Bargmann-Wigner equations in terms of the operators
$(\vslash \pm 1)$ rather than $(\Pslash \pm M)$.

Imposing the equations (\ref{17}) and (\ref{18}) reduces the multispinor
to a product representation of $SU(2)$. We then use the charge
conjugation matrix  $C$ to lower the upper indices. Then choosing a
given symmetry of the indices we pick out a particular irreducible
representation of $SU(2)$.

Following such a procedure for the mesonic projector in eq. (\ref{3}) leads
to the following general form
\be
\chi_{\alpha}\,^{\beta}(v,k) =
\sqrt{M}[\frac{1+\vslash}{2}\G (k) \frac{1-\vslash}{2}]_{\alpha}\,^{\beta}\, ,
\label{19}
\ee
where $\G (k)$ is a positive parity matrix. The $\sqrt{M}$
factor is a normalization factor which we have introduced here for later
convenience when we go to heavy hadrons, where one wants
to factor out the heavy mass scale. The overall parity
of $\chi$  is negative and is fixed by the Bargmann-Wigner equations
\cite{sscm}.

Now to take the direct product of ${\cal L}\otimes O(3,1)$
we require $\G (k)$ to transform appropriately under $L_{\m \n}$,
eq. (\ref{10}). To do this we go to the rest frame $(v=(1,\underline{0}))$,
and look for the appropriate eigenstates of $\vec{L}^{2}$ with $\vec{L}$
defined as in eq. (\ref{16}) or more specifically
\be
L_{1} = i(k_{2}\frac{\partial}{\partial k_{3}} -
k_{3}\frac{\partial}{\partial k_{2}})\, , {\rm etc.}
\label{20}
\ee
Thus for s-waves,  we require $\vec{L}^{2}\G (k) = 0$ which means that
$\G (k)$ can at most depend on $k$ in the form $k_{\bot}^{2}$, where
\be
k_{\bot}^{\m} = k^{\m} - v\cdot k \, v^{\m}\, , \label{21}
\ee
such that
\be
v\cdot k_{\bot} = 0. \label{22}
\ee
Now expanding $\G$ in terms of the 16 gamma matrices we find that they
all reduce to just two non-vanishing terms (as $v^{2} = 1$),
\be
\G (k) = \g_{5}\e_{5}
\label{23}
\ee
or
\be
\G (k) = \g^{\m}\e_{\m}\, , \label{24}
\ee
with $v^{\m}\e_{\m} = 0$.
Eq. (\ref{23}) describes the $0^{-}$ bound state of a $Q\bar{q}$ pair.
$\e_{5}$ is the ``polarisation" pseudoscalar\footnote{For
pseudoscalars and scalars we shall omit these $\e$'s later. In general
we could have an unknown function of $k_{\bot}^{2}$ in eqs. (\ref{23})
and (\ref{24}) but these can be absorbed in the unknown functions
$A_{\alpha}\,^{\delta}$ in eq. (\ref{3})} . (For a
point particle this would just be the pseudoscalar field e.g. the pion field.)
Eq. (\ref{24}) describes the $1^{-}$ bound state with $\e_{\m}$ identified as
the polarisation vector.

For the p-waves we require
\be
\vec{L}^{2}\G (k) = 2\G (k)\, . \label{25}
\ee
This is easily solved by
\be
\G (k) = k_{\bot}\cdot \bar{\G}(k_{\bot}^{2})\, , \label{26}
\ee
where $\bar{\G}_{\m}$ can at most be a vector function of the Lorentz scalar
$k_{\bot}^{2}$. $\bar{\G}_{\m}$ is also a bispinor in Dirac space. Observe that
the vector $k_{\bot}^{\m}$ satisfying $v\cdot k_{\bot}=0$ has the correct
degrees of freedom to describe an angular momentum one object. Constructing
all possible forms we get the p-wave reduced spin wave functions
$\bar{\G}_{\m}$ listed in Table 1\footnote{Again these $\bar{\G}_{\m}$'s
can, in general, be multiplied by unknown functions of $k_{\bot}^{2}$
which will be absorbed in the $A_{\alpha}\,^{\delta}$ in eq. (\ref{3}).
This remark applies to all subsequent constructions of higher spin meson
and baryon projection operators in the rest of this paper and will not
be repeated again.}. This table is reproduced
from \cite{bhkt}. Note that we cannot use $v_{\m}$ in the construction
as this annihilates on $k_{\bot}$ in eq. (\ref{26}).
\begin{table}[htbp]
\centering
\begin{tabular}{ccl} \hline\hline
{\bf state} & ${\bf J^{PC}}$   & $\bf {\bar{\G}_{\m}} $              \\ \hline
${}^{3}P_{0}$ & $0^{++}$ & $\g_{\m}/\sqrt{3}$        \\
${}^{3}P_{1}$ & $1^{++}$ &$ -\frac{i}{\sqrt{2}}\g_{5}\s^{\n}\,_{\m}\e_{\n}$  \\
${}^{3}P_{2}$ & $2^{++}$ & $\g^{\n}\e_{\n \m}$
\\
${}^{1}P_{1}$ & $1^{+-}$ & $\g_{5}\e_{\m}$    \\                    \hline
\end{tabular}
\caption{Reduced spin wave functions $\bar{\G}$ for mesonic p-wave
states.}
\end{table}

$\e_{\m}$ and $\e_{\m \n}$ are the spin 1 and spin 2 polarisation
tensors satisfying $v^{\m}\e_{\m}=v^{\m}\e_{\m \n}=0$ with $\e_{\m \n}$
symmetric and traceless. The Bargmann-Wigner equations ensure that
\be
\frac{1+\vslash}{2}v\cdot\bar{\G}\frac{1-\vslash}{2}= 0\, . \label{27}
\ee

The above states correspond to an LS coupling scheme. In \cite{bhkt} we
have shown how this can be transformed to another basis, the jj basis,
more suitable for heavy quark physics, i.e. where the spin of the light
quark and the orbital angular momentum are added first to get the total
light quark angular momentum with respect to the heavy quark. The heavy
quark spin is added later. We will come back to the jj basis when we
consider the general L-wave case.

 The covariant wave functions, eqs. (\ref{19}) and (\ref{26}),
with $\bar{\G}$ as in Table 1, for p-waves were first constructed
for heavy $(Q\bar{Q})$ quarkonium systems \cite{kks}. However, we now see
that these have nothing intrinsically to do with heavy quarkonium systems
but are rather consequences of the reduction of ${\cal L}\otimes O(3,1)$.

We can now proceed to construct projection operators for
d-waves by looking for solutions of
\be
\vec{L}^{2}\G (k) = 6\G (k)\, .
\ee
This is solved immediately by
\be
\G (k) =(k_{\bot}^{\m}k_{\bot}^{\n}-\frac{1}{3}k_{\bot}^{2}g_{\bot}^{\m \n})
 \bar{\G}_{\m \n}\, , \label{28}
\ee
where
\be
g_{\bot}^{\m \n} = g^{\m \n}-v^{\m}v^{\n}\, .
\label{29}
\ee
$\bar{\G}_{\m \n}$ is a tensor spinor such that
\be
\frac{1+\vslash}{2}v^{\m}\bar{\G}_{\m \n}\frac{1-\vslash}{2}=0\, .
\label{30}
\ee
Thus, all the states $2^{-+};3^{--},2^{--},1^{--}$ arising from
$(0,1)\otimes 2$ of $0(3)_{spin}\otimes 0(3)_{orbital}$ can be immediately
listed as in Table 2.
\begin{table}[htbp]
\centering
\begin{tabular}{ccl} \hline\hline
\bf state & $\bf  J^{PC}$   & $\bf {\bar{\G}_{\m\n}}$                \\ \hline
&& \\
${}^{3}D_{1}$ &$ 1^{--}$ &$ \sqrt{\frac{3}{5}}\g_{\m}\e_{\n} $       \\
&& \\
${}^{3}D_{2}$ &$ 2^{--}$ &$ -\sqrt{\frac{2}{3}}i\g_{5}\s^{\lambda}\,_{\m}
\e_{\lambda \n} $ \\
&& \\
${}^{3}D_{3}$ &$ 3^{--}$ &$ \g^{\lambda}\e_{\lambda \m \n}$        \\
&& \\
${}^{1}D_{2}$ &$ 2^{-+} $& $\g_{5}\e_{\m \n}$                \\
&& \\ \hline
\end{tabular}
\caption{Reduced spin wave functions $\bar{\G}$ for mesonic d-wave
states.}
\end{table}

The $\e$'s are the usual symmetric traceless polarisation tensors satisfying
$v^{\m}\e_{\m}=v^{\m}\e_{\m \n}=v^{\m}\e_{\m \n \lambda}=0$. These are
normalised to $2M$ with respect to the traceless tensor
$\frac{1}{2}(g_{\bot}^{\m \m'}g_{\bot}^{\n
\n'}+g_{\bot}^{\m \n'}g_{\bot}^{\n \m'}-
\frac{2}{3}g_{\bot}^{\m \n}g_{\bot}^{\m' \n'})$ (See Appendix A).

One can now immediately generalise to any higher orbital angular
momentum. For the general orbital angular momentum $L$ case we need the
solution to
\be
\vec{L}^{2}\G (k) = L(L+1)\G (k)\, .
\ee
This is solved by
\be
\G (k) = N^{\m_{1}\m_{2}\ldots\m_{L}}(k)\bar{\G}_{\m_{1}\m_{2}\ldots\m_{L}}\,.
\label{31}
\ee
$N^{\m_{1}\m_{2}\ldots\m_{L}}(k)$ is the symmetric, traceless tensor
product of $L\;k_{\bot}^{\m}$'s satisfying
\be
v_{\m_{i}}N^{\m_{1}\ldots\m_{i}\ldots\m_{L}}(k)=0
\label{32}
\ee
for all $i$.
Specifically
\bea
N^{\m_{1}\m_{2}\ldots\m_{L}}(k) & = & k_{\bot}^{\m_{1}}k_{\bot}^{\m_{2}}
\ldots k_{\bot}^{\m_{L}}-\frac{k_{\bot}^{2}}{2L-1}\sum_{i<j}
g_{\bot}^{\m_{i}\m_{j}}\prod_{n\neq i,j}k_{\bot}^{\m_{n}} \nonumber \\
& + & \frac{k_{\bot}^{4}}{(2L-1)(2L-3)}\sum_{i<j,i<k<l}g_{\bot}^{\m_{i}\m_{j}}
g_{\bot}^{\m_{k}\m_{l}}\prod_{n\neq i,j,k,l}k_{\bot}^{\m_{n}} -\ldots
\label{33}
\eea
We also have
\be
\frac{1+\vslash}{2}v^{\m_{i}}\bar{\G}_{\m_{1} \ldots \m_{i} \ldots \m_{L}}
\frac{1-\vslash}{2}= 0\, .
\label{34}
\ee
for all $i$.

In table 3 we list the $\bar{\G}_{\m_{1}\m_{2} \ldots \m_{L}}$ for the
states $L-1,L,L=1;L$ arising from $(1,0)\otimes L$ of
$O(3)_{spin}\otimes O(3)_{orbital}$.
\begin{table}[htbp]
\centering
\begin{tabular}{ccl} \hline\hline
\bf state & $\bf J^{PC}$&$\bf {\bar{\G}_{\m_{1}\ldots\m_{L}}}  $
  \\ \hline
&& \\
${}^{3}L_{L-1}$ &$ (L-1)^{(-)^{L+1}(-)^{L+1}}$ &
$\sqrt{\frac{2L-1}{2L+1}}\g_{\m_{1}}\e_{\m_{2} \ldots \m_{L}} $ \\
&& \\
${}^{3}L_{L}$ &$ L^{(-)^{L+1}(-)^{L+1}}$ &
$-\sqrt{\frac{L}{L+1}}i\g_{5}\s^{\lambda}\,_{\m_{1}}
\e_{\lambda \m_{2} \ldots \m_{L}} $ \\
&& \\
${}^{3}L_{L+1}$ &$ (L+1)^{(-)^{L+1}(-)^{L+1}}$ &
$\g^{\lambda}\e_{\lambda \m_{1} \ldots \m_{L}}       $ \\
&& \\
${}^{1}L_{L}$ &$ L^{(-)^{L+1}(-)^{L}}$ &$ \g_{5}\e_{\m_{1} \ldots \m_{L}} $ \\
&& \\ \hline
\end{tabular}
\caption{Reduced spin wave functions $\bar{\G}$ for mesonic L-wave
states.}
\end{table}

Here the $\e$'s are the usual fully symmetric, traceless, transverse
polarisation tensors. See Appendix A for the normalisation of these
wave functions.

Having constructed the projection operators we now recall that the B-S
amplitudes for the heavy meson are obtained by substituting these
operators in eq. (\ref{3}). Using Tables 2 and 3, we can now easily consider
heavy to heavy semi-leptonic decays involving d- and
higher waves. This will be done in the next section.

The wave functions as developed upto now are quite general and can be
used in the context of both heavy and light mesons. However as pointed out
in \cite{bhkt} in the case of heavy mesons it is convenient to use a
different basis than that used for the states in Tables 2 and 3. These
states are in the LS
coupling scheme and are eigenstates of $C$-parity. We need to move to
the jj scheme. In this scheme, for heavy mesons, the degenerate
states in the d-wave case are the pairs $(1_{3/2}^{-},2_{3/2}^{-})$ and
$(2_{5/2}^{-},3_{5/2}^{-})$ where the subscripts $3/2$ and $5/2$ indicate the
total angular momentum of the light quark system $(2\otimes 1/2 = 5/2
\oplus 3/2)$. The two states in each pair are related to each other by
flipping the heavy quark spin i.e. they are degenerate because of the
heavy quark spin symmetry. This can be demonstrated by applying the spin
flip operator $\epsslash \g_{5}$ to the appropriate $\bar{\G}$.  The $1^{-}$
and $3^{-}$ states are unchanged whereas the
$2^{-}$ states in this new basis are linear combinations of the LS
states
\bea
\vert 2_{3/2}^{-} \rangle & = & \sqrt{\frac{3}{5}} \vert 2^{--} \rangle
+\sqrt{\frac{2}{5}}\vert 2^{-+} \rangle \nonumber \\
\vert 2_{5/2}^{-} \rangle & = & -\sqrt{\frac{2}{5}}\vert 2^{--} \rangle
+\sqrt{\frac{3}{5}}\vert 2^{-+} \rangle\, .
\label{37}
\eea

In Table 4, we list the $\bar{\G}$'s for the appropriate degenerate
states of the d-wave heavy meson.
\begin{table}[htbp]
\centering
\begin{tabular}{cl} \hline\hline
\bf state & $\bf  {\bar{\G}_{\m \n}}$                     \\ \hline
& \\
$1_{3/2}^{-}$ &$ \sqrt{\frac{3}{5}}\g_{\m}\e_{\n}     $   \\
& \\
$2_{3/2}^{-}$ &$ \sqrt{\frac{2}{5}}\g_{5}\g^{\lambda}\g_{\m}\e_{\lambda \n} $
\\
&\\ \hline
& \\
$2_{5/2}^{-}$ &$ \frac{1}{\sqrt{15}}
\g_{5}(5g_{\m}^{\lambda}-2\g^{\lambda}\g_{\m})
\e_{\lambda \n}$        \\
& \\
$3_{5/2}^{-}$ &$ \g_{\lambda}\e_{\lambda \m \n}$  \\
& \\ \hline
\end{tabular}
\caption{The d-wave heavy meson reduced spin wave functions}
\end{table}

Similarly in the general case one has to go to the heavy quark basis
states
$\vert L_{L-1/2}^{(-)^{L+1}}\rangle$ and $\vert L
_{L+1/2}^{(-)^{L+1}}\rangle$. Recall that total light quark angular
momentum is either $L-1/2$ or $L+1/2$. Thus
\bea
\vert L_{L-1/2}^{(-)^{L+1}}\rangle & = & \sqrt{\frac{L+1}{2L+1}}\vert
L^{(-)^{L+1}(-)^{L+1}}\rangle + \sqrt{\frac{L}{2L+1}}\vert
L^{(-)^{L+1}(-)^{L}}\rangle \nonumber \\
\vert L_{L+1/2}^{(-)^{L+1}}\rangle & = & -\sqrt{\frac{L}{2L+1}}\vert
L^{(-)^{L+1}(-)^{L+1}} \rangle + \sqrt{\frac{L+1}{2L+1}}\vert
L^{(-)^{L+1}(-)^{L}}\rangle \, .
\label{38}
\eea

In table 5, we list the $\bar{\G}$'s for the appropriate pairs of degenerate
states of the L-wave heavy meson.
\begin{table}[htbp]
\centering
\begin{tabular}{cl} \hline\hline
\bf state & $\bf  {\bar{\G}_{\m_{1}\ldots\m_{L}}} $                 \\ \hline
& \\
$(L-1)_{L-1/2}^{(-)^{L+1}}$ &
$\sqrt{\frac{2L-1}{2L+1}}\g_{\m_{1}}\e_{\m_{2} \ldots \m_{L}}     $   \\
& \\
$L_{L-1/2}^{(-)^{L+1}}$ &
$\sqrt{\frac{L}{2L+1}}\g_{5}\g^{\lambda}\g_{\m_{1}}
\e_{\lambda \m_{2} \ldots \m_{L}} $ \\
&\\ \hline
& \\
$L_{L+1/2}^{(-)^{L+1}}$ & $\frac{1}{\sqrt{(L+1)(2L+1)}}\g_{5}((2L+1)
g_{\m_{1}}^{\lambda}-L\g^{\lambda}\g_{\m_{1}})\e_{\lambda \m_{2} \ldots
\m_{L}}  $      \\
& \\
$(L+1)_{L+1/2}^{(-)^{L+1}}$ &$ \g_{\lambda}\e_{\lambda \m_{1} \ldots \m_{L}}$
  \\
& \\
\hline
\end{tabular}
\caption{The L-wave heavy meson reduced spin wave functions}
\end{table}

Tables 2-5 are some of the main results of this paper.
Notice the simplicity and elegance of the wave functions. Heavy meson wave
functions have also been constructed by
Falk \cite{f} following a different route, which always explicitly
involves the heavy quark spinor and
always requires projection operators to go to particular states. Notice
that in our construction there is no sign of the heavy quark spinor. Our
construction is directly in terms of the appropriate polarisation
tensors of the heavy meson state itself. The forms of the wave functions
for heavy mesons given in \cite{f} can be transformed into the forms
given above.

In the above we have listed just the ``pure" spin states characterized by
particular orbital angular momentum. Of course physical states could in
general be mixtures of these states preserving the heavy quark spin symmetry.

\section{Heavy Baryon Wave Functions of Arbitrary Orbital Angular Momentum}

We repeat the procedure of sec. {\bf 2} for the baryon spin projector
$\chi_{\alpha \beta \g}(p_{1},p_{2},p_{3})$ in eq. (\ref{4}). Here
$p_{1},p_{2},p_{3}$ are the quark momenta with $P=Mv=p_{1}+p_{2}+p_{3}$.
We construct the s-,p- and d-wave baryon projection operators in detail and
then indicate how to
generalise to an arbitrary orbital excitation. Recall that the $\chi_{\alpha
\beta \g}$ satisfy the Bargmann-Wigner equations and that the heavy
baryon B-S amplitude is obtained from eq. (\ref{4}) by substituting with
the appropriate $\chi$.

Here the infinitesimal generators of the Lorentz transformations acting
on $\chi_{\alpha \beta \g}(p_{1},p_{2},p_{3})$ are
\be
J_{\m \n} = S_{\m \n} + i\sum_{j=1}^{3}(p_{j \m}\frac{\partial}{\partial
p_{j}^{\n}}-p_{j \n}\frac{\partial}{\partial p_{j}^{\m}}) \, , \label{42}
\ee
where $S_{\m \n}$ are the generators of the Lorentz group acting on the
space\newline
$[(\frac{1}{2},0)\oplus(0,\frac{1}{2})]^{3}$ or
\be
S_{\m \n} = \frac{1}{2}\s_{\m \n}\otimes {\bf 1}\otimes {\bf 1}+{\bf1}\otimes
\frac{1}{2}\s_{\m \n}\otimes {\bf 1} + {\bf 1}\otimes {\bf 1}\otimes
\frac{1}{2}\s_{\m \n}\, , \label{43}
\ee
with the $S_{\m \n}$ acting on the Dirac labels $\alpha$,$\beta$ and
$\g$. As before we reduce $\chi_{\alpha \beta \g}$ in terms of
representations of ${\cal L}\otimes O(3,1)$.

Defining two independent relative four momenta as
\be
k_{3} = \frac{1}{2}(p_{1}-p_{2})
\label{44}
\ee
and
\be
K_{3} = \frac{1}{3}(p_{1}+p_{2}-2p_{3})
\label{45}
\ee
we can write
\be
J_{\m \n} = S_{\m \n} + L_{\m \n}^{(k_{3})} + L_{\m \n}^{(K_{3})}+
i(P_{\m}\frac{\partial}{\partial P^{\n}} -
P_{\n}\frac{\partial}{\partial P^{\m}}) \,.  \label{46}
\ee
Here
\be
L_{\m \n}^{(k)} = i(k_{\m}\frac{\partial}{\partial k^{\n}} -
k_{\n}\frac{\partial}{\partial k^{\m}})\, .
\label{47}
\ee
$L^{k_{3}}_{\m\n}$ is the angular momentum operator for the
relative orbital angular momentum of the quark pair, $q_{1}q_{2}$, while
$L^{K_{3}}_{\m\n}$ is the angular momentum operator for the relative
orbital angular momentum between the centre of mass (c.m.) of the pair
$q_{1}q_{2}$ and
the third quark $q_{3}$.
 Of course, one could also decompose $J_{\m \n}$ in terms of
$k_{1},k_{2}$ and $K_{1},K_{2}$ where
\be
k_{1} = \frac{1}{2}(p_{2}-p_{3})\,,\;\; k_{2} = \frac{1}{2}(p_{3}-p_{1})
\nonumber
\ee
and
\be
K_{1} = \frac{1}{3}(p_{2}+p_{3}-2p_{1}),\,\, K_{2} =
\frac{1}{3}(p_{3}+p_{1}-2p_{2})\, .
\label{48}
\ee

However out of the six momenta, $k_{i}$ and $K_{i}$, only two, plus the
total momentum $P$, are
independent and we choose $k_{3}$ and $K_{3}$. Recall that the third
quark momenta $p_{3}$ is associated with the Dirac label $\g$.

As in the mesonic case the last term in eq. (\ref{46}) describes the
overall orbital angular momentum of the c.m. of the system with respect
to an external origin and drops out of the Pauli-Lubansky operator. Thus
\be
W_{\m} = \frac{1}{2}\e_{\m \n \k \lambda}P^{\n}J^{\k \lambda} =
 \frac{1}{2}\e_{\m \n \k \lambda}P^{\n}M^{\k \lambda} \,,  \label{49}
\ee
with
\be
M_{\m \n} = S_{\m \n} + L_{\m \n}^{(k_{3})} + L_{\m \n}^{(K_{3})}\, .
\label{50}
\ee

\subsection{S-wave Baryon Projection Operators}
\label{sub-swave}

The s-wave baryon projection operators are disposed of quite easily.
The $\chi_{\alpha \beta \g}$ for s-waves can only be functions of
$k_{3 \bot}^{2}$ or $K_{3 \bot}^{2}$ and must satisfy the Bargmann-Wigner
equations on each label. Thus one can in general have two
possibilities \cite{sds,sdrs,hkt}, either antisymmetric or symmetric in
the $\alpha \beta$ indices\footnote{From now on we shall often use the
usual bracket notation to indicate symmetry properties of indices.
Square brackets,[], around indices will represent antisymmetrisation of
the indices enclosed whereas curly brackets,\{\}, around indices indicates
symmetrisation.}
\be
\chi_{[\alpha \beta]\g}^{\Lambda} = \chi_{\alpha\beta}^{0}\Psi_{\g}
\label{51}
\ee
or
\be
\chi_{\{\alpha \beta\}\g}^{\Sigma} = \chi_{\alpha \beta}^{1,\m}
\Psi_{\m,\g}\label{52}
\ee
where $\chi_{\alpha \beta}^{0} \equiv [(\vslash+1)\g_{5}C]_{\alpha \beta}$
and
$\chi_{\alpha \beta}^{1,\m} = [(\vslash +1)\g^{\m}C]_{\alpha \beta}$
with
\be
v^{\m} \Psi_{\m}= 0
\nonumber
\ee
and
\be
(\vslash-1)\Psi = (\vslash -1)\Psi_{\m} = 0\, .
\label{53}
\ee
We have put the superscripts $\Lambda$ and $\Sigma$ because we will soon
show that these correspond to the projection operators for the $\Lambda$
and $\Sigma$-type baryons respectively. These are the only two
independent possibilities because any other like
\mbox{$[(\vslash+1)\g_{5}C]_{\beta \g}\Psi_{\alpha}$} can always be recast in
the forms given in eqs. (\ref{51}) and (\ref{52}) by splitting into $\alpha
\beta$ symmetric and antisymmetric pieces. Specifically
\be
[(\vslash+1)\g_{5}C]_{\beta \g}\Psi_{\alpha} =
-\frac{1}{2}[(\vslash+1)\g_{5}C]_{\alpha
\beta}\Psi_{\g}+\frac{1}{2}[(\vslash+1)\g^{\m}C]_{\alpha
\beta}[(\g_{\m}+v_{\m})\g_{5}\Psi]_{\g}
\label{54}
\ee
and
\be
[(\vslash+1)\g^{\m}C]_{\beta \g}\Psi_{\m,\alpha} =
-\frac{1}{2}[(\vslash+1)\g^{\m}C]_{\alpha
\beta}\Psi_{\m,\g}+\frac{1}{2}[(\vslash+1)\g_{5}C]_{\alpha
\beta}(\g^{\m}\g_{5}\Psi_{\m})_{\g} \, .
\label{55}
\ee
 The meaning of $\chi_{\alpha \beta}^{0}$ and $\chi_{\alpha
\beta}^{1,\m}$ is obvious. $\chi_{\alpha \beta}^{0}$ represents the
antisymmetric spin zero, $S_{q_{1}q_{2}} = 0$ state of the quarks 1 and
2, whereas the transverse part of $\chi_{\alpha \beta}^{1,\m}$,
$\chi_{\bot\alpha \beta}^{1,\m}, (v_{\m}\chi_{\bot\alpha \beta}^{1,
\m}=0$), represents the symmetric spin 1, $S_{q_{1}q_{2}} = 1$, state of
the first two quarks. Because of the relations (\ref{54})
and (\ref{55}) one can always express the spin of any  pair of
quarks in terms of the 1,2 pair.
 One notes also that $\chi_{\bot}^{1,\m} = [(\vslash +
1)(\g^{\m}-v^{\m})C]$. But since $v^{\m}\Psi_{\m}$ in eq. (\ref{52}) and
in subsequent wave functions, one can use interchangeably $\chi^{1,\m}$
or $\chi_{\bot}^{1,\m}$. However it is important to remember the
transversality. In summary, $\chi_{\alpha \beta}^{0}$ and
$\chi_{\bot\alpha \beta}^{1,\m}$ represent the $0^{+}$ and $1^{+}$ spin states
of the $q_{1}q_{2}$ pair.

We can now differentiate the $\Lambda_{Q}$ and $\Sigma_{Q}$-type of
baryons by considering flavour symmetry. The $\Lambda_{Q}$-type baryons
are antisymmetric in flavour for the first two quarks. Hence to preserve
overall
symmetry we see that the $\chi_{[\alpha \beta] \g}^{\Lambda}$ in eq. (\ref{51})
is
the correct, ${\frac{1}{2}}^{+}$, $\Lambda$ wave function with
\be
\Psi_{\g} = u_{\g}\,, \label{56}
\ee
the usual Dirac spinor.

Similarly, $\chi_{\{\alpha \beta\} \g}^{\Sigma}$ describes the symmetric
$\Sigma_{Q}$-type baryons. Here it is easy to see that $\Psi_{\m,\g}$
decomposes uniquely as
\bea
{\frac{3}{2}}^{+}: \;\;\;\;\;   \Psi_{\m,\g} & = & u_{\m,\g} \nonumber \\
{\frac{1}{2}}^{+}: \;\;\;\;\;   \Psi_{\m,\g} & = &\frac{1}{\sqrt{3}}
(\g_{\bot\m}\g_{5}u)_{\g}  \nonumber \\
& = &\frac{1}{\sqrt{3}}
[(v_{\m}+\g_{\m})\g_{5}u]_{\g}\, . \label{57}
\eea
where $\g_{\bot\m}=\g_{\m}-\vslash v_{\m}$.
The particular form of the spin $1/2$ in eq. (\ref{57}) is required by the
conditions (\ref{53}).

\subsection{P-wave Baryon Projection Operators}
\label{sub-pwave}

We now proceed to construct the projection operators for p-wave baryons.
In contrast to the mesonic case, there are now two ways to get p-waves.
Let us denote by $L_{k_{3}}$ the orbital angular momentum of the quark
pair $q_{1}q_{2}$ and by $L_{K_{3}}$ the orbital angular momentum
between the centre of mass of the pair $q_{1}q_{2}$ and the third quark
$q_{3}$. P-wave baryons are obtained by taking either $(L_{k_{3}}=1,
L_{K_{3}}=0)$ or $(L_{k_{3}}=0, L_{K_{3}}=1)$.
Hence we require that $\chi_{\alpha \beta \g}$ should be of the form
\be
\chi_{\alpha \beta \g} = k_{3 \bot}^{\m}\bar{\chi}_{\m,\alpha \beta \g}\;\;\;
{\rm or}\;\;\; K_{3 \bot}^{\m}\bar{\chi}_{\m,\alpha \beta \g}\, ,
\label{58}
\ee
with $\bar{\chi}_{\m,\alpha \beta \g}$ satisfying the Bargmann-Wigner
equations on all the indices $\alpha, \beta, \g$.  Note also that although
both $k_{3}$ and $K_{3}$ are of mixed symmetry ( under the interchange of
$p_{1},p_{2},p_{3}$), $k_{3}$ is antisymmetric
under $p_{1}\leftrightarrow p_{2}$ whereas $K_{3}$ is symmetric. We
construct $\bar{\chi}_{\m,\alpha \beta \g}$ using $\chi^{0}_{\alpha\beta}$
and $\chi^{1,\m}_{\alpha \beta}$ and ensuring
overall symmetry with respect to $flavour\otimes spin\otimes space$ for
the quarks 1 and 2.

Without worrying about the choice of $k_{3 \bot}^{\m}$ or$K_{3
\bot}^{\m}$, it is easy to construct $\bar{\chi}_{\m,\alpha \beta \g}$.
With $\chi_{\alpha \beta}^{0}$ we have simply
\be
\bar{\chi}_{\m,[\alpha \beta]\g} = \chi_{\alpha \beta}^{0}\Psi_{\m,\g}\, ,
\label{59}
\ee
with $\Psi_{\m,\g}$ satisfying
\be
(\vslash-1)\Psi_{\m} =0 \;\; {\rm and}\;\; v^{\m}\Psi_{\m} = 0 \, .
\label{60}
\ee
Thus, $\Psi_{\m}$ decomposes uniquely into the ${\frac{3}{2}}^{-}$ and
${\frac{1}{2}}^{-}$ states as in eq. (\ref{57}) \footnote{The negative
parity is obvious from physical reasons. Also it can be seen from
eqs. (\ref{58}) that since $\chi_{\alpha \beta \g}$ is always of positive
parity, $\bar{\chi}_{\m,\alpha \beta \g}$ must necessarily be odd as
$k_{3\bot}^{\m}(K_{3 \bot}^{\m})$ is odd. Thus from eq. (\ref{59}) we see
that $\Psi_{\m,\g}$ is also odd since $\chi_{\bot \alpha \beta}^{0}$
transforms like $0^{+}$. The parity of the subsequent states eqs.
(\ref{61})-(\ref{67}) is odd for a similar reason since $\chi_{\alpha
\beta}^{1,\m}$ transforms like $1^{+}$. We have continued to use the
usual spinors $u,(u_{\m},u_{\m \n})$, satisfying the Dirac equation
$(\vslash-1)u=0$ but always remembering that they describe negative parity
particles. One could as well use $\g_{5}v$.}. These two states arise from
$(S_{q_{1}q_{2}}=0\otimes L=1) \otimes S_{q_{3}} =
{\frac{3}{2}}^{-}, {\frac{1}{2}}^{-}$ where $S_{q_{3}}=1/2$ is the spin of
the third quark (later we will identify the third quark with the heavy
quark). Since $\chi_{\alpha \beta}^{0}$ is antisymmetric under
$\alpha\leftrightarrow \beta$, to ensure overall symmetry for the
$\Lambda$-type excitation (flavour antisymmetric) we would have $K_{3
\bot}^{\m}\bar{\chi}_{\m,[\alpha \beta]\g}$, i.e. the $L_{K_{3}}=1$ orbital
angular momentum. On the other hand
for the $\Sigma$-type we would have $k_{3 \bot}^{\m}
\bar{\chi}_{\m,[\alpha \beta]\g}$, i.e. the $L_{k_{3}}=1$ orbital angular
momentum.

With $\chi_{\bot \alpha \beta}^{1,\m}$ we have three possibilities. Here
we combine the $S_{q_{1}q_{2}}=1$ with $L=1$ to give
$J_{q_{1}q_{2}}=2,1,0$ where $J_{q_{1}q_{2}}$ is the total angular
momenta of quarks 1 and 2 with respect to the third quark. These three
different sets of states can be written as ($k$ is either $k_{3}$ or
$K_{3}$):

$1.\;\;J_{q_{1}q_{2}}=2$
\be
\frac{1}{2}(k_{\bot}^{\m}\chi_{\bot}^{1,\n}+k_{\bot}^{\n}\chi_{\bot}^{1,\m}
-\frac{2}{3}k_{\bot}\cdot\chi_{\bot}g_{\bot}^{\m \n})_{\alpha \beta}\Psi_{\m
\n ,\g} \, , \label{61}
\ee
with $\Psi_{\m \n,\g}$ a symmetric, traceless tensor-spinor satisfying
\be
(\vslash-1)\Psi_{\m \n}=0\;\; {\rm and}\;\; v^{\m}\Psi_{\m \n}=0\, .
\label{62}
\ee
$\Psi_{\m\n}$ can easily be decomposed as (for normalisations see
Appendix A)
\bea
{\frac{5}{2}}^{-}: \;\;\;\;\;  \Psi_{\m \n} & = & u_{\m \n}\label{63a}
\\
{\frac{3}{2}}^{-}: \;\;\;\;\;  \Psi_{\m \n} & = &
\frac{1}{\sqrt{10}}(\g_{\bot \m}\g_{5}u_{\n}
+\g_{\bot \n}\g_{5}u_{\m})\, . \label{63b}
\eea
Here one could also have a term like $g_{\bot}^{\m\n}u$. However this
would annihilate when folded into the expresssion (\ref{61}) because the
angular
momentum tensor of the $q_{1}q_{2}$ pair is traceless. Such a term will
be called a non-propagating mode.

Thus the two nonvanishing structures correspond exactly to the two
possibilities expected from
$(J_{q_{1}q_{2}}=2)\otimes
(S_{q_{3}}=1/2)={\frac{5}{2}}^{-},\;{\frac{3}{2}}^{-}$. $u_{\m \n}$ is
the usual $5/2$ generalised, symmetric  Rarita-Schwinger tensor-spinor
and $u_{\m}$ is the $3/2$ Rarita-Schwinger spinor.

$2.\;\;J_{q_{1}q_{2}}=1$
\be
\frac{1}{2}(k_{\bot}^{\m}\chi_{\bot}^{1,\n}
-k_{\bot}^{\n}\chi_{\bot}^{1,\m})_{\alpha \beta}\Psi_{\m \n,\g}\,.
\label{64}
\ee
Here $\Psi_{\m \n,\g}$ is an antisymmetric tensor-spinor satisfying
eqs. (\ref{62}). $\Psi_{\m \n}$ is easily decomposed as
\bea
{\frac{3}{2}}^{-}: \;\;\;\;\; \Psi_{\m \n} & = &
\frac{1}{\sqrt{2}}(\g_{\bot \m}\g_{5}u_{\n}
- \g_{\bot \n}\g_{5}u_{\m}) \nonumber \\
{\frac{1}{2}}^{-}:\;\;\;\;\; \Psi_{\m \n} & = &
\frac{1}{2\sqrt{6}}[\g_{\bot \m},\g_{\bot \n}]u\,. \label{65}
\eea

These correspond exactly to the two possibilities expected from
$(J_{q_{1}q_{2}}=1)\otimes(S_{q_{3}=\frac{1}{2}})=
{\frac{3}{2}}^{-},\;{\frac{1}{2}}^{-}$.

$3.\;\;J_{q_{1}q_{2}}=0$
\be
\frac{1}{\sqrt{3}}(k_{\bot}.\chi_{\bot}^{1})_{\alpha \beta}g_{\bot}^{\m
\n}\Psi_{\m \n, \g}\, . \label{66}
\ee
Again $\Psi_{\m \n}$ is a traceless, symmetric tensor satisfying
eqs. (\ref{62}). Here $\Psi_{\m \n}$ decomposes uniquely as
\be
{\frac{1}{2}}^{-}: \;\;\;\;\;\Psi_{\m \n}=\frac{1}{\sqrt{3}}g_{\bot\m \n}u\,.
\label{67}
\ee
This state corresponds to the one state
expected from $(J_{q_{1}q_{2}}=0)\otimes(S_{q_{3}}=\frac{1}{2})=
{\frac{1}{2}}^{-}$.

The Lorentz structure of the states given by the
eqs. (\ref{61})-(\ref{67}) is valid for both $\Lambda_{q_{3}}$ and
$\Sigma_{q_{3}}$ type baryons. Since these wave functions
(\ref{61})-(\ref{67}) are symmetric under $\alpha \leftrightarrow
\beta$, to get $\Lambda$-type baryons we take $k=k_{3}$ and for
$\Sigma$-type we take $k=K_{3}$ in these equations to ensure overall symmetry.

\subsection{Baryon Projection Operators for the First Positive
Parity Excitations}
\label{sub-dwave}

There are three possibilities in constructing the first positive
parity excitations. In the subsequent equations of this subsection, $L$
refers to the total orbital angular momentum of the quark pair, $q_{1}$ and
$q_{2}$. By first positive parity excitations we mean all those
excitations with $L_{K_{3}}+L_{k_{3}}=2$.

$(1)\,\,\, L_{k_{3}}=2,\, L_{K_{3}}=0,\, L=2$

Such an orbital state is represented by the symmetric traceless tensor
$N^{\m_{1}\m_{2}}(k_{3})$ as defined in eq. (\ref{33}).

$(2)\,\,\, L_{k_{3}}=0,\, L_{K_{3}}=2,\, L=2$

Such an orbital state is represented by the symmetric traceless tensor
$N^{\m_{1}\m_{2}}(K_{3})$.

$(3)\,\,\, L_{k_{3}}=1,\, L_{K_{3}}=1$

$\,\,\,(a)\, L=2$

Here the orbital state is represented by the tensor
\be
N^{\{\m_{1}\m_{2}\}}(K_{3},k_{3})=
\frac{1}{2}(K_{3\bot}^{\m_{1}}k_{3\bot}^{\m_{2}}
+K_{3\bot}^{\m_{2}}k_{3\bot}^{\m_{1}}-\frac{2}{3}K_{3\bot}\cdot k_{3\bot}
g_{\bot}^{\m_{1}\m_{2}})\,.\label{dw1}
\ee

$\,\,\,(b)\, L=1$

Here the orbital state is represented by the tensor
\be
N^{[\m_{1}\m_{2}]}(K_{3},k_{3})=
\frac{1}{2}(K_{3\bot}^{\m_{1}}k_{3\bot}^{\m_{2}}
-K_{3\bot}^{\m_{2}}k_{3\bot}^{\m_{1}})\,.\label{dw2}
\ee

$\,\,\,(c)\, L=0$

Here the orbital state is represented by the scalar
\be
N(K_{3},k_{3})=\frac{1}{\sqrt{3}}K_{3\bot}\cdot k_{3\bot}
g_{\bot}^{\m_{1}\m_{2}}\,.\label{dw3}
\ee
The orbital states  (1) and (2) are symmetric under interchange of
$p_{1}$ and $p_{2}$ while the orbital states in (3) are antisymmetric
under the same interchange. Using these symmetry properties we now
construct the projection operators for the $\Lambda$-type and
$\Sigma$-type baryons.

\subsubsection{$\Lambda$-type first positive parity excitation projection
operators}

The $\Lambda$-type baryon is in a flavour antisymmetric state for the
quarks labelled 1 and 2. Because of the requirement of overall symmetry,
the orbital states (1) and (2) above can only combine with the antisymmetric
spin zero,  $S_{q_{1}q_{2}}=0$, state of the quarks 1 and 2, i.e. with
$\chi^{0}_{\alpha\beta}$, leading to the total light quark angular
momentum of $J_{q_{1}q_{2}}=2$. This in turn leads to the total baryon angular
momentum states $\frac{5}{2}^{+}$ and $\frac{3}{2}^{+}$. The physical
states will in general be mixtures of the states arising from the two
orbital states (1) and (2). We list these projection operators below:

\underline{$\;(1)\,\,\, L_{K_{3}}=2,\, L_{k_{3}}=0,\, L=2$ or
$\;(2)\,\,\, L_{K_{3}}=0,\, L_{k_{3}}=2,\, L=2$}

In both cases we have
\be
N^{\m_{1}\m_{2}}\chi^{0}_{\alpha\beta}\Psi_{\m_{1}\m_{2},\g}\,,
\label{dw4}
\ee
where $N^{\m_{1}\m_{2}}$ is either $N^{\m_{1}\m_{2}}(K_{3})$ or
$N^{\m_{1}\m_{2}}(k_{3})$.
$\Psi_{\m_{1}\m_{2}}$ decomposes into its $\frac{5}{2}^{+}$ and
$\frac{3}{2}^{+}$ components just as in eqs. (\ref{62}) and
(\ref{63a}, \ref{63b}),
\bea
{\frac{5}{2}}^{+}: \;\;\;\;\;  \Psi_{\m_{1} \m_{2}} & = &
u_{\m_{1} \m_{2}}\label{dw4a}        \\
{\frac{3}{2}}^{+}: \;\;\;\;\;  \Psi_{\m_{1} \m_{2}} & = &
\frac{1}{\sqrt{10}}(\g_{\bot \m_{1}}\g_{5}u_{\m_{2}}
+\g_{\bot \m_{2}}\g_{5}u_{\m_{1}})\, . \label{dw4b}
\eea

In contrast, the orbital states (3), being antisymmetric with respect
to the interchange of $p_{1}$ and $p_{2}$, can only combine with the
symmetric, spin one, $S_{q_{1}q_{2}}=1$, state of the quarks 1 and 2,
i.e. with $\chi^{1,\m}_{\alpha\beta}$. These will give rise to the
states whose projection operators are discussed in the rest of this
subsection. All the $\Psi$s which are about to appear are transverse
with respect to $v_{\m}$ and satisfy the Dirac equation. Furthermore
they are traceless in the vector labels. The symmetry of these labels
will be specified as we proceed.

\underline{$\;(3)\,(a)\, L_{K_{3}}=1,\, L_{k_{3}}=1,\, L=2$}

The total orbital angular momentum, $L=2$, here combines with the spin,
$S_{q_{1}q_{2}}=1$, to give total light quark angular momenta
$J_{q_{1}q_{2}}=3,2,1$, for which the projection operators are listed
below.

$\;\;\;(i)\, J_{q_{1}q_{2}}=3$
\be
\frac{1}{3}[(\chi_{\bot}^{1,\m}N^{\{\m_{1}\m_{2}\}}
-\frac{2}{5}g_{\bot}^{\m\m_{1}}\chi_{\bot\n}N^{\{\n\m_{2}\}})
+symmetrised\;in\;\m,\m_{1},\m_{2}]_{\alpha\beta}\Psi_{\m\m_{1}\m_{2},\g}
\label{dw5}
\ee
with $\Psi_{\m\m_{1}\m_{2},\g}$  fully symmetric. $\Psi_{\m\m_{1}\m_{2}}$ can
easily be decomposed as
\bea
{\frac{7}{2}}^{+}: \;\;\;\;\;  \Psi_{\m \m_{1}\m_{2}} & = & u_{\m\m_{1}\m_{2}}
\label{dw6}        \\
{\frac{5}{2}}^{+}: \;\;\;\;\;  \Psi_{\m \m_{1}\m_{2}} & = &
\frac{1}{\sqrt{21}}(\g_{\bot \m}\g_{5}u_{\m_{1}\m_{2}}
+\g_{\bot
\m_{1}}\g_{5}u_{\m_{2}\m}+\g_{\bot\m_{2}}\g_{5}u_{\m\m_{2}})\nonumber\\
& = & \sqrt{\frac{3}{7}}\g_{\bot \m}\g_{5}u_{\m_{1}\m_{2}}\, ,
\label{dw7}
\eea
where in the last step we have used the fact that
$\Psi_{\m\m_{1}\m_{2}}$ is folded into a symmetric tensor in eq. (\ref{dw5}).
(There are, of course, as usual non-propagating modes just as in the
p-wave case. From now we shall just ignore them.)  The two non-vanishing states
correspond exactly to the two expected from
$(J_{q_{1}q_{2}}=3)\otimes(S_{q_{3}}=\frac{1}{2})=\frac{7}{2}^{+},
\frac{5}{2}^{+}$. $u_{\m\m_{1}\m_{2}}$ and $u_{\m_{1}\m_{2}}$ are the
generalised $\frac{7}{2}$ and $\frac{5}{2}$ Rarita-Schwinger spinors.

$\;\;\;(ii)\, J_{q_{1}q_{2}}=2$
\be
\frac{1}{3}[\chi_{\bot}^{1,\m}N^{\{\m_{1}\m_{2}\}}
-\chi_{\bot}^{1,\m_{1}}N^{\{\m\m_{2}\}}
-\frac{1}{2}g_{\bot}^{\m\m_{2}}\chi_{\bot\n}N^{\{\n\m_{1}\}}
+\frac{1}{2}g_{\bot}^{\m_{1}\m_{2}}
\chi_{\bot\n}N^{\{\n\m\}}]_{\alpha\beta}\Psi_{[\m\m_{1}]\m_{2},\g}\,,
\label{dw8}
\ee
with $\Psi_{[\m\m_{1}]\m_{2},\g}$  mixed symmetric.
$\Psi_{[\m\m_{1}]\m_{2}}$ can easily
be decomposed as (see Appendix B for the construction and Appendix A for
the normalisation)
\bea
{\frac{5}{2}}^{+}: \;\;\;  \Psi_{[\m \m_{1}]\m_{2}} & = &
\frac{1}{\sqrt{2}}(\g_{\bot\m}\g_{5}u_{\m_{1}\m_{2}}
-\g_{\bot\m_{1}}\g_{5}u_{\m\m_{2}})
\label{dw9}        \\
{\frac{3}{2}}^{+}: \;\;\;  \Psi_{[\m \m_{1}]\m_{2}} & = &
\frac{1}{3\sqrt{5}}\{[\g_{\bot \m},\g_{\bot\m_{1}}]u_{\m_{2}}
+(\g_{\bot \m}\g_{\bot\m_{2}}u_{\m_{1}}
-\g_{\bot\m_{1}}\g_{\bot\m_{2}}u_{\m})\}\, .\nonumber\\
&&
\label{dw10}
\eea

$\;\;\;(iii)\, J_{q_{1}q_{2}}=1$
\be
[\chi_{\bot\n}^{1}N^{\{\n\m_{2}\}}]_{\alpha\beta}\Psi_{\m_{2},\g}\,.
\label{dw11}
\ee
$\Psi_{\m_{2}}$ can easily be decomposed as
\bea
{\frac{3}{2}}^{+}: \;\;\;\;\;  \Psi_{\m_{2}} & = & u_{\m_{2}}
\label{dw12}        \\
{\frac{1}{2}}^{+}: \;\;\;\;\;  \Psi_{\m_{2}} & = &
\frac{1}{\sqrt{3}}\g_{\bot \m_{2}}\g_{5}u\, .
\label{dw13}
\eea

\underline{$\;(3)\,(b)\, L_{K_{3}}=1,\, L_{k_{3}}=1,\, L=1$}

Here the total angular momentum of quarks 1 and 2 is either 2,1 or
0, i.e. $(L=1)\otimes (S_{q_{1}q_{2}}=1)=(J_{q_{1}q_{2}}=2,1,0)$. The
appropriate projectors are listed below. (See Appendix B for details of
the construction)

$\;\;\;(i)\, J_{q_{1}q_{2}}=2$
\be
[\chi_{\bot}^{1,\m}N^{[\m_{1}\m_{2}]}
-\frac{1}{2}g_{\bot}^{\m\m_{1}}\chi_{\bot\n}^{1}
N^{[\n\m_{2}]}+\frac{1}{2}g_{\bot}^{\m\m_{2}}\chi_{\bot\n}^{1}
N^{[\n\m_{1}]}]_{\alpha\beta}\Psi_{\m[\m_{1}\m_{2}],\g}\,,
\label{dw14}
\ee
with $\Psi_{\m[\m_{1}\m_{2}],\g}$ mixed symmetric. It can easily
be decomposed as
\bea
{\frac{5}{2}}^{+}: \;\;\;  \Psi_{\m [\m_{1}\m_{2}]} & = &
\frac{1}{\sqrt{2}}(\g_{\bot\m_{1}}\g_{5}u_{\m\m_{2}}
-\g_{\bot\m_{2}}\g_{5}u_{\m\m_{1}})
\label{dw15}        \\
{\frac{3}{2}}^{+}: \;\;\;  \Psi_{\m [\m_{1}\m_{2}]} & = &
\frac{1}{3\sqrt{5}}\{[\g_{\bot \m_{1}},\g_{\bot\m_{2}}]u_{\m}
+(\g_{\bot \m_{1}}\g_{\bot\m}u_{\m_{2}}-\g_{\bot\m_{2}}\g_{\bot\m}u_{\m_{1}})\}
.\nonumber\\
&&
\label{dw16}
\eea

Note that these $\frac{5}{2}^{+}$ and $\frac{3}{2}^{+}$ are slightly
different from the corresponding states in (3)(a)(ii), eqs. (\ref{dw9} and
\ref{dw10}). For the details of this construction, please see the
Appendix B.

$\;\;\;(ii)\, J_{q_{1}q_{2}}=1$
\be
(\chi_{\bot\n}^{1}N^{[\n\m_{2}]})_{\alpha\beta}\Psi_{\m_{2},\g}
\label{dw17}
\ee
$\Psi_{\m_{2}}$ can easily
be decomposed as
\bea
\frac{3}{2}^{+}: \;\;\;\;\;  \Psi_{\m_{2}} & = & u_{\m_{2}}
\label{dw18}        \\
{\frac{1}{2}}^{+}: \;\;\;\;\;  \Psi_{\m_{2}} & = &
\frac{1}{\sqrt{3}}\g_{\bot \m_{2}}\g_{5}u\, .
\label{dw19}
\eea

$\;\;\;(iii)\, J_{q_{1}q_{2}}=0$
\be
\frac{1}{3}(\chi_{\bot}^{1,\m}N^{[\m_{1}\m_{2}]}
+\chi_{\bot}^{1,\m_{1}}N^{[\m_{2}\m]}
+\chi_{\bot}^{1,\m_{2}}N^{[\m\m_{1}]})_{\alpha\beta}\Psi_{[\m\m_{1}\m_{2}],\g}\,,
\label{dw19a}
\ee
with $\Psi_{[\m\m_{1}\m_{2}]}$  fully antisymmetric. This decomposes uniquely
as
\bea
\frac{1}{2}^{+}:\;\Psi_{[\m\m_{1}\m_{2}]} & =&
\frac{1}{6\sqrt{6}}\{[\g_{\bot\m},\g_{\bot\m_{1}}]\g_{\bot\m_{2}}
+[\g_{\bot\m_{1}},\g_{\bot\m_{2}}]\g_{\bot\m}
+[\g_{\bot\m_{2}},\g_{\bot\m}]\g_{\bot\m_{1}}\}\g_{5}u\nonumber\\
&=&\frac{1}{\sqrt{6}}\g_{\bot\m}\g_{\bot\m_{1}}\g_{\bot\m_{2}}\g_{5}u\,.
\label{dw20}        \\
\eea
In the last step we have used the fact that $\Psi_{[\m\m_{1}\m_{2}]}$ is
folded into a fully antisymmetric tensor in eq. (\ref{dw19a}).

\underline{$\;(3)\,(c)\, L_{K_{3}}=1,\, L_{k_{3}}=1,\, L=0$}

Here the $L=0$ orbital angular momentum of the quarks 1 and 2 combines
with the $S_{q_{1}q_{2}}=1$ spin state to give a total angular momentum
$J_{q_{1}q_{2}}=1$ for the quark pair. The projection operator is
\be
(K_{3\bot}\cdot k_{3\bot}\chi_{\bot}^{1,\m})_{\alpha\beta}\Psi_{\m,\g}\,.
\label{dw21}
\ee
$\Psi_{\m}$ decomposes in the usual manner as
\bea
\frac{3}{2}^{+}: \;\;\;\;\;  \Psi_{\m} & = & u_{\m}
\label{dw22}        \\
{\frac{1}{2}}^{+}: \;\;\;\;\;  \Psi_{\m} & = &
\frac{1}{\sqrt{3}}\g_{\bot \m}\g_{5}u\, .
\label{dw23}
\eea

We have now come to the end of enumerating the projection operators for
the first positive parity excited $\Lambda$-type baryons.

\subsubsection{$\Sigma$-type first positive parity excitation
projection operators}

These are in some sense easier to write down. Here because the flavour
state of quarks 1 and 2 is symmetric, the symmetric orbital angular
momentum $L=2$ states (1) and (2) above combine with the symmetric spin
one state, $S_{q_{1}q_{2}}=1$, $\chi_{\bot\alpha\beta}^{1,\m}$, to
preserve overall symmetry. This gives rise to the total angular momenta,
$J_{q_{1}q_{2}}=3,2,1$, of the quark pair $q_{1}$ and $q_{2}$ with respect
to the third quark. We treat cases (1)
and (2) together. In contrast the antisymmetric orbital states (3), with
total orbital angular momentum $L=2,1$ or $0$, arising from
$L_{K_{3}}=1, L_{k_{3}}=1$ necessarily combine with the antisymmetric
spin zero state to give total $J_{q_{1}q_{2}}=2,1$ or $0$.

\underline{$\;(1)\, L_{K_{3}}=2,\, L_{k_{3}}=0,\, L=2$ or
$\;(2)\, L_{K_{3}}=0,\, L_{k_{3}}=2,\, L=2$}

$\;\;(1)\,(a)\, J_{q_{1}q_{2}}=3 $
\be
\frac{1}{3}[(\chi_{\bot}^{1,\m}N^{\m_{1}\m_{2}}
-\frac{2}{5}g_{\bot}^{\m\m_{1}}\chi_{\bot\n}^{1}
N^{\n\m_{2}})+symmetrised\;in\;\m,\m_{1},\m_{2}]_{\alpha\beta}
\Psi_{\m\m_{1}\m_{2},\g}\,,
\label{sdw1}
\ee
with $\Psi_{\m\m_{1}\m_{2},\g}$  fully symmetric.It can easily
be decomposed as
\bea
{\frac{7}{2}}^{+}: \;\;\;\;\;  \Psi_{\m \m_{1}\m_{2}} & = &
u_{\m\m_{1}\m_{2}}
\label{sdw2}        \\
{\frac{5}{2}}^{+}: \;\;\;\;\;  \Psi_{\m \m_{1}\m_{2}} & = &
\frac{1}{\sqrt{21}}\g_{\bot \{\m}\g_{5}u_{\m_{1}\m_{2}\}}\nonumber\\
& = & \sqrt{\frac{3}{7}}\g_{\bot\m}\g_{5}u_{\m_{1}\m_{2}}\, .
\label{sdw3}
\eea
Again in the last step we have used the fact that in eq. (\ref{sdw1})
$\Psi_{\m\m_{1}\m_{2}}$ is traced into a fully symmetric tensor.

$\;\;(1)\,(b)\, J_{q_{1}q_{2}}=2 $
\be
\frac{1}{3}[\chi_{\bot}^{1,\m}N^{\m_{1}\m_{2}}
-\chi_{\bot}^{1,\m_{1}}N^{\m\m_{2}}
-\frac{1}{2}g_{\bot}^{\m\m_{2}}\chi_{\bot\n}^{1}
N^{\n\m_{1}}+\frac{1}{2}g_{\bot}^{\m_{1}\m_{2}}\chi_{\bot\n}^{1}
N^{\n\m}]_{\alpha\beta}\Psi_{[\m\m_{1}]\m_{2},\g}
\label{sdw4}
\ee
with $\Psi_{[\m\m_{1}]\m_{2},\g}$  mixed symmetric. As usual
it decomposes as (see Appendix B for construction)
\bea
{\frac{5}{2}}^{+}: \;\;\;  \Psi_{[\m \m_{1}]\m_{2}} & = &
\frac{1}{\sqrt{2}}(\g_{\bot\m}\g_{5}u_{\m_{1}\m_{2}}
-\g_{\bot\m_{1}}\g_{5}u_{\m\m_{2}})
\label{sdw5}        \\
{\frac{3}{2}}^{+}: \;\;\;  \Psi_{[\m \m_{1}]\m_{2}} & = &
\frac{1}{3\sqrt{5}}\{[\g_{\bot \m},\g_{\bot\m_{1}}]u_{\m_{2}}
+(\g_{\bot\m}\g_{\bot\m_{2}}u_{\m_{1}}
-\g_{\bot\m_{1}}\g_{\bot\m_{2}}u_{\m})\}\, .\nonumber\\
&&
\label{sdw6}
\eea

$\;\;(1)\,(c)\, J_{q_{1}q_{2}}=1 $
\be
[\chi_{\bot\n}^{1}N^{\n\m}]_{\alpha\beta}\Psi_{\m,\g}\,,
\label{sdw7}
\ee
with $\Psi_{\m,\g}$ being written as usual as
\bea
{\frac{3}{2}}^{+}: \;\;\;\;\;  \Psi_{\m} & = & u_{\m}
\label{sdw8}        \\
{\frac{1}{2}}^{+}: \;\;\;\;\;  \Psi_{\m} & = &
\frac{1}{\sqrt{3}}\g_{\bot \m}\g_{5}u \, .
\label{sdw9}
\eea

\underline{$\;(3)\, L_{K_{3}}=1,\, L_{k_{3}}=1$}

$\;\;(3)\,(a)\,L=2,\, J_{q_{1}q_{2}}=2$
\be
[\chi^{0}N^{\{\m_{1}\m_{2}\}}]_{\alpha\beta}\Psi_{\m_{1}\m_{2},\g}
\label{sdw10}
\ee
with $N^{\{\m_{1}\m_{2}\}}$ as defined in eq. (\ref{dw1}).
$\Psi_{\m_{1}\m_{2}}$ is symmetric and is as usual given by
\bea
{\frac{5}{2}}^{+}: \;\;\;\;\;  \Psi_{\m_{1}\m_{2}} & = & u_{\m_{1}\m_{2}}
\label{sdw11}        \\
{\frac{3}{2}}^{+}: \;\;\;\;\;  \Psi_{\m_{1}\m_{2}} & = &
\frac{1}{\sqrt{10}}(\g_{\bot \m_{1}}\g_{5}u_{\m_{2}}
+\g_{\bot \m_{2}}\g_{5}u_{\m_{1}})\, .
\label{sdw12}
\eea

$\;\;(3)\,(b)\,L=1,\, J_{q_{1}q_{2}}=1$
\be
(\chi^{0}N^{[\m_{1}\m_{2}]})_{\alpha\beta}\Psi_{[\m_{1}\m_{2}],\g}\,,
\label{sdw13}
\ee
with $N^{[\m_{1}\m_{2}]}$ as defined in eq. (\ref{dw2}).
$\Psi_{[\m_{1}\m_{2}]}$ is antisymmetric and is given by
\bea
{\frac{3}{2}}^{+}: \;\;\;\;\;  \Psi_{[\m_{1}\m_{2}]} & = &
\frac{1}{\sqrt{2}}(\g_{\bot\m_{1}}\g_{5}u_{\m_{2}}
-\g_{\bot\m_{2}}\g_{5}u_{\m_{1}})
\label{sdw14}        \\
{\frac{1}{2}}^{+}: \;\;\;\;\;  \Psi_{[\m_{1}\m_{2}]} & = &
\frac{1}{2\sqrt{6}}[\g_{\bot \m_{1}},\g_{\bot\m_{2}}]u\, .
\label{sdw15}
\eea

$\;\;(3)(c)L=0, J_{q_{1}q_{2}}=0$
\be
(\chi^{0}K_{3\bot}\cdot k_{3\bot})_{\alpha\beta}\Psi_{\g}\,,
\label{sdw16}
\ee
with $\Psi$  given by
\be
{\frac{1}{2}}^{+}: \;\;\;\;\;  \Psi  = u\,. \label{sdw{17}}
\ee

\subsection{$(-1)^{L}$ Parity Excitation Baryon Projection Operators}

By $(-1)^{L}$ parity excitations we mean all those states with
$L_{K_{3}}+L_{k_{3}}=L$. These contain total orbital angular
momenta starting from $L$ down to $1$ if $L$ is odd and $0$ if $L$ is
even. The situation for the general L-wave resonances becomes rather
complicated because of the large number of possible states due to the
the $L+1$ partitions of L, arising from
the choice of the angular momenta $L_{K_{3}}$ and $L_{k_{3}}$. We will
not do the full construction here but just indicate how one goes about
it.

Let us take $L_{k_{3}}=n$. Then for the $(-1)^{L}$ parity excitations we
must have $L_{K_{3}}=L-n$. A particular partition of $L$, i.e. a
particular choice of $n,\,0<n<L$, signifies taking the direct product of
the symmetric tensor $N^{\m_{1}\ldots\m_{n}}(k_{3})$ with the other
symmetric tensor $N^{\m_{1}\ldots\m_{L-n}}(K_{3})$. This direct product
can be decomposed into the direct sum of orbital angular momenta
ranging from $L$ to $\vert L-2n\vert$ corresponding to the various Young
tableaux present in the reduction of the product of two fully symmetric
Young tableaux. All these different irreducible, traceless product
tensors will have symmetry $(-)^{n}$ under the interchange
$p_{1}\leftrightarrow p_{2}$, i.e. they will be symmetric if $n$ is even
and antisymmetric if $n$ is odd. Thus one can differentiate the states
arising from $n$ even or odd.

\underline{(i)(a) $n$ even. $\Lambda$-type baryon}

In this case to preserve overall symmetry of the wave function the
irreducible orbital angular momentum tensors would have to be combined
with the antisymmetric spin zero, $S_{q_{1}q_{2}}=0$, state of the
$q_{1}q_{2}$ pair represented by $\chi_{\alpha \beta}^{0}$. This gives
rise to the following total angular momenta, $J_{q_{1}q_{2}}$, of the
quark pair, $q_{1}q_{2}$, with respect to the third quark,
\be
J_{q_{1}q_{2}}=L,L-1, \ldots \vert L-2n\vert\,.
\ee
Then combining these with the spin of the third quark
$S_{q_{3}}=\frac{1}{2}$ gives the total angular momentum (spin) of the baryon.

\underline{(i)(b) $n$ even. $\Sigma$-type baryon}

In this case to preserve overall symmetry the orbital angular momentum
tensors would have to be combined with the symmetric spin one,
$S_{q_{1}q_{2}}=1$, state of the quark pair $q_{1}q_{2}$ represented by
$\chi_{\bot\alpha \beta}^{1,\m}$. Such an operation will give rise to
the following groupings of the total angular momentum of the pair
$q_{1}q_{2}$ with respect to the third quark,
\be
J_{q_{1}q_{2}}=(L+1,L,L-1);\,(L,L-1,L-2);\,\ldots (\vert
L-2n\vert+1,\vert L-2n\vert ,\vert L-2n\vert -1)\,.
\ee
Again the total angular momentum (spin) of the baryon is obtained by
combining finally with $S_{q_{3}}=\frac{1}{2}$.

\underline{(ii)(a) $n$ odd. $\Lambda$-type baryon}

Now the situation is reversed. To preserve overall symmetry the orbital
tensors have to be combined with $\chi_{\alpha \beta}^{1,\m}$ in the
usual manner as in (i)(b) above.

\underline{(ii)(b) $n$ odd. $\Sigma$-type baryon}

Here the orbital tensors are combined with $\chi_{\alpha \beta}^{0}$ to
preserve symmetry as in (i)(a) above.

All these tensors representing the different values of $J_{q_{1}q_{2}}$
will have $L$ or $L+1$ indices of various symmetry types. Now to
construct the projection operators one has to multiply them into tensor
spinors $\Psi_{\m_{1}\ldots\m_{N},\g}$ (where $N$ is either $L$ or
$L+1)$ of the corresponding symmetry. Because of the Bargmann-Wigner
equations these must satisfy the Dirac equation and the transversality
condition
\be
v^{\m_{1}}\Psi_{\m_{1}\ldots\m_{N}}=0\,
\ee
on each index.

These are easy to construct, in any particular case, by noting that if
$\f$ satisfies the Dirac equation then so do $\g_{\bot\m}\g_{5}\f$ and
$\g_{\bot\m_{1}}\g_{\bot\m_{2}}\f$ plus obviously being transverse. One
then has to use appropriately symmetrised chains of $\g_{\bot\m}\g_{5}$
and/or $\g_{\bot\m_{1}}\g_{\bot\m_{2}}$ along with generalised traceless,
symmetric Rarita-Schwinger spinors $u_{\m_{1}\ldots\m_{m}}$ to
construct the projection operators.

We have illustrated now how to construct any arbitrary orbital excited
baryonic projection operator. As usual they will come in pairs which
correspond to the heavy quark basis. Note that in
the case $L=2$, treated above, the two partitions
$L_{K_{3}}=2,\,\, L_{k_{3}}=0$
and $L_{K_{3}}=0,\,\,L_{k_{3}}=2$ give rise to the same Lorentz
structure and to the same kind of states. Similarly from the $L+1$
partitions in the L-wave case only a certain number can give rise to
possible different kinds of states. When L is odd there are $(L+1)/2$
partitions giving rise to possible different Lorentz structures, whereas
when L is even there are $(L+2)/2$ possibilities.

Writing down all the projection operators in the general case is thus
rather cumbersome so we shall not pursue it further. However the highest
weight L-wave states are relatively easier to write down.
We will not list them here but will show them in the next subsection
when we put everything together in the final form of the Bethe-Salpeter
amplitudes.

\subsection{Heavy Baryon Bethe-Salpeter Amplitudes}

The heavy baryon B-S amplitudes are now obtained from
eq. (\ref{4}) by substituting with the appropriate projection operators
developed in subsections {\bf \ref{sub-swave}} to {\bf \ref{sub-dwave}} above.
The $B_{\alpha
\beta \g}$ now satisfy the Bargmann-Wigner equation on one label $\g$.
We identify the third quark with the heavy quark. Of course,we are
listing just the ``pure" spin states characterized by definite values of
the orbital angular momentum. In general the physical mass
eigenstates  may be mixtures of states preserving the heavy quark spin
symmetry.

\subsubsection{S-wave Heavy Baryons}
\label{sub-sheavy}

We define the following quantities
\bea
S_{\alpha \beta}^{0} & = & \chi_{\delta \r}^{0}A_{\alpha \beta}^{\delta
\r}\, , \nonumber \\
S_{\alpha \beta}^{\m} & = & \chi_{\bot \delta \r}^{1, \m}A_{\alpha
\beta}^{\delta \r}\, .
\label{shbs1}
\eea
$S^{0}$ and $S^{\m}$ are the projections of the $0^{+}$ and
$1^{+}$ total ($J_{q_{1}q_{2}}$) angular momenta of the light quark
pair, $q_{1},q_{2}$ with repect to the heavy quark. These are functions of
$K_{3}$ and $k_{3}$. We can
now write the s-wave heavy B-S amplitude in compact notation, dropping the
Dirac
indices $\alpha, \beta$ and $\g$. However remember that the light quark
indices are on the $S$'s whereas the heavy quark index $\g$ is always on
the explicit spinor structure. Thus
\bea
\Lambda-type:  &      &       \nonumber    \\
                       & {\frac{1}{2}}^{+}:   & S^{0}u \nonumber  \\
\Sigma-type:   &      &       \nonumber    \\
                       & {\frac{3}{2}}^{+}:  &  S^{\m}u_{\m} \nonumber \\
                       & {\frac{1}{2}}^{+}: &
S^{\m}\frac{1}{\sqrt{3}}\g_{\bot \m}\g_{5}u\,.   \label{shbs2}
\eea

\subsubsection{P-wave Heavy Baryons}
\label{sub-pheavy}

For the p-wave case, we define the following quantities:
\bea
P_{k;\alpha \beta}^{\m} & = & k_{\bot}^{\m}\chi^{0}_{\delta\r}
A^{\delta \r}_{\alpha\beta}\, ,\nonumber \\
P_{k;\alpha \beta}^{\{\m \n \}} & = &
\frac{1}{2}(k_{\bot}^{\m}\chi_{\bot}^{1,\n} +
k_{\bot}^{\n}\chi_{\bot}^{1,\m} -
\frac{2}{3}k_{\bot}\cdot\chi_{\bot}g_{\bot}^{\m \n})_{\delta \r}A^{\delta
\r}_{\alpha \beta}\, ,  \nonumber \\
P_{k;\alpha \beta}^{[\m \n ]} & = &
\frac{1}{2}(k_{\bot}^{\m}\chi_{\bot}^{1,\n}
- k_{\bot}^{\n}\chi_{\bot}^{1,\m})_{\delta \r}A^{\delta \r}_{\alpha
\beta}\, ,
\nonumber \\
P_{k;\alpha \beta} & = & (k_{\bot}\cdot\chi_{\bot}^{1})_{\delta
\r}A^{\delta \r}_{\alpha \beta}\, .
\label{phbs1}
\eea
$P_{k}^{\m},P_{k}^{\{\m\n\}},P_{k}^{[\m\n]}$ and $P_{k}$ are the
projections of the $1^{-};2^{-},1^{-}$ and $0^{-}$ total angular
momenta respectively of the light quark pair. Of course
these $P$'s are functions of both $k_{3}$ and $K_{3}$. The subscript
$k$ in the $P$'s is to be understood as a label indicating which
$k$ $(k_{3}$ or $K_{3})$ we use in the definitions (\ref{phbs1}).

The pairs of states listed below are separately degenerate because of
the heavy quark spin symmetry.

$\Lambda-type$:

\underline{(i) $L_{K_{3}}=1,\,S_{q_{1}q_{2}}=0,\,J_{q_{1}q_{2}}=1$.}
\bea
      & {\frac{3}{2}}^{-}:  & P_{K_{3}}^{\m}u_{\m} \nonumber \\
      & {\frac{1}{2}}^{-}:  & P_{K_{3}}^{\m}
\frac{1}{\sqrt{3}}\g_{\bot \m}\g_{5}u \label{72}
\eea

\underline{(ii) $L_{k_{3}}=1,\,S_{q_{1}q_{2}}=1,\,J_{q_{1}q_{2}}=2,1,0$.}

\hspace{.5cm}(a) $J_{q_{1}q_{2}} =2$
\bea
     &  {\frac{5}{2}}^{-}:  & P_{k_{3}}^{\{\m \n\}}u_{\m
\n}\nonumber \\
     &  {\frac{3}{2}}^{-}:  & P_{k_{3}}^{\{\m \n\}}\frac{1}{\sqrt{10}}
(\g_{\bot \m}\g_{5}u_{\n}+\g_{\bot \n}\g_{5}u_{\m})
\label{73}
\eea

\hspace{.5cm}(b) $J_{q_{1}q_{2}}=1$.
\bea
     &  {\frac{3}{2}}^{-}:  & P_{k_{3}}^{[\m \n]}\frac{1}{\sqrt{2}}
(\g_{\bot \m}\g_{5}u_{\n}-\g_{\bot \n}\g_{5}u_{\m}) \nonumber \\
     &  {\frac{1}{2}}^{-}:  & P_{k_{3}}^{[\m \n ]}\frac{1}{2\sqrt{6}}
[\g_{\bot \m},\g_{\bot \n}]u
\label{74}
\eea

\hspace{.5cm}(c) $J_{q_{1}q_{2}}=0$.
\bea
   & {\frac{1}{2}}^{-}:   & P_{k_{3}}u\,.
\label{75}
\eea

For the $\Sigma$-type heavy baryons one has to simply interchange
$k_{3}$ and $K_{3}$ in eqs. (\ref{72}) - (\ref{75}).

\subsubsection{First Positive Parity Excited Heavy Baryons}
\label{sub-dheavy}

{\bf A. First Positive Parity Excited Heavy $\Lambda$}

For the first positive parity excited heavy $\Lambda$ the projections
of the possible total angular momenta of the light quark pair are
represented by the following matrices:

\begin{equation}
\begin{array}{llcl}
2^{+}& D_{k;\alpha\beta}^{\m_{1}\m_{2}} & = &
N^{\m_{1}\m_{2}}(k)\chi^{0}_{\delta\r}A^{\delta\r}_{\alpha\beta}, \\
3^{+}& D_{\Lambda;\alpha\beta}^{\{\m\m_{1}\m_{2}\}} & =&
\frac{1}{3}[(\chi_{\bot}^{1,\m}N^{\{\m_{1}\m_{2}\}}
-\frac{2}{5}g_{\bot}^{\m\m_{1}}\chi_{\bot\n}^{1}N^{\{\n\m_{2}\}})+
symmetrised\;(\m,\m_{1},\m_{2})]_{\delta\r}A^{\delta\r}_{\alpha\beta}, \\
2^{+}& D_{\Lambda;\alpha\beta}^{[\m\m_{1}]\m_{2}} & =&
\frac{1}{3}[\chi_{\bot}^{1,\m}N^{\{\m_{1}\m_{2}\}}
-\chi_{\bot}^{1,\m_{1}}N^{\{\m\m_{2}\}}\\
&&&-\frac{1}{2}g_{\bot}^{\m\m_{2}}\chi_{\bot\n}^{1}N^{\{\n\m_{1}\}}
+\frac{1}{2}g_{\bot}^{\m_{1}\m_{2}}\chi_{\bot\n}^{1}
N^{\{\n\m\}}]_{\delta\r}A^{\delta\r}_{\alpha\beta} , \\
1^{+}& D_{S;\alpha\beta}^{\m} & = & [\chi_{\bot\n}^{1}N^{\{\n\m\}}]_{\delta\r}
A^{\delta\r}_{\alpha\beta}, \\
2^{+}& D_{\Lambda;\alpha\beta}^{\m[\m_{1}\m_{2}]} & = &
[\chi_{\bot}^{1,\m}N^{[\m_{1}\m_{2}]}
-\frac{1}{2}g_{\bot}^{\m\m_{1}}\chi_{\bot\n}^{1}N^{[\n\m_{2}]}
+\frac{1}{2}g_{\bot}^{\m\m_{2}}\chi_{\bot\n}^{1}
N^{[\n\m_{1}]}]_{\delta\r}A^{\delta\r}_{\alpha\beta}\, ,\\
2^{+}& D_{A;\alpha\beta}^{\m} & = & [\chi_{\bot\n}^{1}N^{[\n\m]}]_{\delta\r}
A^{\delta\r}_{\alpha\beta}, \\
0^{+}& D_{\alpha\beta}^{[\m\m_{1}\m_{2}]} & =&
\frac{1}{3}[\chi_{\bot}^{1,\m}N^{[\m_{1}\m_{2}]}+
\chi_{\bot}^{1,\m_{1}}N^{[\m_{2}\m]}
+\chi_{\bot}^{1,\m_{2}}N^{[\m\m_{1}]}]_{\delta\r}
A^{\delta\r}_{\alpha\beta}, \\
1^{+}& D_{T;\alpha\beta}^{\m} & =&  K_{3\bot}\cdot
k_{3\bot}\chi_{\bot\delta\r}^{1,\m}
A^{\delta\r}_{\alpha\beta}\,.
\end{array}
\end{equation}

With the help of these matrices, we can now write down the B-S
amplitudes for the pairs of
degenerate first positive parity excited heavy  $\Lambda$ states.

$\;(1)\,\,\, L_{K_{3}}=2,\, L_{k_{3}}=0,\, L=2,
\,S_{q_{1}q_{2}}=0,\,J_{q_{1}q_{2}}=2$
\bea
      & {\frac{5}{2}}^{+}:  & D_{K_{3}}^{\m_{1}\m_{2}}u_{\m_{1}\m_{2}}
\nonumber \\
      & {\frac{3}{2}}^{+}:  & D_{K_{3}}^{\m_{1}\m_{2}}
\frac{1}{\sqrt{10}}(\g_{\bot
\m_{1}}\g_{5}u_{\m_{2}}+\g_{\bot\m_{2}}\g_{5}u_{\m_{1}})
 \label{dhbs1}
\eea

$\;(2)\,\,\, L_{K_{3}}=0,\, L_{k_{3}}=2,\, L=2,
\,S_{q_{1}q_{2}}=0,\,J_{q_{1}q_{2}}=2$
\bea
      & {\frac{5}{2}}^{+}:  & D_{k_{3}}^{\m_{1}\m_{2}}u_{\m_{1}\m_{2}}
\nonumber \\
      & {\frac{3}{2}}^{+}:  & D_{k_{3}}^{\m_{1}\m_{2}}
\frac{1}{\sqrt{10}}(\g_{\bot
\m_{1}}\g_{5}u_{\m_{2}}+\g_{\bot\m_{2}}\g_{5}u_{\m_{1}})
 \label{dhbs2}
\eea

$\;(3)\,(a)\,(i)\,\,\, L_{K_{3}}=1,\, L_{k_{3}}=1,\, L=2,
\,S_{q_{1}q_{2}}=1,\,J_{q_{1}q_{2}}=3$
\bea
      & {\frac{7}{2}}^{+}:  &
D_{\Lambda}^{\{\m\m_{1}\m_{2}\}}u_{\m\m_{1}\m_{2}} \nonumber \\
      & {\frac{5}{2}}^{+}:  & D_{\Lambda}^{\{\m\m_{1}\m_{2}\}}
\sqrt{\frac{3}{7}}\g_{\bot\m}\g_{5}u_{\m_{1}\m_{2}}
 \label{dhbs3}
\eea

$\;(3)\,(a)\,(ii)\,\,\, L_{K_{3}}=1,\, L_{k_{3}}=1,\, L=2,
\,S_{q_{1}q_{2}}=1,\,J_{q_{1}q_{2}}=2$
\bea
      & {\frac{5}{2}}^{+}:  & D_{\Lambda}^{[\m\m_{1}]\m_{2}}
\frac{1}{\sqrt{2}}(\g_{\bot\m}\g_{5}u_{\m_{1}\m_{2}}-\g_{\bot\m_{1}}
\g_{5}u_{\m\m_{2}}) \nonumber \\
      & {\frac{3}{2}}^{+}:  & D_{\Lambda}^{[\m\m_{1}]\m_{2}}
\frac{1}{3\sqrt{5}}\{[\g_{\bot\m},\g_{\bot\m_{1}}]u_{\m_{2}}+
(\g_{\bot\m}\g_{\bot\m_{2}}u_{\m_{1}}
-\g_{\bot\m_{1}}\g_{\bot\m_{2}}u_{\m})\}
 \label{dhbs4}
\eea

$\;(3)\,(a)\,(iii)\,\,\, L_{K_{3}}=1,\, L_{k_{3}}=1,\, L=2,
\,S_{q_{1}q_{2}}=1,\,J_{q_{1}q_{2}}=1$
\bea
      & {\frac{3}{2}}^{+}:  & D_{S}^{\m}u_{\m} \nonumber \\
      & {\frac{1}{2}}^{+}:  & D_{S}^{\m}\frac{1}{\sqrt{3}}\g_{\bot\m}\g_{5}u
 \label{dhbs5}
\eea

$\;(3)\,(b)\,(i)\,\,\, L_{K_{3}}=1,\, L_{k_{3}}=1,\, L=1,
\,S_{q_{1}q_{2}}=1,\,J_{q_{1}q_{2}}=2$
\bea
      & {\frac{5}{2}}^{+}:  & D_{\Lambda}^{\m[\m_{1}\m_{2}]}
\frac{1}{\sqrt{2}}(\g_{\bot\m_{1}}\g_{5}u_{\m\m_{2}}-\g_{\bot\m_{2}}
\g_{5}u_{\m\m_{1}}) \nonumber \\
      & {\frac{3}{2}}^{+}:  & D_{\Lambda}^{\m[\m_{1}\m_{2}]}
\frac{1}{3\sqrt{5}}\{[\g_{\bot\m_{1}},\g_{\bot\m_{2}}]u_{\m}+
(\g_{\bot\m_{1}}\g_{\bot\m}u_{\m_{2}}
-\g_{\bot\m_{2}}\g_{\bot\m}u_{\m_{1}})\}
 \label{dhbs6}
\eea

$\;(3)\,(b)\,(ii)\,\,\, L_{K_{3}}=1,\, L_{k_{3}}=1,\, L=1,
\,S_{q_{1}q_{2}}=1,\,J_{q_{1}q_{2}}=1$
\bea
      & {\frac{3}{2}}^{+}:  & D_{A}^{\m}u_{\m} \nonumber \\
      & {\frac{1}{2}}^{+}:  & D_{A}^{\m}\frac{1}{\sqrt{3}}\g_{\bot\m}\g_{5}u
 \label{dhbs7}
\eea

$\;(3)\,(b)\,(iii)\,\,\, L_{K_{3}}=1,\, L_{k_{3}}=1,\, L=1,
\,S_{q_{1}q_{2}}=1,\,J_{q_{1}q_{2}}=0$
\bea
          {\frac{1}{2}}^{+}:  & D^{[\m\m_{1}\m_{2}]}\frac{1}{6\sqrt{6}}
\{[\g_{\bot\m},\g_{\bot\m_{1}}]\g_{\bot\m_{2}}
+[\g_{\bot\m_{1}},\g_{\bot\m_{2}}]\g_{\bot\m}
+[\g_{\bot\m_{2}},\g_{\bot\m}]\g_{\bot\m_{1}}\}\g_{5}u &  \nonumber \\
          &=D^{[\m\m_{1}\m_{2}]}
 \frac{1}{\sqrt{6}}\g_{\bot\m}\g_{\bot\m_{1}}\g_{\bot\m_{2}}\g_{5}u &
\label{dhbs8}
\eea

$\;(3)\,(c)\,\,\, L_{K_{3}}=1,\, L_{k_{3}}=1,\, L=0,
\,S_{q_{1}q_{2}}=1,\,J_{q_{1}q_{2}}=1$
\bea
      & {\frac{3}{2}}^{+}:  & D_{T}^{\m}u_{\m} \nonumber \\
      & {\frac{1}{2}}^{+}:  & D_{T}^{\m}\frac{1}{\sqrt{3}}\g_{\bot\m}\g_{5}u\,.
 \label{dhbs9}
\eea

{\bf B. First Positive Parity Excited Heavy $\Sigma$}

For the first positive parity heavy $\Sigma$ the projections of
the possible total angular momenta of the light quark pair are represented
by the following matrices:
\be
\begin{array}{llcl}
3^{+}&D_{\Sigma;k;\alpha\beta}^{\m\m_{1}\m_{2}}&=&
\frac{1}{3}[(\chi_{\bot}^{1,\m}N^{\m_{1}\m_{2}}(k)
-\frac{2}{5}g_{\bot}^{\m\m_{1}}\chi_{\bot\n}^{1}N^{\n\m_{2}}(k))+
symmetrised\;(\m,\m_{1},\m_{2})]_{\delta\r}A^{\delta\r}_{\alpha\beta}, \\
2^{+}& D_{\Sigma;k;\alpha\beta}^{[\m\m_{1}]\m_{2}} & =&
\frac{1}{3}[\chi_{\bot}^{1,\m}N^{\m_{1}\m_{2}}(k)
-\chi_{\bot}^{1,\m_{1}}N^{\m\m_{2}}(k)\\
&&&-\frac{1}{2}g_{\bot}^{\m\m_{2}}\chi_{\bot\n}^{1}N^{\n\m_{1}}(k)
+\frac{1}{2}g_{\bot}^{\m_{1}\m_{2}}\chi_{\bot\n}^{1}
N^{\n\m}(k)]_{\delta\r}A^{\delta\r}_{\alpha\beta} , \\
1^{+}& D_{\Sigma;k;\alpha\beta}^{\m} & = &
[\chi_{\bot\n}^{1}N^{\n\m}(k)]_{\delta\r}
A^{\delta\r}_{\alpha\beta}, \\
2^{+}& D_{\Sigma;\alpha\beta}^{\{\m_{1}\m_{2}\}}&=&
N^{\{\m_{1}\m_{2}\}}\chi^{0}_{\delta\r}A^{\delta\r}_{\alpha\beta}, \\
1^{+}& D_{\Sigma;\alpha\beta}^{[\m_{1}\m_{2}]}&=&
N^{[\m_{1}\m_{2}]}\chi^{0}_{\delta\r}A^{\delta\r}_{\alpha\beta}, \\
0^{+}& D_{\Sigma;\alpha\beta}&=& K_{3\bot}\cdot k_{3\bot}
\chi^{0}_{\delta\r}A^{\delta\r}_{\alpha\beta}.
\end{array}
\ee

One can now write down the B-S amplitudes for the pairs of degenerate
first positive parity excited heavy $\Sigma$ baryons.

$\;(1)\,(a)\,\,\, L_{K_{3}}=2,\, L_{k_{3}}=0,\, L=2,
\,S_{q_{1}q_{2}}=1,\,J_{q_{1}q_{2}}=3$
\bea
      & {\frac{7}{2}}^{+}:  & D_{\Sigma;K_{3}}^{\m\m_{1}\m_{2}}
u_{\m\m_{1}\m_{2}} \nonumber \\
      & {\frac{5}{2}}^{+}:  & D_{\Sigma;K_{3}}^{\m\m_{1}\m_{2}}
\sqrt{\frac{3}{7}}\g_{\bot\m}\g_{5}u_{\m_{1}\m_{2}}
 \label{dhbs10}
\eea

$\;(1)\,(b)\,\,\, L_{K_{3}}=2,\, L_{k_{3}}=0,\, L=2,
\,S_{q_{1}q_{2}}=1,\,J_{q_{1}q_{2}}=2$
\bea
      & {\frac{5}{2}}^{+}:  & D_{\Sigma;K_{3}}^{[\m\m_{1}]\m_{2}}
\frac{1}{\sqrt{2}}(\g_{\bot\m}\g_{5}u_{\m_{1}\m_{2}}-\g_{\bot\m_{1}}
\g_{5}u_{\m\m_{2}}) \nonumber \\
      & {\frac{3}{2}}^{+}:  & D_{\Sigma;K_{3}}^{[\m\m_{1}]\m_{2}}
\frac{1}{3\sqrt{5}}\{[\g_{\bot\m},\g_{\bot\m_{1}}]u_{\m_{2}}+
(\g_{\bot\m}\g_{\bot\m_{2}}u_{\m_{1}}
-\g_{\bot\m_{1}}\g_{\bot\m_{2}}u_{\m})\}
 \label{dhbs11}
\eea

$\;(1)\,(c)\,\,\, L_{K_{3}}=2,\, L_{k_{3}}=0,\, L=2,
\,S_{q_{1}q_{2}}=1,\,J_{q_{1}q_{2}}=1$
\bea
      & {\frac{3}{2}}^{+}:  & D_{\Sigma;K_{3}}^{\m}u_{\m} \nonumber \\
      & {\frac{1}{2}}^{+}:  &
D_{\Sigma;K_{3}}^{\m}\frac{1}{\sqrt{3}}\g_{\bot\m}\g_{5}u\,.
 \label{dhbs12}
\eea

The states arising from the orbital angular momentum case (2),
$L_{K_{3}}=0,\,L_{k_{3}}=2$ are identical to the above except for the
replacement of $K_{3}$ by $k_{3}$ in the D's.

$\;(3)\,(a)\,\,\, L_{K_{3}}=1,\, L_{k_{3}}=1,\, L=2,
\,S_{q_{1}q_{2}}=0,\,J_{q_{1}q_{2}}=2$
\bea
      & {\frac{5}{2}}^{+}:  & D_{\Sigma}^{\{\m_{1}\m_{2}\}}u_{\m_{1}\m_{2}}
\nonumber \\
      & {\frac{3}{2}}^{+}:  & D_{\Sigma}^{\{\m_{1}\m_{2}\}}
\frac{1}{\sqrt{10}}(\g_{\bot
\m_{1}}\g_{5}u_{\m_{2}}+\g_{\bot\m_{2}}\g_{5}u_{\m_{1}})
 \label{dhbs13}
\eea

$\;(3)\,(b)\,\,\, L_{K_{3}}=1,\, L_{k_{3}}=1,\, L=1,
\,S_{q_{1}q_{2}}=0,\,J_{q_{1}q_{2}}=1$
\bea
      & {\frac{3}{2}}^{+}:  & D_{\Sigma}^{[\m_{1}\m_{2}]}
\frac{1}{\sqrt{2}}(\g_{\bot\m_{1}}\g_{5}u_{\m_{2}}
-\g_{\bot\m_{2}}\g_{5}u_{\m_{1}}) \nonumber \\
      & {\frac{1}{2}}^{+}:  & D_{\Sigma}^{[\m_{1}\m_{2}]}
\frac{1}{2\sqrt{6}}[\g_{\bot\m_{1}},\g_{\bot\m_{2}}]u
 \label{dhbs14}
\eea

$\;(3)\,(c)\,\,\, L_{K_{3}}=1,\, L_{k_{3}}=1,\, L=0,
\,S_{q_{1}q_{2}}=0,\,J_{q_{1}q_{2}}=0$
\bea
      & {\frac{1}{2}}^{+}:  & D_{\Sigma}u\,.
 \label{dhbs15}
\eea
\subsubsection{L-wave Heavy Baryons}
\label{sub-Lheavy}

Although, as we have pointed out above, the general case for any arbitrary
higher excitation is rather involved, we can easily generalise these results
to the ``highest weight" states of any orbital excitation. In
such a case, the total orbital angular momentum $L$ will be represented by the
transverse, symmetric, traceless product of $L$ momenta (both $k_{3 \bot}$ or
$K_{3 \bot}$ can appear). This will be an obvious generalisation of eq.
(\ref{33}).
It does not matter which of the different, $L+1$,
possibilities we choose as long as we are only concerned with the
Lorentz structure and not with the flavour. Let us continue to call such
a product of $L$ momenta, $k$, as $N^{\m_{1}\m_{2}\ldots\m_{L}}$. This
orbital angular momenta has then to be combined, on the light side, with
the total spin of the light quarks, which is either $0$ or $1$ and which
is represented as before by $\chi_{\alpha \beta}^{0}$ or $\chi_{\bot,
\alpha \beta}^{1,\m}$.

Thus one will get the following states, all of parity $(-1)^{L}$ :

\be
\begin{array}{rll}
((S_{q_{1}q_{2}}=0\otimes L) =(J_{q_{1}q_{2}}=L))\otimes (S_{Q}=1/2)&=&
L+\frac{1}{2}, L-\frac{1}{2} \\
&& \\
((S_{q_{1}q_{2}}=1\otimes L)
=(J_{q_{1}q_{2}}=L+1,L,L-1))\otimes(S_{Q}=1/2)&=&       \\
(L+\frac{3}{2},L+\frac{1}{2})\oplus(L+\frac{1}{2},L-\frac{1}{2})
\oplus(L-\frac{1}{2},L-\frac{3}{2})\,.& &
\end{array}
\ee

The total light angular momentum projections $(J_{q_{1}q_{2}}=L+1,L,L-1)$
arising from the orbital angular momentum, $L$, and the light quark
spin $1$ , is obtained in the usual way by taking the symmetric,
antisymmetric product and trace of $N^{\m_{1}\m_{2}\ldots\m_{L}}$ with
$\chi_{\bot}^{1,\m}$. For details of the mixed symmetric tensor, see
Appendix B. Define
\bea
\f_{\alpha \beta}^{\m_{1}\m_{2}\ldots\m_{L}} & = &
N^{\m_{1}\m_{2}\ldots\m_{L}}\chi_{\delta \r}^{0}A_{\alpha \beta}^{\delta
\r}\label{78a} \\
\f_{\alpha \beta}^{\{\m
\m_{1}\m_{2}\dots\m_{L}\}}&=&[\chi_{\bot}^{1,\m}N^{\m_{1}\m_{2}\ldots\m_{L}}
+\sum_{i=1}^{L}\chi_{\bot}^{1,\m_{i}}N^{\m_{1}\ldots\m_{i-1}\m\m_{i=1}\ldots\m_{L}}
\nonumber \\
&-&\frac{2}{2L+1}\sum_{i=1}^{L}g_{\bot}^{\m
\m_{i}}\chi_{\bot\n}^{1}N^{\n\m_{1}\ldots\m_{i-1}
\m_{i+1}\ldots\m_{L}}\nonumber \\
&-&\frac{2}{2L+1}\sum_{i,j}g_{\bot}^{\m_{i}\m_{j}}\chi_{\bot\n}^{1}
N^{\m_{1}\ldots\m_{i-1}\m\m_{i+1}\ldots\m_{j-1}\n\m_{j+1}\ldots\m_{L}}
]_{\delta \r}A_{\alpha \beta}^{\delta
\r}\nonumber\\
&&
\label{78b}\\
\f_{\alpha
\beta}^{[\m\m_{1}]\m_{2}\ldots\m{L}}&=&\frac{L}{2(L+1)}[\chi_{\bot}^{1,\m}
N^{\m_{1}\ldots\m_{L}}-\chi_{\bot}^{1,\m_{1}}N^{\m\m_{2}\ldots\m_{L}}
\nonumber \\
&-&\frac{2}{L}\sum_{i=2}^{L}g_{\bot}^{\m\m_{i}}\chi_{\bot\n}^{1}
N^{\n\m_{1}\ldots\m_{i-1}\m_{i+1}\ldots\m_{L}}\nonumber \\
&+&\frac{2}{L}\sum_{i=2}^{L}g_{\bot}^{\m_{1}\m_{i}}\chi_{\bot\n}^{1}
N^{\n\m\m_{2}\ldots\m_{i-1}\m_{i+1}\ldots\m_{L}}]_{\delta \r}
A_{\alpha \beta}^{\delta \r} \label{78c}\\
\f_{\alpha
\beta}^{\m_{2}\ldots\m_{L}}&=&(\chi_{\bot\m_{1}}^{1}N^{\m_{1}\m_{2}\ldots\m_{L}}
)_{\delta \r}A_{\alpha \beta}^{\delta \r}\,.
\label{78d}
\eea
These $\f$'s contain the ``brown muck" information, or, better, lack of
information. They represent respectively the following total angular
momentum and parity of the light quarks:

\be
\begin{array}{cc}
\f^{\m_{1}\m_{2}\ldots\m_{L}}:    &  L^{(-)^{L}} \\
\f^{\{\m \m_{1}\m_{2}\ldots\m_{L}\}}: & (L+1)^{(-)^{L}}\\
\f^{[\m\m_{1}]\m_{2}\ldots\m_{L}}: & L^{(-)^{L}}\\
\f^{\m_{2}\ldots\m_{L}}: & (L-1)^{(-)^{L}}\,.
\end{array}
\ee

We can now write down the B-S amplitudes for the L-wave heavy baryons of
parity $(-1)^{L}$.
Each pair of states listed is separately degenerate because of the heavy
quark spin symmetry.

\underline{(i) $L=L,\,S_{q_{1}q_{2}}=0,\,J_{q_{1}q_{2}}=L$.}
\bea
 &L+\frac{1}{2}:& \f^{\m_{1}\ldots\m_{L}}u_{\m_{1}\ldots\m_{L}} \nonumber \\
 &L-\frac{1}{2}:&
\f^{\m_{1}\ldots\m_{L}}\frac{1}{\sqrt{L(2L+1)}}[\g_{\bot \m_{1}}
\g_{5}u_{\m_{2}\ldots\m_{L}}\nonumber \\
&&+\g_{\bot \m_{2}}\g_{5}u_{\m_{1}\m_{3}\ldots\m_{L}}+\cdots
\g_{\bot \m_{L}}\g_{5}u_{\m_{1}\ldots\m_{L-1}}]\nonumber \\
&&=\f^{\m_{1}\ldots\m_{L}}\sqrt{\frac{L}{2L+1}}\g_{\bot \m_{1}}\g_{5}
u_{\m_{2}\ldots\m_{L}}\,. \label{79}
\eea

\underline{(ii) $L=L,\,S_{q_{1}q_{2}}=1,\,J_{q_{1}q_{2}}=L+1,L,L-1$.}

\hspace{.5cm}(a)
$J_{q_{1}q_{2}}=L+1$.
\bea
 &L+\frac{3}{2}:&\f^{\{\m\m_{1}\m_{2}\ldots\m_{L}\}}u_{\m\m_{1}\ldots\m_{L}}
\nonumber \\
 &L+\frac{1}{2}:&\f^{\{\m\m_{1}\ldots\m_{L}\}}\sqrt{\frac{L+1}{2L+3}}
\g_{\bot\m}\g_{5}u_{\m_{1}\ldots\m_{L}}\,. \label{80}
\eea

This last can also be written in fully symmetrised form as in
(\ref{79}).

\hspace{.5cm}(b) $J_{q_{1}q_{2}}=L$.

\bea
 &L+\frac{1}{2}:&\f^{[\m\m_{1}]\m_{2}\ldots\m_{L}}\frac{1}{\sqrt{2}}
\{\g_{\bot \m}\g_{5}u_{\m_{1}\ldots\m_{L}}-\g_{\bot \m_{1}}
\g_{5}u_{\m\m_{2}\ldots\m_{L}}\}\, , \nonumber \\
 &L-\frac{1}{2}:&\f^{[\m\m_{1}]\m_{2}\ldots\m_{L}}\frac{1}{L+1}
\sqrt{\frac{L}{2(2L+1)}}\{[\g_{\bot
\m},\g_{\bot\m_{1}}]u_{\m_{2}\ldots\m_{L}}
\nonumber \\
&&+(L-1)(\g_{\bot \m}\g_{\bot \m_{2}}u_{\m_{1}\m_{3}\ldots\m_{L}}-
\g_{\bot \m_{1}}\g_{\bot \m_{2}}u_{\m\m_{3}\ldots\m_{L}})\}\,.\label{81}
\eea

\hspace{.5cm}(c)
$J_{q_{1}q_{2}}=L-1$.

\bea
 &L-\frac{1}{2}:&\f^{\m_{2}\ldots\m_{L}}u_{\m_{2}\ldots\m_{L}}\,,
\nonumber \\
 &L-\frac{3}{2}:&\f^{\m_{2}\ldots\m_{L}}\sqrt{\frac{L-1}{2L-1}}
\g_{\bot \m_{2}}\g_{5}u_{\m_{3}\ldots\m_{L}}\, . \label{82}
\eea

The $u_{\m_{1}\ldots\m_{L}}$in eqs. (\ref{79}) - (\ref{82}) are the
appropriate generalised symmetric, traceless Rarita-Schwinger spinors
satisfying
\bea
v^{\m_{1}}u_{\m_{1}\ldots\m_{L}} &=&0\, , \nonumber \\
\g^{\m_{1}}u_{\m_{1}\ldots\m_{L}}&=&0\, , \nonumber \\
(\vslash-1)u_{\m_{1}\ldots\m_{L}}&=&0\, . \label{83}
\eea

These spinors represent bound states with parity $(-)^{L}$. (See footnote
on page 13).

As already remarked, Falk \cite{f} has followed a different route to
construct wave functions of heavy baryons with arbitrary spin. However,
in his construction he only has the symmetric wave functions
(\ref{79}), (\ref{80}) and (\ref{82}). We see that, in general, there
will always be mixed symmetry type wave functions like (\ref{81}). In
fact, this observation has important consequences for the decay matrix
elements as we shall see in a subsequent section.

Flavour is easily accounted for as we have already discussed. If one
requires $\Lambda$-type
heavy baryons, antisymmetric under flavour, overall symmetry symmetry is
assured by taking the appropriate number of $k_{3\bot}$'s and
$K_{3\bot}$'s to ensure that $L^{\m_{1}\ldots\m_{L}}$ in eq. (\ref{78a})
is symmetric under $p_{1}\leftrightarrow p_{2}$ and antisymmetric under
$p_{1}\leftrightarrow p_{2}$ in eqs. (\ref{78b}) - (\ref{78d}). For
$\Sigma$-type heavy baryons we have the opposite situation.

For the construction of the p- and higher wave baryons we were inspired
by the seminal work, \cite{sdrs}, of Salam, Delbourgo, Rashid and
Strathdee. In that work they had constructed the mixed symmetry tensor
which we have used in describing the p-wave baryons. However, they had
the wrong particle(spin and parity) content because they did not include
internal momenta. Once one does that it is easy to show that the mixed
symmetry, $\tilde{U}(12)$ representation, contains nothing but the set of
p-wave baryons
which we have constructed above.

\section{Heavy Meson Transitions}

In this section we calculate the current induced flavour changing
transitions between ground state heavy mesons and ground and excited
state mesons of different flavour. Exemplary processes are the
$b\rightarrow c$ transitions $B \rightarrow D, D^{*},D^{**}\ldots.$

We define the transition matrix elements as
\be
M_{\lambda} = \langle M_{2}(v_{2})\vert J_{\lambda}^{V-A}\vert
M_{1}(v_{1})\rangle\, . \label{84}
\ee
Here $M_{1}$ and $M_{2}$ are the initial and final mesons with
velocities $v_{1}$ and $v_{2}$, respectively.
Closely following the presentation in \cite{bhkt} the transition
matrix element can be written in the form of a trace.

First we introduce tensor-valued reduced meson projectors
$\chi_{\m_{1}\ldots\m_{L}}$ in analogy to the traceless,
symmetric tensors $\bar{\G}_{\m_{1}\ldots\m_{L}}$ in eq. (\ref{31}). Thus
using eq. (\ref{31}), one can write the $\chi$ of eq. (\ref{19}) as
\be
\chi=N^{\m_{1}\ldots\m_{L}}\chi_{\m_{1}\ldots\m_{L}}\,  .
\label{85}
\ee
Specifically,
\be
\chi_{\m_{1}\ldots\m_{L}} =
\frac{1+\vslash}{2}\bar{\G}_{\m_{1}\ldots\m_{L}}\frac{1-\vslash}{2}\, .
\label{85a}
\ee

The transition matrix element (\ref{84}) can then be written as
\be
M_{\lambda}=
Tr\{\bar{\chi}_{2(L\pm)\m_{1}\ldots\m_{L}}\g_{\lambda}(1-\g_{5})
\chi_{1}\}N(v_{1})^{\m_{1}\ldots\m_{L}}\xi_{L\pm}(\omega)\, ,
\label{86}
\ee
where $N^{\m_{1}\ldots\m_{L}}(v_{1})$ is identical to the symmetric
traceless tensor $N^{\m_{1}\ldots\m_{L}}$ defined in (\ref{33}) except
for the replacement $k_{\bot}\rightarrow v_{1 \bot}=v_{1}-\omega
v_{2}$ where $\omega = v_{1}.v_{2}$. $\chi_{1}$ is the
incoming reduced spin wave function for ground state mesons,
eqs. (\ref{19}, \ref{23}, and \ref{24}). The subscript $L\pm$ refers to the
two pairs of degenerate states in table (5) labelled by the subscripts
$L+\frac{1}{2}$ and $L-\frac{1}{2}$, respectively.
$\xi_{L\pm}(\omega)$ are two independent form factor functions of the
variable $\omega$ defined above.

 The last two factors in eq. (\ref{86}) characterize the covariance
structure of the light side transition which is parametrised according
to \cite{bhkt}
\be
\int d^{4}k_{1}d^{4}k_{2}A_{1}(k_{1}){\cal T}(k_{1},k_{2};v_{1},v_{2})
\bar{A}_{2}(k_{2})N^{\m_{1}\ldots\m_{L}}(k_{2})=\xi_{L\pm}(\omega)
N^{\m_{1}\ldots\m_{L}}(v_{1})\, . \label{87}
\ee

The single covariant $N^{\m_{1}\ldots\m_{L}}(v_{1})$ in eq. (\ref{86}) is
the most general covariant that can be written down for the light-side
transition given by the overlap integral on the l.h.s. of eq. (\ref{87})
when taken ``between the projectors" in the trace eq. (\ref{86}). At first
glance one would also write down a second covariant
\bea
N^{\prime\m_{1}\ldots\m_{L}}(v_{1})& = & \sum_{i=1}^{L}\g_{\bot_{2}}^{\m_{i}}
N^{\m_{1}\ldots\m_{i-1}\m_{i+1}\ldots\m_{L}}(v_{1}) \nonumber \\
& - &\frac{2}{2L-1}\sum_{i<j}^{L}g_{\bot_{2}}^{\m_{i}\m_{j}}\g_{\bot_{2}\n}
N^{\n\m_{1}\ldots\m_{i-1}\m_{i+1}\ldots\m_{j-1}\m_{j+1}\ldots\m_{L}}(v_{1})\, .
\label{88}
\eea
To avoid confusion from now on we add a subscript to the symbol $\bot$
to indicate transversality with respect either to the incoming velocity
$v_{1}$ or the final velocity $v_{2}$. The symbol $\bot_{i}$ hence means
perpendicular to the velocity $v_{i}$. However, the covariant
$N^{\prime\m_{1}\ldots\m_{L}}(v_{1})$ is not linearly independent of
$N^{\m_{1}\ldots\m_{L}}(v_{1})$ when taken in the trace eq. (\ref{86}). In
fact one can easily check that $N^{\prime\m_{1}\ldots\m_{L}}$ is
linearly related to $N^{\m_{1}\ldots\m_{L}}$ for the transitions to the
$(L-\frac{1}{2})$ states and identical to zero for transitions to the
$(L+\frac{1}{2})$ states.
 We are now in a position to write down the trace expressions. Using the
expressions for $\bar{\G}$ from Table 5 in eq. (\ref{86}) one can write the
matrix elements for the transitions from the ground state s-wave mesons
to the L-wave mesons as (all the matrix elements are multiplied by a common
factor $\sqrt{M_{1}M_{2}}$):

$(a)\;\;(0^{-},1^{-})\rightarrow (L-1)_{L-\frac{1}{2}}$
\be
-\sqrt{\frac{2L-1}{2L+1}}(1+\omega)Tr\{[v_{1}^{\m_{2}}
+(1-\omega)\frac{L-1}{2L-1}\g^{\m_{2}}]\frac{1+\vslash_{2}}{2}
\g_{\lambda}(1-\g_{5})\chi_{1}\}
v_{1}^{\m_{3}}\ldots v_{1}^{\m_{L}}
\e_{\m_{2}\ldots\m_{L}}^{*}\xi_{L-}(\omega) \label{89}
\ee

$(b)\;\;(0^{-},1^{-})\rightarrow L_{L-\frac{1}{2}}$
\be
\sqrt{\frac{L}{2L+1}}(1+\omega)Tr\{\g^{\m_{1}}\g_{5}
\frac{1+\vslash_{2}}{2}\g_{\lambda}(1-\g_{5})\chi_{1}\}
v_{1}^{\m_{2}}\ldots v_{1}^{\m_{L}}\e_{\m_{1}\ldots\m_{L}}^{*}\xi_{L-}
(\omega) \label{90}
\ee

$(c)\;\;(0^{-},1^{-})\rightarrow L_{L+\frac{1}{2}}$
\bea
&-\frac{1}{\sqrt{(L+1)(2L+1)}}Tr\{[L(1+\omega)
\g^{\m_{1}}+(2L+1)v_{1}^{\m_{1}}]& \nonumber \\
\times& \g_{5}\frac{1+\vslash_{2}}{2}
\g_{\lambda}(1-\g_{5})\chi_{1}\}v_{1}^{\m_{2}}\ldots v_{1}^{\m_{L}}
\e_{\m_{1}\ldots\m_{L}}^{*}\xi_{L+}(\omega)& \label{91}
\eea

$(d)\;\;(0^{-},1^{-})\rightarrow (L+1)_{L+\frac{1}{2}}$
\be
Tr\{\g^{\n}\frac{1+\vslash_{2}}{2}\g_{\lambda}(1-\g_{5})\chi_{1}\}
v_{1}^{\m_{1}}\ldots
v_{1}^{\m_{L}}\e_{\n\m_{1}\ldots\m_{L}}^{*}\xi_{L+}(\omega) \label{92}
\ee

Notice that all these amplitudes vanish at $\omega=1$ where
$v_{1}=v_{2}$.
 To summarize we see that to describe all the heavy s- to L-wave transitions
we need just two form factors. This is a tremendous simplification.

As an obvious first application of our formalism we consider the
contributions of the current
induced transitions
$0^{-}\rightarrow~(L-1)_{L-\frac{1}{2}}$, $0^{-}\rightarrow~
L_{L-\frac{1}{2}}$ and \newline $0^{-}\rightarrow~L_{L+\frac{1}{2}}$,
$0^{-}\rightarrow~(L+1)_{L+\frac{1}{2}}$ to the Bjorken sum
rule \cite{bj}. Technically the easiest route to do this is to first
calculate the longitudinal helicity amplitude, as done in \cite{bhkt},
and then to square it in order to obtain the contribution of a given
excitation to the Bjorken sum rule. This is an elegant device that
avoids the tedium of having to do spin sums in squared covariant
matrix elements. To this end one needs an explicit representation of the
helicity $0$ component of the spin $j$ polarisation tensor in terms of
products of the spin $1$ polarisation vector. One has
\be
\e_{\m_{1}\ldots\m_{j}}(0) =
\sqrt{\frac{j!}{(2j-1)!!}}\e_{\m_{1}}(0)\ldots\e_{\m_{j}}(0)+\ldots,
\label{93}
\ee
where the ellipsis stand for terms involving transverse spin $1$
polarisation vectors. A glance at the structure of the transition matrix
elements eqs. (\ref{89}-\ref{92}) shows that these extra transverse terms do
not contribute to the longitudinal helicity amplitudes and thus their
explicit form is not needed here. It is then rather straightforward to
calculate the longitudinal helicity amplitudes. Upon squaring the
longitudinal amplitudes and dividing out the longitudinal
structure function, \cite{bhkt}, $K_{L}=4M_{1}M_{2}[\omega(M_{1}^{2}+
M_{2}^{2})-2M_{1}M_{2}]/q^{2}$, one obtains
\be
1=\frac{\omega+1}{2}\{\vert\xi(\omega)\vert^{2}+\sum_{L\geq 1}
\frac{L!(\omega^{2}-1)^{L}}{(2L+1)!!}(L\vert\xi_{L-}(\omega)\vert^{2}
+(L+1)\vert\xi_{L+}(\omega)\vert^{2})\} \, , \label{94}
\ee
where $\xi(\omega)$ denotes the reduced form factor function of the
ground-state to ground-state transition. Compared to \cite{bhkt} we no
longer absorb a factor $(\omega+1)$ in the reduced form factor
$\xi_{L-}(\omega)$, i.e. one has
$\xi_{\frac{1}{2}}^{*}(\omega)=(\omega+1)\xi_{1-}(\omega)$ where
$\xi_{\frac{1}{2}}(\omega)$ is the s-wave to p-wave reduced form factor
introduced in \cite{bhkt}. If one identifies
$\xi(\omega)=\xi_{0+}(\omega)$ one can rewrite eq. (\ref{94}) as a single
sum, i.e. one has
\be
1=\frac{\omega+1}{2}\sum_{L\geq 0}\frac{L!}{(2L+1)!!}(\omega^{2}-1)^{L}
(L\vert\xi_{L-}(\omega)\vert^{2}+(L+1)\vert\xi_{L+}(\omega)\vert^{2})\, .
\label{95}
\ee

\section{Heavy Baryon Transitions}
 Using the heavy baryon wave functions developed in section {\bf 3}, we now
calculate the current induced flavour changing transitions between
ground state heavy baryons and heavy baryon orbital excitations of
different flavours. The procedure has already been outlined in the
previous section in the case of mesons. The transitions between s-wave
baryons has already been treated in \cite{hkkt} and \cite{hlkkt}. We
shall not repeat them here.
As before the transition matrix element can be written in terms of a
trace \cite{hkkt}, \cite{hlkkt}.

\subsection{Decay of heavy s-wave baryons to heavy p-wave baryons}
 We first consider the decays of s- to p-wave heavy baryons using the
wave functions eqs. (\ref{shbs2}-\ref{75}). We present separately the results
for
decays of $\Lambda$ type and $\Sigma (\Omega)$ type baryons.

\subsubsection{Heavy $\Lambda_{Q}$ to heavy p-wave $\Lambda_{Q^{\prime}}$.}

We shall first demonstrate the general structure of the matrix element
for the heavy s-wave $\Lambda_{Q}$ to the heavy p-wave
$\Lambda_{Q^{\prime}}$'s. We will
then see that we need only two independent form factors to describe the
decays of the $\Lambda_{Q}(\frac{1}{2})$ to all the p-wave
$\Lambda_{Q^{\prime}}$'s
$(\frac{5}{2}^{-},\frac{3}{2}^{-}; \frac{3}{2}^{-},\frac{1}{2}^{-};
\frac{3}{2}^{-},\frac{1}{2}^{-}; \frac{1}{2}^{-})$. We list the matrix
elements according to the various final p-wave states developed in
section {\bf 3}.

\underline{(i) $\Lambda_{Q},\frac{1}{2}^{+}\rightarrow$ p-wave
$\Lambda_{Q^{\prime}}$ with
$S_{q_{1}q_{2}}=0$.}
\bea
&\frac{1}{2}^{+}\rightarrow\frac{3}{2}^{-}:&
\bar{u}_{\m}\g_{\lambda}(1-\g_{5})u M^{\m} \nonumber
\\
&\frac{1}{2}^{+}\rightarrow\frac{1}{2}^{-}:&
\frac{1}{\sqrt{3}}\bar{u}\g_{5}\g_{\bot_{2}\m}\g_{\lambda}(1-\g_{5})uM^{\m}\,.
\label{96}
\eea
Here $M^{\m}$ is the overlap integral
\be
M^{\m}=Tr\int d\G \bar{P}_{K_{3}^{(2)}}^{\m}(k_{3}^{(2)},K_{3}^{(2)})
{\cal T}(k_{3}^{(1)},K_{3}^{(1)},k_{3}^{(2)},K_{3}^{(2)};v_{1},v_{2})
S(k_{3}^{(1)},K_{3}^{(1)}) \,, \label{97}
\ee
where
$d\G=d^{4}k_{3}^{(1)}d^{4}K_{3}^{(1)}d^{4}k_{3}^{(2)}d^{4}K_{3}^{(2)}$
and $k_{3}^{(1)},K_{3}^{(1)} (k_{3}^{(2)},K_{3}^{(2)})$ are the two
pairs of initial and final relative momenta. $v_{1}, v_{2}$ are the
velocities of the initial and final baryons, respectively.
In eq. (\ref{97}), we have explicitly displayed the arguments of $\bar{P}$
and $S$.
We will not display these in subsequent equations. Clearly $M^{\m}$ is a
vector function of the velocities, $v_{1}$ and $v_{2}$, satisfying
$v_{2\m}M^{\m}=0$. Hence the most general nonvanishing covariant is
$v_{1\bot}^{\m}$ leading to just one form factor $f_{1}^{(1)}$, so that
eqs. (\ref{96}) can be written as:
\bea
&\frac{1}{2}^{+}\rightarrow\frac{3}{2}^{-}:&
f_{1}^{(1)}v_{1}^{\m}\bar{u}_{\m}\g_{\lambda}(1-\g_{5})u
\nonumber \\
&\frac{1}{2}^{+}\rightarrow\frac{1}{2}^{-}:&
\frac{f_{1}^{(1)}}{\sqrt{3}}\bar{u}\g_{5}(\omega+\vslash_{1})\g_{\lambda}(1-\g_{5})u\,.
\label{98}
\eea

\underline{(ii) $\Lambda_{Q},\frac{1}{2}^{+}\rightarrow$ to p-wave
$\Lambda_{Q^{\prime}}$ with $S_{q_{1}q_{2}}=1$.}

\hspace{.5cm}(a) $S_{q_{1}q_{2}}=1, J_{q_{1}q_{2}}=2$
\bea
&\frac{1}{2}^{+}\rightarrow\frac{5}{2}^{-}:& \bar{u}_{\m
\n}\g_{\lambda}(1-\g_{5})uM^{\{\m \n\}}
\nonumber \\
&\frac{1}{2}^{+}\rightarrow\frac{3}{2}^{-}:&
\sqrt{\frac{2}{5}}\bar{u}_{\m}\g_{5}\g_{\bot_{2}\n}
\g_{\lambda}(1-\g_{5})uM^{\{\m \n\}}  \label{99}\,,
\eea
with
\be
M^{\{\m \n\}}=Tr \int d\G \bar{P}_{k_{3}^{(2)}}^{\{\m \n\}}{\cal T}S
\label{100}
\ee
a negative parity symmetric, traceless tensor satisfying
$v_{2\m}M^{\{\m\n\}}=0$. Such a tensor does not exist. Hence
$\Lambda_{Q}$ does not decay to this pair of $\frac{5}{2}^{-}$ and
$\frac{3}{2}^{-}$ states.

\hspace{.5cm}(b) $S_{q_{1}q_{2}}=1, J_{q_{1}q_{2}}=1$
\bea
&\frac{1}{2}^{+}\rightarrow\frac{3}{2}^{-}:&
\sqrt{2}\bar{u}_{\n}\g_{5}\g_{\bot_{2}\m}\g_{\lambda}(1-\g_{5})uM^{[\m\n]}
\nonumber \\
&\frac{1}{2}^{+}\rightarrow\frac{1}{2}^{-}:&
\frac{1}{2}\sqrt{\frac{2}{3}}\bar{u}\g_{\bot_{2}\n}\g_{\bot_{2}\m}\g_{\lambda}
(1-\g_{5})uM^{[\m \n]}\,,
\label{101}
\eea
with
\be
M^{[\m\n]}=Tr\int d\G \bar{P}_{k_{3}^{(2)}}^{[\m\n]}{\cal T}S
\label{102}
\ee
a negative parity antisymmetric tensor satisfying $v_{2\m}M^{[\m\n]}=0$. Such a
tensor does exist uniquely,
namely $i\e^{\m\n\r\k}v_{1\r}v_{2\k}$, leading to one form factor,
$f_{2}^{(1)}$, for this pair of transitions. Thus the matrix elements
eq. (\ref{101}) can be written as:
\bea
&\frac{1}{2}^{+}\rightarrow\frac{3}{2}^{-}:&
\sqrt{2}if_{2}^{(1)}\e^{\m\n\r\k}v_{1\r}v_{2\k}\bar{u}_{\n}\g_{5}\g_{\m}
\g_{\lambda}(1-\g_{5})u \nonumber \\
&\frac{1}{2}^{+}\rightarrow\frac{1}{2}^{-}:&\frac{1}{2}
\sqrt{\frac{2}{3}}if_{2}^{(1)}\e^{\m\n\r\k}v_{1\r}v_{2\k}\bar{u}\g_{\n}\g_{\m}
\g_{\lambda}(1-\g_{5})u\,.\label{103}
\eea
On using standard $\e^{\m\n\k\lambda}$ identities, these matrix elements
can also be written as:
\bea
&\frac{1}{2}^{+}\rightarrow\frac{3}{2}^{-}:&
\sqrt{2}f_{2}^{(1)}v_{1}^{\m}\bar{u}_{\m}\g_{\lambda}(1-\g_{5})u \nonumber \\
&\frac{1}{2}^{+}\rightarrow\frac{1}{2}^{-}:&
\sqrt{\frac{2}{3}}f_{2}^{(1)}\bar{u}\g_{5}(\omega+\vslash_{1})\g_{\lambda}
(1-\g_{5})u\,.\label{105}
\eea

\hspace{.5cm}(c)$S_{q_{1}q_{2}}=1, J_{q_{1}q_{2}}=0$
\bea
&\frac{1}{2}^{+}\rightarrow\frac{1}{2}^{-};&
\bar{u}\g_{\lambda}(1-\g_{5})uM \label{106}\,,
\eea
with
\be
M=Tr\int d\G \bar{P}_{k_{3}^{(2)}}{\cal T}S\nonumber
\ee
a pseudoscalar. It is not possible to construct such a pseudoscalar with
the two vectors available to us, $v_{1}$ and $v_{2}$. Hence the
$\Lambda_{Q}, \frac{1}{2}^{+}$ does not decay to this $\frac{1}{2}^{-}$.

 To summarise we see that all the decays of the ground state heavy
$\Lambda$ to p-wave heavy $\Lambda$'s are controlled by just two form
factors which we have called $f_{1}^{(1)}$ and $f_{2}^{(1)}$.

\subsubsection{Heavy $\Sigma_{Q}(\Omega_{Q})$ to heavy p-wave
$\Sigma_{Q^{\prime}}(\Omega_{Q^{\prime}})$.}

Using the B-S amplitudes developed in earlier sections we can also
immediately write down the matrix elements for the semi-leptonic decays
of the $\Sigma(\Omega)$ type ground state heavy baryons to the
corresponding p-wave heavy baryons. The initial ground state is either
$\frac{3}{2}^{+}$ or
$\frac{1}{2}^{+}$. We list below, in obvious matrix form, the weak
current matrix elements for the transition to the various p-wave states.

\underline{(i) $\Sigma_{Q}(\Omega_{Q})(\frac{3}{2}^{+},
\frac{1}{2}^{+})\rightarrow$ to
p-wave $\Sigma_{Q^{\prime}}(\Omega_{Q^{\prime}})$ with $(S_{q_{1}q_{2}}=0;
\frac{3}{2}^{-},\frac{1}{2}^{-})$.}
\be
\left( \begin{array}{c}
\bar{u}_{\m}\\ \frac{1}{\sqrt{3}}\bar{u}\g_{5}\g_{\bot_{2}\m}
\end{array}\right)
\g_{\lambda}(1-\g_{5})\left( \begin{array}{c}
u_{\n}\\ \frac{1}{\sqrt{3}}\g_{\bot_{1}\n}\g_{5}u \end{array}\right)M^{\m\n}\,,
\label{107}
\ee
with
\be
M^{\m\n}=Tr\int d\G \bar{P}_{k_{3}^{(2)}}^{\m}{\cal T}S^{\n}\nonumber
\ee
a negative parity tensor satisfying $v_{2\m}M^{\m\n}=v_{1\n}M^{\m\n}=0$. The
general unique form of this tensor is $i\e^{\m\n\r\k}v_{1\r}v_{2\k}$
leading to just one form factor, $g_{1}^{(1)}$. The matrix elements
eq. (\ref{107}) can then be written as
\be
ig_{1}^{(1)}\e^{\m\n\r\k}v_{1\r} v_{2\k}\left( \begin{array}{c}
\bar{u}_{\m}\\ \frac{1}{\sqrt{3}}\bar{u}\g_{5}\g_{\m} \end{array}\right)
\g_{\lambda}(1-\g_{5})\left( \begin{array}{c}
u_{\n}\\ \frac{1}{\sqrt{3}}\g_{\n}\g_{5}u \end{array}\right)\,.\label{108}
\ee

\underline{(ii)
$\Sigma_{Q}(\Omega_{Q}),(\frac{3}{2}^{+},\frac{1}{2}^{+})\rightarrow$ to p-wave
$\Sigma_{Q^{\prime}}(\Omega_{Q^{\prime}})$ with $S_{q_{1}q_{2}}=1$.}

\hspace{.5cm}(a)
$(\frac{3}{2}^{+},\frac{1}{2}^{+})\rightarrow(J_{q_{1}q_{2}}=2;\frac{5}{2}^{-},
\frac{3}{2}^{-})$
\be
\left( \begin{array}{c}
\bar{u}_{\m\n}\\ \sqrt{\frac{2}{5}}\bar{u}_{\m}\g_{5}\g_{\bot_{2}\n}
\end{array}\right)
\g_{\lambda}(1-\g_{5})\left( \begin{array}{c}
u_{\k}\\ \frac{1}{\sqrt{3}}\g_{\bot_{1}\k}\g_{5}u
\end{array}\right)M^{\{\m\n\}\k}\,,
\label{109}
\ee
with
\be
M^{\{\m\n\}\k}=Tr\int d\G\bar{P}_{K_{3}^{(2)}}^{\{\m\n\}}{\cal T}S^{\k}
\nonumber
\ee
a negative parity, traceless and symmetric in $\{\m,\n\}$, tensor
satisfying
\be
v_{2\m}M^{\{\m\n\}\k}=v_{1\k}M^{\{\m\n\}\k}=0\,.\nonumber
\ee
Thus there are two form factors as $M^{\{\m\n\}\k}$ can in general be
written as
\be
M^{\{\m\n\}\k}=g_{2}^{(1)}(v_{1\bot}^{\m}v_{1\bot}^{\n}
-\frac{1}{3}v_{1\bot}^{2}g_{\bot_{2}}^{\m\n})v_{2\bot}^{\k}+
g_{3}^{(1)}(v_{1\bot}^{\m}g_{\bot_{2}\bot_{1}}^{\n\k}
+v_{1\bot}^{\n}g_{\bot_{2}\bot_{1}}^{\m\k}
-\frac{2}{3}g_{\bot_{2}}^{\m\n}v_{1\bot\r}g_{\bot_{2}\bot_{1}}^{\r\k})\,,
\nonumber
\ee
where
\be
v_{1\bot}=v_{1}-\omega v_{2},\,\,\, v_{2\bot}=v_{2}-\omega v_{1}\,
\label{vperp}
\ee
and
\be
g_{\bot_{2}\bot_{1}}^{\n\k}=g^{\n\k}-(v_{1}^{\n}v_{1}^{\k}+v_{2}^{\n}v_{2}^{\k})
+\omega v_{1}^{\k}v_{2}^{\n}\,,\label{gpp}
\ee
so that
$v_{2\n}g_{\bot_{2}\bot_{1}}^{\n\k}=v_{1\k}g_{\bot_{2}\bot_{1}}^{\n\k}=0$.
Thus the matrix element eq. (\ref{109}) becomes
\be
\{g_{2}^{(1)}v_{1}^{\m}v_{1}^{\n}v_{2}^{\k}
+g_{3}^{(1)}(v_{1}^{\m}g^{\n\k}+v_{1}^{\n}g^{\m\k})\}\left( \begin{array}{c}
\bar{u}_{\m\n}\\ \sqrt{\frac{2}{5}}\bar{u}_{\m}\g_{5}\g_{\bot_{2}\n}
\end{array}\right)
\g_{\lambda}(1-\g_{5})\left( \begin{array}{c}
u_{\k}\\ \frac{1}{\sqrt{3}}\g_{\bot_{1}\k}\g_{5}u
\end{array}\right)\,.\label{110}
\ee

\hspace{.5cm}(b)
$(\frac{3}{2}^{+},\frac{1}{2}^{+})\rightarrow(J_{q_{1}q_{2}}=1;\frac{3}{2}^{-},
\frac{1}{2}^{-})$
\be
\left( \begin{array}{c}
\sqrt{2}\bar{u}_{\n}\g_{5}\g_{\bot_{2}\m}\\
\frac{1}{2}\sqrt{\frac{2}{3}}\bar{u}\g_{\bot_{2}\n}
\g_{\bot_{2}\m} \end{array}\right)
\g_{\lambda}(1-\g_{5})\left( \begin{array}{c}
u_{\k}\\ \frac{1}{\sqrt{3}}\g_{\bot_{1}\k}\g_{5}u
\end{array}\right)M^{[\m\n]\k}\,,\label{111}
\ee
with
\be
M^{[\m\n]\k}=Tr\int d\G\bar{P}_{K_{3}^{(2)}}^{[\m\n]}{\cal T}S^{\k}
\nonumber
\ee
a $[\m\n]$ antisymmetric, negative parity tensor such that
\be
v_{2\m}M^{[\m\n]\k}=v_{1\k}M^{[\m\n]\k}=0\,.\nonumber
\ee
The unique form for such a tensor is
\bea
M^{[\m\n]\k}&=&g_{4}^{(1)}(g_{\bot_{2}\bot_{1}}^{\m\k}v_{1\bot}^{\n}
-g_{\bot_{2}\bot_{1}}^{\n\k}v_{1\bot}^{\m})\nonumber\\
&=&g_{4}^{(1)}(g_{\bot_{2}}^{\m\k}v_{1\bot}^{\n}
-g_{\bot_{2}}^{\n\k}v_{1\bot}^{\m})
\eea
Hence the matrix elements eq. (\ref{111}) become
\be
g_{4}^{(1)}(g^{\m\k}v_{1}^{\n}-g^{\n\k}v_{1}^{\m})\left( \begin{array}{c}
\sqrt{2}\bar{u}_{\n}\g_{5}\g_{\bot_{2}\m}\\
\frac{1}{2}\sqrt{\frac{2}{3}}\bar{u}\g_{\bot_{2}\n}
\g_{\bot_{2}\m} \end{array}\right)
\g_{\lambda}(1-\g_{5})\left( \begin{array}{c}
u_{\k}\\ \frac{1}{\sqrt{3}}\g_{\bot_{1}\k}\g_{5}u
\end{array}\right)\label{112}
\ee

\hspace{.5cm}(c)
$(\frac{3}{2}^{+},\frac{1}{2}^{+})\rightarrow(J_{q_{1}q_{2}}=0;\frac{1}{2}^{-})$
\be
\bar{u}\g_{\lambda}(1-\g_{5})\left( \begin{array}{c}
u_{\m}\\ \frac{1}{\sqrt{3}}\g_{\bot_{1}\m}\g_{5}u
\end{array}\right)M^{\m}\,,
\label{113}
\ee
with
\be
M^{\m}=Tr\int d\G \bar{P}_{K_{3}^{(2)}}{\cal T}S^{\m}\nonumber
\ee
a vector satisfying $v_{1\m}M^{\m}=0$. Such a vector is uniquely
$v_{2\bot}^{\m}$, giving rise to just one form factor $g_{5}^{(1)}$. The matrix
element can be written as
\be
g_{5}^{(1)}\bar{u}\g_{\lambda}(1-\g_{5})\left( \begin{array}{c}
v_{2}^{\m}u_{\m}\\ \frac{1}{\sqrt{3}}(\vslash_{2}+\omega)\g_{5}u
\end{array}\right)\,.
\label{114}
\ee
Thus all the ground state heavy $\Sigma(\Omega)$ transitions to all the
heavy p-wave $\Sigma(\Omega)$ states are described by just five form factors.

\subsection{Decay of s-wave heavy baryons to first positive parity
excited heavy baryons}

In this subsection, we present the counting of form factors for the
weak decays of s-wave heavy baryons to the first positive parity excited
heavy baryons.

\subsubsection{Heavy $\Lambda_{Q}$ to first positive parity excited
$\Lambda_{Q^{\prime}}$.}
We list the matrix elements in terms of the final states.

$\;(1)\,\,\, L_{K_{3}}=2,\, L_{k_{3}}=0,\, L=2,
\,S_{q_{1}q_{2}}=0,\,J_{q_{1}q_{2}}=2$

or

$\;(2)\,\,\, L_{K_{3}}=0,\, L_{k_{3}}=2,\, L=2,
\,S_{q_{1}q_{2}}=0,\,J_{q_{1}q_{2}}=2$
\bea
&\frac{1}{2}^{+}\rightarrow\frac{5}{2}^{+}:& \bar{u}_{\m_{1}\m_{2}}
\g_{\lambda}(1-\g_{5})u M^{\m_{1}\m_{2}} \nonumber
\\
&\frac{1}{2}^{+}\rightarrow\frac{3}{2}^{+}:&
\sqrt{\frac{2}{5}}\bar{u}_{\m_{1}}\g_{5}\g_{\bot_{2}\m_{2}}
\g_{\lambda}(1-\g_{5})uM^{\m_{1}\m_{2}}\,.
\label{115}
\eea
Here $M^{\m_{1}\m_{2}}$ is a positive parity, symmetric, traceless tensor
transverse to $v_{2}$.\footnote{From now on we will not write down
the full expressions for the matrices
$M$ which appear in the transition amplitudes. In the case of the
transitions to the next excited positive parity baryons they
all have the general form
\be
M = Tr \int d\G \bar{D}{\cal T}S
\ee
where the $D$'s and the $S$'s are as defined in
subsections~\ref{sub-dheavy} and~\ref{sub-sheavy} respectively.}
The only non-vanishing form for this tensor is
$M^{\m_{1}\m_{2}}=f_{1}^{(2)}N^{\m_{1}\m_{2}}(v_{1})$ with
$N^{\m_{1}\m_{2}}$ defined as in eq. (\ref{33}), but with argument
$v_{1\bot}$. Although the Lorentz structure is the same, we will in
general have two independent form factors arising from the cases (1) and
(2). Let us call these form factors $f_{1}^{(2)}$ and $f_{1}^{\prime(2)}$.
Hence we can write the matrix elements (\ref{115}) in terms of one of
these form factors as
\bea
&\frac{1}{2}^{+}\rightarrow\frac{5}{2}^{+}:&
f_{1}^{(2)}v_{1}^{\m_{1}}v_{1}^{\m_{2}}\bar{u}_{\m_{1}\m_{2}}
\g_{\lambda}(1-\g_{5})u \nonumber
\\
&\frac{1}{2}^{+}\rightarrow\frac{3}{2}^{+}:&
\sqrt{\frac{2}{5}}f_{1}^{(2)}v_{1}^{\m_{1}}\bar{u}_{\m_{1}}\g_{5}(\vslash_{1}+\omega)
\g_{\lambda}(1-\g_{5})u\,.
\label{116}
\eea
The same structure holds with the other form factor.

$\;(3)\,(a)\,(i)\,\,\, L_{K_{3}}=1,\, L_{k_{3}}=1,\, L=2,
\,S_{q_{1}q_{2}}=1,\,J_{q_{1}q_{2}}=3$
\bea
&\frac{1}{2}^{+}\rightarrow\frac{7}{2}^{+}:& \bar{u}_{\m\m_{1}\m_{2}}
\g_{\lambda}(1-\g_{5})u M^{\m\m_{1}\m_{2}} \nonumber
\\
&\frac{1}{2}^{+}\rightarrow\frac{5}{2}^{+}:&
\sqrt{\frac{3}{7}}\bar{u}_{\m_{1}\m_{2}}\g_{5}\g_{\bot_{2}\m}
\g_{\lambda}(1-\g_{5})uM^{\m\m_{1}\m_{2}}\,.
\label{117}
\eea
Here $M^{\m\m_{1}\m_{2}}$ is a positive parity, symmetric, traceless
tensor transverse with respect to $v_{2}$. It is impossible to construct
such a tensor with the two vectors, $v_{1}$ and $v_{2}$ available to us.
Hence this pair of decays are not allowed.

$\;(3)\,(a)\,(ii)\,\,\, L_{K_{3}}=1,\, L_{k_{3}}=1,\, L=2,
\,S_{q_{1}q_{2}}=1,\,J_{q_{1}q_{2}}=2$
\bea
&\frac{1}{2}^{+}\rightarrow\frac{5}{2}^{+}:& \sqrt{2}\bar{u}_{\m_{1}\m_{2}}
\g_{5}\g_{\bot_{2}\m}\g_{\lambda}(1-\g_{5})u M^{[\m\m_{1}]\m_{2}} \nonumber
\\
&\frac{1}{2}^{+}\rightarrow\frac{3}{2}^{+}:&
\frac{2}{3\sqrt{5}}(\bar{u}_{\m_{2}}\g_{\bot_{2}\m_{1}}
+\bar{u}_{\m_{1}}\g_{\bot_{2}\m_{2}})\g_{\bot_{2}\m}\g_{\lambda}
(1-\g_{5})uM^{[\m\m_{1}]\m_{2}}\,.
\label{118}
\eea
$M^{[\m\m_{1}]\m_{2}}$ is a positive parity, mixed symmetry tensor
transverse to $v_{2}$. There is a unique nonvanishing form for this
tensor giving rise to just one new form factor, $f_{2}^{(2)}$, i.e.
\be
M^{[\m\m_{1}]\m_{2}}=
if_{2}^{(2)}\e^{\m\m_{1}\r\k}v_{1\r}v_{2\k}v_{1\bot}^{\m_{2}}\,.
\ee
Thus the matrix elements (\ref{118}) can be written as
\bea
&\frac{1}{2}^{+}\rightarrow\frac{5}{2}^{+}:& \sqrt{2}if_{2}^{(2)}
\e^{\m\m_{1}\r\k}v_{1\r}v_{2\k}v_{1}^{\m_{2}}\bar{u}_{\m_{1}\m_{2}}
\g_{5}\g_{\m}\g_{\lambda}(1-\g_{5})u \nonumber
\\
&\frac{1}{2}^{+}\rightarrow\frac{3}{2}^{+}:&
\frac{2}{3\sqrt{5}}if_{2}^{(2)}
\e^{\m\m_{1}\r\k}v_{1\r}v_{2\k}\{v_{1}^{\m_{2}}\bar{u}_{\m_{2}}\g_{\m_{1}}
+\bar{u}_{\m_{1}}(\vslash_{1}-\omega)\}\g_{\m}\g_{\lambda}
(1-\g_{5})u\,.\nonumber\\
&&
\label{119}
\eea
Using $\e$ identities one can write these matrix elements in simpler
form as
\bea
&\frac{1}{2}^{+}\rightarrow\frac{5}{2}^{+}:& \sqrt{2}f_{2}^{(2)}
v_{1}^{\m_{1}}v_{1}^{\m_{2}}\bar{u}_{\m_{1}\m_{2}}
\g_{\lambda}(1-\g_{5})u \nonumber
\\
&\frac{1}{2}^{+}\rightarrow\frac{3}{2}^{+}:&
\frac{2}{\sqrt{5}}f_{2}^{(2)}
v_{1}^{\m}\bar{u}_{\m}\g_{5}(\vslash_{1}+\omega)\g_{\lambda}
(1-\g_{5})u\,.\nonumber\\
&&
\eea

$\;(3)\,(a)\,(iii)\,\,\, L_{K_{3}}=1,\, L_{k_{3}}=1,\, L=2,
\,S_{q_{1}q_{2}}=1,\,J_{q_{1}q_{2}}=1$
\bea
&\frac{1}{2}^{+}\rightarrow\frac{3}{2}^{+}:& \bar{u}_{\m_{2}}
\g_{\lambda}(1-\g_{5})u M^{\m_{2}} \nonumber
\\
&\frac{1}{2}^{+}\rightarrow\frac{1}{2}^{+}:&
\frac{1}{\sqrt{3}}\bar{u}\g_{5}\g_{\bot_{2}\m_{2}}
\g_{\lambda}(1-\g_{5})uM^{\m_{2}}\,.
\label{120}
\eea
$M^{\m_{2}}$ is an axial vector transverse with respect to $v_{2}$. It
is not possible to construct such a vector from the two vectors at our
disposal. Hence these decays are forbidden.

$\;(3)\,(b)\,(i)\,\,\, L_{K_{3}}=1,\, L_{k_{3}}=1,\, L=1,
\,S_{q_{1}q_{2}}=1,\,J_{q_{1}q_{2}}=2$
\bea
&\frac{1}{2}^{+}\rightarrow\frac{5}{2}^{+}:& \sqrt{2}\bar{u}_{\m\m_{2}}
\g_{5}\g_{\bot_{2}\m_{1}}\g_{\lambda}(1-\g_{5})u M^{\m[\m_{1}\m_{2}]} \nonumber
\\
&\frac{1}{2}^{+}\rightarrow\frac{3}{2}^{+}:&
\frac{2}{3\sqrt{5}}(\bar{u}_{\m}\g_{\bot_{2}\m_{2}}\g_{\bot_{2}\m_{1}}
+\bar{u}_{\m_{2}}\g_{\bot_{2}\m}\g_{\bot_{2}\m_{1}})
\g_{\lambda}(1-\g_{5})uM^{\m[\m_{1}\m_{2}]}\,.\nonumber\\
&&
\label{121}
\eea
Here $M^{\m[\m_{1}\m_{2}]}$ is a positive parity, mixed symmetry tensor,
transverse with respect to $v_{2}$. There is a unique non vanishing
tensor, of this type, giving rise to one form factor, $f_{3}^{(2)}$:
\be
M^{\m[\m_{1}\m_{2}]}=
if_{3}^{(2)}\e^{\m_{1}\m_{2}\r\k}v_{1\r}v_{2\k}v_{1\bot}^{\m}\,.
\ee
Hence the matrix elements (\ref{121}) can be written as
\bea
&\frac{1}{2}^{+}\rightarrow\frac{5}{2}^{+}:& \sqrt{2}if_{3}^{(2)}
\e^{\m_{1}\m_{2}\r\k}v_{1\r}v_{2\k}v_{1}^{\m}\bar{u}_{\m\m_{2}}
\g_{5}\g_{\m_{1}}\g_{\lambda}(1-\g_{5})u  \nonumber
\\
&\frac{1}{2}^{+}\rightarrow\frac{3}{2}^{+}:&\frac{2}{3\sqrt{5}}
if_{3}^{(2)}\e^{\m_{1}\m_{2}\r\k}v_{1\r}v_{2\k}
\{v_{1}^{\m}\bar{u}_{\m}\g_{\m_{2}}\g_{\m_{1}}
+\bar{u}_{\m_{2}}(\vslash_{1}-\omega)\g_{\m_{1}}\}
\g_{\lambda}(1-\g_{5})u\,.\nonumber\\
&&
\label{122}
\eea
Again using $\e$ identities these can further be simplified to
\bea
&\frac{1}{2}^{+}\rightarrow\frac{5}{2}^{+}:& \sqrt{2}f_{3}^{(2)}
v_{1}^{\m_{1}}v_{1}^{\m_{2}}\bar{u}_{\m_{1}\m_{2}}
\g_{\lambda}(1-\g_{5})u  \nonumber\\
&\frac{1}{2}^{+}\rightarrow\frac{3}{2}^{+}:&\frac{2}{\sqrt{5}}
f_{3}^{(2)}v_{1}^{\m}\bar{u}_{\m}\g_{5}(\vslash_{1}+\omega)
\g_{\lambda}(1-\g_{5})u\,.\nonumber\\
&&
\eea

$\;(3)\,(b)\,(ii)\,\,\, L_{K_{3}}=1,\, L_{k_{3}}=1,\, L=1,
\,S_{q_{1}q_{2}}=1,\,J_{q_{1}q_{2}}=1$
\bea
&\frac{1}{2}^{+}\rightarrow\frac{3}{2}^{+}:& \bar{u}_{\m_{2}}
\g_{\lambda}(1-\g_{5})u M^{\m_{2}} \nonumber
\\
&\frac{1}{2}^{+}\rightarrow\frac{1}{2}^{+}:&
\frac{1}{\sqrt{3}}\bar{u}\g_{5}\g_{\bot_{2}\m_{2}}
\g_{\lambda}(1-\g_{5})uM^{\m_{2}}\,.
\label{123}
\eea
We see immediately that these transition amplitudes vanish as it is not
possible to construct an axial vector $M^{\m_{2}}$, transverse to
$v_{2}$, from the two velocity vectors available to us.

$\;(3)\,(b)\,(iii)\,\,\, L_{K_{3}}=1,\, L_{k_{3}}=1,\, L=1,
\,S_{q_{1}q_{2}}=1,\,J_{q_{1}q_{2}}=0$
\bea
&\frac{1}{2}^{+}\rightarrow\frac{1}{2}^{+}:&
\frac{1}{\sqrt{6}}\bar{u}\g_{\bot_{2}\m}\g_{\bot_{2}\m_{1}}\g_{\bot_{2}\m_{2}}
\g_{\lambda}(1-\g_{5})uM^{[\m\m_{1}\m_{2}]}\,.
\label{124}
\eea
Here $M^{[\m\m_{1}\m_{2}]}$ is a totally antisymmetric, positive parity
tensor, transverse with respect to $v_{2}$. There is a unique
nonvanishing form for this tensor and hence we get one form factor,
$f_{4}^{(2)}$:
\be
M^{[\m\m_{1}\m_{2}]}=if_{4}^{(2)}(\omega^{2}-1)\e^{\m\m_{1}\m_{2}\r}v_{2\r}\,.
\label{f42}
\ee
In writing the coupling (\ref{f42}) we have exhibited an explicit
dynamical threshold factor which would arise, for example, in a dynamical
calculation from the different mode structures of excited and
ground state baryons in overlap integrals.  Hence the matrix element
(\ref{124}) becomes
\bea &\frac{1}{2}^{+}\rightarrow\frac{1}{2}^{+}:&
\frac{1}{\sqrt{6}}if_{4}^{(2)}(\omega^{2}-1)\e^{\m\m_{1}\m_{2}\r}v_{2\r}
\bar{u}\g_{\m}\g_{\m_{1}}\g_{\m_{2}}
\g_{\lambda}(1-\g_{5})u\,.
\label{125}
\eea
Once more this can be further simplified to
\bea
&\frac{1}{2}^{+}\rightarrow\frac{1}{2}^{+}:&
-\sqrt{6}f_{4}^{(2)}(\omega^{2}-1)\bar{u}\g_{\lambda}(1-\g_{5})u\,.
\eea

$\;(3)\,(c)\,\,\, L_{K_{3}}=1,\, L_{k_{3}}=1,\, L=0,
\,S_{q_{1}q_{2}}=1,\,J_{q_{1}q_{2}}=1$
\bea
&\frac{1}{2}^{+}\rightarrow\frac{3}{2}^{+}:&\bar{u}_{\m}\g_{\lambda}
(1-\g_{5})uM^{\m}\nonumber \\
&\frac{1}{2}^{+}\rightarrow\frac{1}{2}^{+}:&
\frac{1}{\sqrt{3}}\bar{u}\g_{5}\g_{\bot_{2}\m}
\g_{\lambda}(1-\g_{5})uM^{\m}\,.
\label{126}
\eea
Again these transition amplitudes vanish because we cannot construct an
axial vector transverse to $v_{2}$.

In conclusion, we see that that the transition amplitudes for the ground
state heavy $\Lambda$ to the next positive parity excited $\Lambda$'s
are described in terms of just five form factors.

\subsubsection{Heavy $\Sigma_{Q}(\Omega_{Q})$ to first positive parity excited
$\Sigma_{Q^{\prime}}(\Omega_{Q^{\prime}})$.}

We list the matrix elements, in terms of the final states in the usual
matrix form as we have done above in the case of decays to p-wave
states.

$\;(1)\,(a)\,\,\, L_{K_{3}}=1,\, L_{k_{3}}=1,\, L=2,
\,S_{q_{1}q_{2}}=0,\,J_{q_{1}q_{2}}=2$

$(\frac{3}{2}^{+},\frac{1}{2}^{+})\rightarrow(\frac{5}{2}^{+},\frac{3}{2}^{+})$
\be
\left( \begin{array}{c}
\bar{u}_{\m_{1}\m_{2}}\\ \sqrt{\frac{2}{5}}
\bar{u}_{\m_{1}}\g_{5}\g_{\bot_{2}\m_{2}}
 \end{array}\right)
\g_{\lambda}(1-\g_{5})\left( \begin{array}{c}
u_{\n}\\ \frac{1}{\sqrt{3}}\g_{\bot_{1}\n}\g_{5}u \end{array}\right)
M^{\{\m_{1}\m_{2}\}\n}\,.\label{127}
\ee
Here $M^{\{\m_{1}\m_{2}\}\n}$ is a positive parity tensor, symmetric and
traceless in $\m_{1},\m_{2}$, satisfying
\be
v_{2\m_{1}}M^{\{\m_{1}\m_{2}\}\n}=v_{1\n}M^{\{\m_{1}\m_{2}\}\n}=0\,.
\ee
There is just one such tensor giving rise to one form factor,
$g_{1}^{(2)}$:
\be
M^{\{\m_{1}\m_{2}\}\n}=
g_{1}^{(2)}i(\e^{\m_{1}\n\r\k}v_{1\r}v_{2\k}v_{1\bot}^{\m_{2}}
+\e^{\m_{2}\n\r\k}v_{1\r}v_{2\k}v_{1\bot}^{\m_{1}})\,.
\ee
Hence the matrix elements (\ref{127}) can be written as
\be
g_{1}^{(2)}i(\e^{\m_{1}\n\r\k}v_{1\r}v_{2\k}v_{1}^{\m_{2}}
+\e^{\m_{2}\n\r\k}v_{1\r}v_{2\k}v_{1}^{\m_{1}})\left( \begin{array}{c}
\bar{u}_{\m_{1}\m_{2}}\\ \sqrt{\frac{2}{5}}
\bar{u}_{\m_{1}}\g_{5}\g_{\bot_{2}\m_{2}}
 \end{array}\right)
\g_{\lambda}(1-\g_{5})\left( \begin{array}{c}
u_{\n}\\ \frac{1}{\sqrt{3}}\g_{\n}\g_{5}u \end{array}\right)
\,.\label{128}
\ee

$\;(1)\,(b)\,\,\, L_{K_{3}}=1,\, L_{k_{3}}=1,\, L=1,
\,S_{q_{1}q_{2}}=0,\,J_{q_{1}q_{2}}=1$

$(\frac{3}{2}^{+},\frac{1}{2}^{+})\rightarrow(\frac{3}{2}^{+},\frac{1}{2}^{+})$
\be
\left( \begin{array}{c}
\sqrt{2}\bar{u}_{\m_{2}}\g_{5}\g_{\bot_{2}\m_{1}}\\
\frac{1}{\sqrt{6}}\bar{u}\g_{\bot_{2}\m_{2}}\g_{\bot_{2}\m_{1}}
 \end{array}\right)
\g_{\lambda}(1-\g_{5})\left( \begin{array}{c}
u_{\n}\\ \frac{1}{\sqrt{3}}\g_{\bot_{1}\n}\g_{5}u \end{array}\right)
M^{[\m_{1}\m_{2}]\n}\,,\label{129}
\ee
where $M^{[\m_{1}\m_{2}]\n}$ is a positive parity tensor antisymmetric
in $\m_{1},\m_{2}$ and satisfies
\be
v_{2\m_{1}}M^{[\m_{1}\m_{2}]\n}=v_{1\n}M^{[\m_{1}\m_{2}]\n}=0\,.
\ee
We can construct just one such tensor leading to one form factor,
$g_{7}^{(2)}$:
\be
M^{[\m_{1}\m_{2}]\n}=
g_{7}^{(2)}i\e^{\m_{1}\m_{2}\r\k}v_{1\r}v_{2\k}v_{2\bot}^{\n}\,.
\ee
Hence, using $\e$ identities, we can write the matrix elements (\ref{129}) as
\be
g_{7}^{(2)}\left( \begin{array}{c}
\sqrt{2}v_{1}^{\m}\bar{u}_{\m}\\
\sqrt{\frac{2}{3}}\bar{u}\g_{5}(\omega+\vslash_{1})
 \end{array}\right)
\g_{\lambda}(1-\g_{5})\left( \begin{array}{c}
v_{2}^{\n}u_{\n}\\ \frac{1}{\sqrt{3}}(\vslash_{2}+\omega)\g_{5}u
\end{array}\right)
\,.\label{130}
\ee

$\;(1)\,(c)\,\,\, L_{K_{3}}=1,\, L_{k_{3}}=1,\, L=0,
\,S_{q_{1}q_{2}}=0,\,J_{q_{1}q_{2}}=0$

$(\frac{3}{2}^{+},\frac{1}{2}^{+})\rightarrow(\frac{1}{2}^{+})$
\be
\bar{u}
\g_{\lambda}(1-\g_{5})\left( \begin{array}{c}
u_{\n}\\ \frac{1}{\sqrt{3}}\g_{\bot_{1}\n}\g_{5}u \end{array}\right)
M^{\n}\,,\label{131}
\ee
where $M^{\n}$ is an axial vector, transverse to $v_{1}$. It is not
possible to construct such an axial vector. Hence this transition is
forbidden.

$\;(2)\,(a)\,\,\, L_{K_{3}}=2,\, L_{k_{3}}=0,\, L=2,
\,S_{q_{1}q_{2}}=1,\,J_{q_{1}q_{2}}=3$

$(\frac{3}{2}^{+},\frac{1}{2}^{+})\rightarrow(\frac{7}{2}^{+},\frac{5}{2}^{+})$
\be
\left( \begin{array}{c}
\bar{u}_{\m\m_{1}\m_{2}}\\ \sqrt{\frac{3}{7}}\bar{u}_{\m_{1}\m_{2}}
\g_{5}\g_{\bot_{2}\m} \end{array}\right)
\g_{\lambda}(1-\g_{5})\left( \begin{array}{c}
u_{\n}\\ \frac{1}{\sqrt{3}}\g_{\bot_{1}\n}\g_{5}u \end{array}\right)
M^{\{\m\m_{1}\m_{2}\}\n}\,.\label{132}
\ee
Here $M^{\{\m\m_{1}\m_{2}\}\n}$ is a positive parity tensor, symmetric
and traceless
in the labels $\m,\m_{1},\m_{2}$ and satisfying the following
transversality conditions:
\be
v_{2\m}M^{\{\m\m_{1}\m_{2}\}\n}=v_{1\n}M^{\{\m\m_{1}\m_{2}\}\n}=0\,.
\ee
The general form for such a tensor is
\bea
M^{\{\m\m_{1}\m_{2}\}\n}&=&g_{2}^{(2)}N^{\m\m_{1}\m_{2}}
(v_{1})v_{2\bot}^{\n}\nonumber\\
&&+g_{3}^{(2)}\{v_{1\bot}^{\{\m}v_{1\bot}^{\m_{1}}
g_{\bot_{2}\bot_{1}}^{\m_{2}\}\n}
-\frac{2}{5}v_{1\bot}^{\{\m}g_{\bot_{2}}^{\m_{1}\m_{2}\}}
v_{1\bot\r}g_{\bot_{2}\bot_{1}}^{\r\n}
-\frac{1}{5}v_{1\bot}^{2}g_{\bot_{2}}^{\{\m_{1}\m_{2}}
g_{\bot_{2}\bot_{1}}^{\m\}\n}\}\,.\nonumber\\
&&
\eea
Here $N^{\m\m_{1}\m_{2}}$ is the traceless, transverse tensor defined in
eq. (\ref{33}). $v_{1\bot},v_{2\bot}$ and $g_{\bot_{2}\bot_{1}}$ are as
defined in eqs. (\ref{vperp}-\ref{gpp}).

Hence the matrix elements, eq. (\ref{132}), for this set of transitions
can be written as
\bea
&\{g_{2}^{(2)}v_{1}^{\m_{1}}v_{1}^{\m_{2}}v_{1}^{\m}v_{2}^{\n}+
g_{3}^{(2)}v_{1}^{\{\m}v_{1}^{\m_{1}}g^{\m_{2}\}\n}\}&\nonumber\\
\times&\left( \begin{array}{c}
\bar{u}_{\m\m_{1}\m_{2}}\\ \sqrt{\frac{3}{7}}\bar{u}_{\m_{1}\m_{2}}
\g_{5}\g_{\bot_{2}\m} \end{array}\right)
\g_{\lambda}(1-\g_{5})\left( \begin{array}{c}
u_{\n}\\ \frac{1}{\sqrt{3}}\g_{\bot_{1}\n}\g_{5}u
\end{array}\right)&\,.\label{133}
\eea

$\;(2)\,(b)\,\,\, L_{K_{3}}=2,\, L_{k_{3}}=0,\, L=2,
\,S_{q_{1}q_{2}}=1,\,J_{q_{1}q_{2}}=2$

$(\frac{3}{2}^{+},\frac{1}{2}^{+})\rightarrow(\frac{5}{2}^{+},\frac{3}{2}^{+})$
\be
\left( \begin{array}{c}
\sqrt{2}\bar{u}_{\m_{1}\m_{2}}\g_{5}\g_{\bot_{2}\m}\\ \frac{2}{3\sqrt{5}}
(\bar{u}_{\m_{2}}\g_{\bot_{2}\m_{1}}\g_{\bot_{2}\m}
+\bar{u}_{\m_{1}}\g_{\bot_{2}\m_{2}}\g_{\bot_{2}\m}) \end{array}\right)
\g_{\lambda}(1-\g_{5})\left( \begin{array}{c}
u_{\n}\\ \frac{1}{\sqrt{3}}\g_{\bot_{1}\n}\g_{5}u \end{array}\right)
M^{[\m\m_{1}]\m_{2}\n}\,,\label{134}
\ee
where $M^{[\m\m_{1}]\m_{2}\n}$ is a positive parity tensor,
antisymmetric with respect to the indices $\m,\m_{1}$ and satisfying the
conditions
\be
v_{2\m}M^{[\m\m_{1}]\m_{2}\n}=v_{2\m_{2}}M^{[\m\m_{1}]\m_{2}\n}=
v_{1\n}M^{[\m\m_{1}]\m_{2}\n}=0\,.
\ee
It is also traceless with respect to the indices $\m,\m_{2}$. There is a
unique tensor satisfying these conditions, giving rise to one form factor,
$g_{4}^{(2)}$:
\bea
M^{[\m\m_{1}]\m_{2}\n}&=&g_{4}^{(2)}[v_{1\bot}^{\m_{2}}(g_{\bot_{2}}^{\m\n}
v_{1\bot}^{\m_{1}}-g_{\bot_{2}}^{\m_{1}\n}v_{1\bot}^{\m}))\nonumber\\
&&-\frac{1}{2}\{v_{1\bot}^{\n}(g_{\bot_{2}}^{\m\m_{2}}v_{1\bot}^{\m_{1}}
-g_{\bot_{2}}^{\m_{1}\m_{2}}v_{1\bot}^{\m})
+v_{1\bot}^{2}(g_{\bot_{2}}^{\m_{1}\m_{2}}g_{\bot_{2}}^{\m\n}
-g_{\bot_{2}}^{\m\m_{2}}g_{\bot_{2}}^{\m_{1}\n})\}]\nonumber\,. \\
&&
\eea
In this case we will not simplify the matrix element (\ref{134}) further.

$\;(2)\,(c)\,\,\, L_{K_{3}}=2,\, L_{k_{3}}=0,\, L=2,
\,S_{q_{1}q_{2}}=1,\,J_{q_{1}q_{2}}=1$

$(\frac{3}{2}^{+},\frac{1}{2}^{+})\rightarrow(\frac{3}{2}^{+},\frac{1}{2}^{+})$
\be
\left( \begin{array}{c}
\bar{u}_{\m}\\ \frac{1}{\sqrt{3}}
\bar{u}\g_{5}\g_{\bot_{2}\m}
 \end{array}\right)
\g_{\lambda}(1-\g_{5})\left( \begin{array}{c}
u_{\n}\\ \frac{1}{\sqrt{3}}\g_{\bot_{1}\n}\g_{5}u \end{array}\right)
M^{\m\n}\,.\label{135}
\ee
Here $M^{\m\n}$ is a positive parity tensor satisfying
\be
v_{2\m}M^{\m\n}=v_{1\n}M^{\m\n}=0\,.
\ee
Clearly there are two such tensors giving rise to two form factors:
\be
M^{\m\n}=g_{5}^{(2)}v_{1\bot}^{\m}v_{2\bot}^{\n}+
g_{6}^{(2)}g_{\bot_{2}\bot_{1}}^{\m\n}\,.
\ee
Hence the matrix elements can be written as
\be
(g_{5}^{(2)}v_{1}^{\m}v_{2}^{\n}+g_{6}^{(2)}g^{\m\n})\left( \begin{array}{c}
\bar{u}_{\m}\\ \frac{1}{\sqrt{3}}
\bar{u}\g_{5}\g_{\bot_{2}\m}
 \end{array}\right)
\g_{\lambda}(1-\g_{5})\left( \begin{array}{c}
u_{\n}\\ \frac{1}{\sqrt{3}}\g_{\bot_{1}\n}\g_{5}u \end{array}\right)
\,.\label{136}
\ee

Case (3) with $L_{K_{3}}=0,L_{k_{3}}=2$ gives rise to the same Lorentz
structure for the form factors as above except that the five form
factors, labelled $g_{i}^{\prime(2)}$ with $i=2\ldots 6$, will in general
 be different from the above five, $g_{i}^{(2)}$, $i=2\ldots 6$, because they
will involve different overlap integrals.

To summarise we have a total of twelve form factors to describe the
transitions of the heavy $\Sigma$ to the first positive parity excited heavy
$\Sigma$'s.

\subsection{Decay of s-wave baryons to L-wave baryons}

We now generalise the results of the previous subsections to the decays of
ground state baryons to the ``highest weight" L-wave excitations using the wave
functions developed in subsection~\ref{sub-Lheavy} .

\subsubsection{Heavy $\Lambda_{Q}$ to L-wave $\Lambda_{Q^{\prime}}$.}

As before we shall list the matrix elements according to the various
final L-wave states with parity $(-1)^{L}$.

\underline{(i) $\Lambda_{Q},\frac{1}{2}^{+}\rightarrow$ L-wave
$\Lambda_{Q^{\prime}}$ with
$S_{q_{1}q_{2}}=0$.}
\bea
&\frac{1}{2}^{+}\rightarrow L+\frac{1}{2}:& \bar{u}_{\m_{1}\ldots\m_{L}}
\g_{\lambda}(1-\g_{5})uM^{\m_{1}\ldots\m_{L}} \nonumber\\
&\frac{1}{2}^{+}\rightarrow L-\frac{1}{2}:& \sqrt{\frac{L}{2L+1}}
\bar{u}_{\m_{2}\ldots\m_{L}}\g_{5}\g_{\bot_{2}\m_{1}}\g_{\lambda}(1-\g_{5})
uM^{\m_{1}\ldots\m_{L}}\,, \label{137}
\eea
with $M^{\m_{1}\ldots\m_{L}}$
an L-th rank, symmetric, traceless tensor, transverse with respect to
$v_{2}$ and with parity $(-1)^{L}$. Such a tensor is uniquely given by
the tensor $N^{\m_{1}\ldots\m_{L}}(v_{1})$ which is defined as in
eq. (\ref{33}) with $k_{\bot}$ replaced by $v_{1\bot}$.
Hence there is just one form factor, here called $f^{(L)}_{1}$, and
the transition matrix elements, eq. (\ref{137}), can be written as
\bea
&\frac{1}{2}^{+}\rightarrow
L+\frac{1}{2}:&f^{(L)}_{1}v_{1}^{\m_{1}}\ldots v_{1}^{\m_{L}}
\bar{u}_{\m_{1}\ldots\m_{L}}
\g_{\lambda}(1-\g_{5})u \nonumber\\
&\frac{1}{2}^{+}\rightarrow L-\frac{1}{2}:& f^{(L)}_{1}\sqrt{\frac{L}{2L+1}}
v_{1}^{\m_{2}}\ldots
v_{1}^{\m_{L}}\bar{u}_{\m_{2}\ldots\m_{L}}\g_{5}(\omega+\vslash_{1})
\g_{\lambda}(1-\g_{5})u\,.\nonumber\\
&&
 \label{138}
\eea

\underline{(ii) $\Lambda_{Q},\frac{1}{2}^{+}\rightarrow$ to L-wave
$\Lambda_{Q^{\prime}}$ with $S_{q_{1}q_{2}}=1$.}

\hspace{.5cm}(a) $S_{q_{1}q_{2}}=1, J_{q_{1}q_{2}}=L+1$
\bea
&\frac{1}{2}^{+}\rightarrow L+\frac{3}{2}:&
\bar{u}_{\m\m_{1}\ldots\m_{L}}\g_{\lambda}(1-\g_{5})u
M^{\{\m\m_{1}\ldots\m_{L}\}}
\nonumber \\
&\frac{1}{2}^{+}\rightarrow L+\frac{1}{2}:& \sqrt{\frac{L+1}{2L+1}}
\bar{u}_{\m_{1}\ldots\m_{L}}\g_{5}\g_{\bot_{2}\m}
\g_{\lambda}(1-\g_{5})uM^{\{\m\m_{1}\ldots\m_{L}\}}\,,  \label{139}
\eea
with $M^{\{\m\m_{1}\ldots\m_{L}\}}$ a $(-1)^{L}$ parity,
symmetric tensor satisfying
$v_{2\m}M^{\{\m\m_{1}\ldots\m_{L}\}}=0$. It is impossible to construct
such an unnatural parity tensor with the two vectors available to us. Hence
$\Lambda_{Q}$ does not decay to this pair of $L+\frac{3}{2}$ and
$L+\frac{1}{2}$ states.

\hspace{.5cm}(b) $S_{q_{1}q_{2}}=1, J_{q_{1}q_{2}}=L$
\bea
&\frac{1}{2}^{+}\rightarrow L+\frac{1}{2}:&
\sqrt{2}\bar{u}_{\m_{1}\ldots\m_{L}}\g_{5}\g_{\bot_{2}\m}\g_{\lambda}
(1-\g_{5})uM^{[\m\m_{1}]\m_{2}\ldots\m_{L}}
\nonumber \\
&\frac{1}{2}^{+}\rightarrow L-\frac{1}{2}:&
\frac{1}{L+1}\sqrt{\frac{2L}{2L+1}}\{\bar{u}_{\m_{2}\ldots\m_{L}}
\g_{\bot_{2}\m_{1}}\g_{\bot_{2}\m})\nonumber \\
&&+(L-1)\bar{u}_{\m_{1}\m_{3}\ldots\m_{L}}
\g_{\bot_{2}\m_{2}}\g_{\bot_{2}\m}\}\g_{\lambda}
(1-\g_{5})uM^{[\m \m_{1}]\m_{2}\ldots\m_{L}},
\label{141}
\eea
with
$M^{[\m\m_{1}]\m_{2}\ldots\m_{L}}$ a $(-1)^{L}$ parity, $L+1$ index mixed
symmetry, traceless tensor
transverse with respect to $v_{2}$.  Such a tensor does exist uniquely,
namely
\bea
&M^{[\m\m_{1}]\m_{2}\ldots\m_{L}}=&
if^{(L)}_{2}[\e^{\m\m_{1}\r\k}v_{1\r}v_{2\k}N^{\m_{2}\ldots\m_{L}}(v_{1})\nonumber\\
&&-\frac{1}{L}\sum_{i=2}^{L}g_{\bot_{2}}^{\m\m_{i}}\e_{\n}\,^{\m_{1}\r\k}v_{1\r}
v_{2\k}N^{\m_{2}\ldots\m_{i-1}\n\ldots\m_{L}}(v_{1})\nonumber\\
&&+\frac{1}{L}\sum_{i=2}^{L}g_{\bot_{2}}^{\m_{1}\m_{i}}\e_{\n}\,^{\m\r\k}v_{1\r}
v_{2\k}N^{\m_{2}\ldots\m_{i-1}\n\ldots\m_{L}}(v_{1})+\dots],
\label{143}
\eea
leading to one form factor, $f^{(L)}_{2}$ for this pair of transitions. The
extra terms indicated by dots in eq. (\ref{143}) are needed to make the
tensor traceless but they do not contribute to the matrix elements.
Using $\e$ identities one can then write the matrix elements eq. (\ref{141}) as
\bea
&\frac{1}{2}^{+}\rightarrow L+\frac{1}{2}:&
\sqrt{2}f_{2}^{(L)}v_{1}^{\m_{1}}\cdots v_{1}^{\m_{L}}
\bar{u}_{\m_{1}\ldots\m_{L}}\g_{\lambda}
(1-\g_{5})u\nonumber \\
&\frac{1}{2}^{+}\rightarrow L-\frac{1}{2}:&f_{2}^{(L)}
\sqrt{\frac{2L}{2L+1}}v_{1}^{\m_{1}}\cdots v_{1}^{\m_{L-1}}
\bar{u}_{\m_{1}\ldots\m_{L-1}}
\g_{5}(\vslash_{1}+\omega)\g_{\lambda}
(1-\g_{5})u\,.\nonumber \\
&&
\eea
\hspace{.5cm}(c)$S_{q_{1}q_{2}}=1, J_{q_{1}q_{2}}=L-1$
\bea
&\frac{1}{2}^{+}\rightarrow L-\frac{1}{2};&
\bar{u}_{\m_{2}\ldots\m_{L}}\g_{\lambda}(1-\g_{5})uM^{\m_{2}\ldots\m_{L}}
\nonumber\\
&\frac{1}{2}^{+}\rightarrow L-\frac{3}{2};&
\sqrt{\frac{L-1}{2L-1}}\bar{u}_{\m_{3}\ldots\m_{L}}\g_{5}\g_{\bot_{2}\m_{2}}
\g_{\lambda}(1-\g_{5})u M^{\m_{2}\ldots\m_{L}}\,,\label{144}
\eea
with $M^{\m_{2}\ldots\m_{L}}$
a $(L-1)$ index symmetric tensor with parity $(-1)^{L}$.
It is not possible to construct such a tensor with
the two vectors available to us, $v_{1}$ and $v_{2}$. Hence the
$\Lambda_{Q}, \frac{1}{2}^{+}$ does not decay to these $(L-\frac{1}{2})$
and $(L-\frac{3}{2})$ states.

  From the analysis given in this subsection it would seem that there
are only two form factors governing the transition of ground state
$\Lambda$'s to L-wave excited $\Lambda$'s. However this is misleading as
in the general case there are many partitions of the total orbital
angular momentum $L$ giving rise to the same Lorentz structure and the
same symmetry properties under $p_{1}\leftrightarrow p_{2}$. For each such
set of excited $\Lambda$ states there will be two form factors.

\subsubsection{Heavy $\Sigma_{Q}(\Omega_{Q})$ to L-wave $\Sigma_{Q^{\prime}}
(\Omega_{Q^{\prime}})$.}

 We now list all the matrix elements for the transitions of the ground state
$\frac{3}{2}^{+},\frac{1}{2}^{+}$ heavy $\Sigma(\Omega)$ states to the
corresponding highest weight L-wave excitations.

\underline{(i) $\Sigma_{Q}(\Omega_{Q})(\frac{3}{2}^{+},
\frac{1}{2}^{+})\rightarrow$ to
L-wave $\Sigma_{Q^{\prime}}(\Omega_{Q^{\prime}})$ with
$S_{q_{1}q_{2}}=0,J_{q_{1}q_{2}}=L; L~+~\frac{1}{2},L~-~\frac{1}{2}$.}
\be
\left( \begin{array}{c}
\bar{u}_{\m_{1}\ldots\m_{L}}\\ \sqrt{\frac{L}{2L+1}}
\bar{u}_{\m_{2}\ldots\m_{L}}\g_{5}\g_{\bot_{2}\m_{1}} \end{array}\right)
\g_{\lambda}(1-\g_{5})\left( \begin{array}{c}
u_{\n}\\ \frac{1}{\sqrt{3}}\g_{\bot_{1}\n}\g_{5}u \end{array}\right)
M^{\m_{1}\ldots\m_{L}\n}\,,\label{145}
\ee
with
$M^{\m_{1}\ldots\m_{L}\n}$
a $(L+1)$ rank, parity $(-1)^{L}$ tensor, symmetric and traceless in the
indices $\m_{1}\ldots\m_{L}$ and satisfying
\be
v_{2\m_{i}}M^{\m_{1}\ldots\m_{i}\ldots\m_{L}\n}=
v_{1\n}M^{\m_{1}\ldots\m_{L}\n}=0\,.\nonumber
\ee
The
general unique form of this tensor is
\bea
M^{\m_{1}\ldots\m_{L}\n}&=&
g_{1}^{(L)}iv_{1\r}v_{2\k}\{\sum_{i=1}^{L}
\e^{\m_{i}\n\r\k}N^{\m_{1}\ldots\m_{i-1}\m_{i+1}\ldots\m_{L}}(v_{1})\nonumber\\
&&-\frac{2}{L+1}\sum_{i<j}g_{\bot_{2}}^{\m_{i}\m_{j}}\e_{\m}\,^{\n\r\k}
N^{\m\m_{1}\ldots\m_{i-1}\m_{i+1}\ldots\m_{j-1}\m_{j+1}\ldots\m_{L}}(v_{1})\}\,,\label{146}
\eea
leading to just one form factor, $g_{1}^{(L)}$. When one substitutes this
in eq. (\ref{145}) we see that the second term does not contribute, and
in fact from the first term only the terms without the traces in the
tensor $N(v_{1})$ survive giving the matrix elements as
\bea
&ig^{(L)}_{1}v_{1\r} v_{2\k}\sum_{i=1}^{L}\e^{\m_{i}\n\r\k}
v_{1}^{\m_{1}}\ldots v_{1}^{\m_{i-1}}v_{1}^{\m_{i+1}}\ldots
v_{1}^{\m_{L}}&\nonumber\\
\times&\left( \begin{array}{c}
\bar{u}_{\m_{1}\ldots\m_{L}}\\ \sqrt{\frac{L}{2L+1}}
\bar{u}_{\m_{2}\ldots\m_{L}}\g_{5}\g_{\m_{1}} \end{array}\right)
\g_{\lambda}(1-\g_{5})\left( \begin{array}{c}
u_{\n}\\ \frac{1}{\sqrt{3}}\g_{\n}\g_{5}u \end{array}\right)\,.&\label{147}
\eea

\underline{(ii)  $\Sigma_{Q}(\Omega_{Q}),\frac{1}{2}^{+}\rightarrow$ to L-wave
$\Sigma_{Q^{\prime}}(\Omega_{Q^{\prime}})$ with $S_{q_{1}q_{2}}=1$.}

\hspace{.5cm}(a)
$(\frac{3}{2}^{+},\frac{1}{2}^{+})\rightarrow J_{q_{1}q_{2}}=L+1;L+\frac{3}{2},
L+\frac{1}{2}$
\be
\left( \begin{array}{c}
\bar{u}_{\m\m_{1}\ldots\m_{L}}\\
\sqrt{\frac{L+1}{2L+3}}\bar{u}_{\m_{1}\ldots\m_{L}}
\g_{5}\g_{\bot_{2}\m} \end{array}\right)
\g_{\lambda}(1-\g_{5})\left( \begin{array}{c}
u_{\n}\\ \frac{1}{\sqrt{3}}\g_{\bot_{1}\n}\g_{5}u
\end{array}\right)M^{\{\m\m_{1}\ldots\m_{L}\}\n}\,,
\label{148}
\ee
with $M^{\{\m\m_{1}\ldots\m_{L}\}\n}$ a parity $(-1)^{L}$ tensor, symmetric
and traceless in $\{\m\m_{1}\ldots\m_{L}\}$, satisfying
\be
v_{2\m}M^{\{\m\m_{1}\ldots\m_{L}\}\n}=
v_{1\n}M^{\{\m\m_{1}\ldots\m_{L}\}\n}=0\,.\nonumber
\ee
Thus there are two form factors, as $M^{\{\m\m_{1}\ldots\m_{L}\}\n}$ can in
general be written as
\bea
M^{\{\m\m_{1}\ldots\m_{L}\}\n}&=&
g_{2}^{(L)}v_{2\bot}^{\n}N^{\m\m_{1\ldots\m_{L}}}(v_{1})\nonumber \\
&&+g_{3}^{(L)}[g_{\bot_{2}\bot_{1}}^{\m\n}N^{\m_{1}\ldots\m_{L}}(v_{1})+
\sum_{i=1}^{L}g_{\bot_{2}\bot_{1}}^{\m_{i}\n}
N^{\m\m_{1}\ldots\m_{i-1}\m_{i+1}\ldots\m_{L}}(v_{1})+\dots]\,.\nonumber\\
&&
\label{149}
\eea
The dots in the above equation stand for terms required to make the
tensor traceless but these do not contribute to the matrix elements.
Thus the matrix elements, eq. (\ref{148}), become
\bea
&\{g_{2}^{(L)}v_{1}^{\m}v_{1}^{\m_{1}}\ldots v_{1}^{\m_{L}}v_{2}^{\n}+
g_{3}^{(L)}v_{1}^{\{\m_{1}}\ldots v_{1}^{\m_{L}}g^{\m\}\n}\}&\nonumber\\
\times&
\left( \begin{array}{c}
\bar{u}_{\m\m_{1}\ldots\m_{L}}\\
\sqrt{\frac{L+1}{2L+3}}\bar{u}_{\m_{1}\ldots\m_{L}}
\g_{5}\g_{\bot_{2}\m} \end{array}\right)
\g_{\lambda}(1-\g_{5})\left( \begin{array}{c}
u_{\n}\\ \frac{1}{\sqrt{3}}\g_{\bot_{1}\n}\g_{5}u
\end{array}\right)\,.&
\label{150}
\eea

\hspace{.5cm}(b)
$(\frac{3}{2}^{+},\frac{1}{2}^{+})\rightarrow J_{q_{1}q_{2}}=L;L+\frac{1}{2},
L-\frac{1}{2}$
\bea
&\left( \begin{array}{c}
\sqrt{2}\bar{u}_{\m_{1}\ldots\m_{L}}\g_{5}\g_{\bot_{2}\m}\\
\frac{1}{L+1}\sqrt{\frac{2L}{2L+1}}\{\bar{u}_{\m_{2}\ldots\m_{L}}
\g_{\bot_{2}\m_{1}}\g_{\bot_{2}\m}
+(L-1)\bar{u}_{\m_{1}\m_{3}\ldots\m_{L}}
\g_{\bot_{2}\m_{2}}\g_{\bot_{2}\m}\} \end{array}\right)&\nonumber\\
\times&\g_{\lambda}(1-\g_{5})\left( \begin{array}{c}
u_{\n}\\ \frac{1}{\sqrt{3}}\g_{\bot_{1}\n}\g_{5}u
\end{array}\right)M^{[\m\m_{1}]\m_{2}\ldots\m_{L}\n}\,,&
\label{151}
\eea
with $M^{[\m\m_{1}]\m_{2}\ldots\m_{L}\n}$ a parity $(-1)^{L}$
tensor, mixed symmetric and traceless in $\m,\m_{1}\ldots\m_{L}$, such that
\be
v_{2\m}M^{[\m\m_{1}]\m_{2}\dots\m_{L}\n}
=v_{2\m_{i}}M^{[\m\m_{1}]\m_{2}\ldots\m_{i}\ldots\m_{L}\n}=
v_{1\n}M^{[\m\m_{1}]\m_{2}\ldots\m_{L}\n}=0\,.\nonumber
\ee
The unique form for such a tensor is
\bea
M^{[\m\m_{1}]\m_{2}\ldots\m_{L}\n}&=&g_{4}^{(L)}[
(g_{\bot_{2}}^{\m\n}v_{1\bot}^{\m_{1}}
-g_{\bot_{2}}^{\m_{1}\n}v_{1\bot}^{\m})N^{\m_{2}\ldots\m_{L}}(v_{1})\nonumber\\
&&-\frac{1}{L}\sum_{i=2}^{L}g_{\bot_{2}}^{\m\m_{i}}(v_{1\bot}^{\m_{1}}
N^{\n\m_{2}\ldots\m_{i-1}\m_{i+1}\ldots\m_{L}}(v_{1})
-g_{\bot_{2}}^{\m_{1}\n}v_{1\r}
N^{\r\m_{2}\ldots\m_{i-1}\m_{i+1}\ldots\m_{L}}(v_{1}))\nonumber\\
&&+\frac{1}{L}\sum_{i=2}^{L}(\m,\m_{1} interchanged)]\,.
\label{152}
\eea
Thus we have one form factor $g_{4}^{(L)}$ to describe these transitions.

\hspace{.5cm}(c)
$(\frac{3}{2}^{+},\frac{1}{2}^{+})\rightarrow J_{q_{1}q_{2}}=L-1;L-\frac{1}{2},
L-\frac{3}{2}$
\be
\left( \begin{array}{c}
\bar{u}_{\m_{2}\ldots\m_{L}}\\
\sqrt{\frac{L-1}{2L-1}}\bar{u}_{\m_{3}\ldots\m_{L}}
\g_{5}\g_{\bot_{2}\m_{2}} \end{array}\right)
\g_{\lambda}(1-\g_{5})\left( \begin{array}{c}
u_{\n}\\ \frac{1}{\sqrt{3}}\g_{\bot_{1}\n}\g_{5}u
\end{array}\right)M^{\m_{2}\ldots\m_{L}\n}\,,
\label{153}
\ee
with $M^{\m_{2}\ldots\m_{L}\n}$ a parity $(-1)^{L}$
tensor, symmetric and traceless in $\m_{2}\ldots\m_{L}$, such that
\be
v_{2\m_{2}}M^{\m_{2}\dots\m_{L}\n}
=v_{1\n}M^{\m_{2}\ldots\m_{L}\n}=0\,.\nonumber
\ee
One can construct two such tensors giving rise to two form factors,
$g_{5}^{(L)}$ and $g_{6}^{(L)}$, i.e.
\bea
M^{\m_{2}\ldots\m_{L}\n}&=&
g_{5}^{(L)}N^{\m_{2}\ldots\m_{L}}(v_{1})v_{2\bot}^{\n}\nonumber\\
&&+g_{6}^{(L)}\{\sum_{i=2}^{L}[g^{\m_{i}\n}
-(v_{1}^{\m_{i}}v_{1}^{\n}+v_{2}^{\m_{i}}v_{2}^{\n})
+\omega v_{1}^{\n}v_{2}^{\m_{i}}]
N ^{\m_{2}\ldots\m_{i-1}\m_{i+1}\ldots\m_{L}}(v_{1})\nonumber\\
&&-\frac{1}{L+1}\sum_{i<j}g_{\bot_{2}}^{\m_{i}\m_{j}}g_{\bot_{1}\r}^{\n}
N^{\r\m_{2}\ldots\m_{i-1}\m_{i+1}\ldots\m_{j-1}\m_{j+1}\ldots\m_{L}}(v_{1})\}\,.
\eea
Here too the total number of form factors is not just six. As in the
$\Lambda_{Q}$
case there are many  partitions of the total orbital angular momentum
$L$ giving rise to the same Lorentz structure and the same symmetry
properties under $p_{1}\leftrightarrow p_{2}$. For each such partition
there will be five form factors describing the transitions.

\subsection{Bjorken sum rule for heavy baryon decays}

In this subsection we give the complete contributions of the s-wave,
p-wave and the first positive parity excitations, as well as of the
highest weight L-wave states, to the Bjorken sum rule
for the decays of the heavy s-wave baryons. Other contributions
can be calculated easily given the wave functions.

\subsubsection{Bjorken sum rule for $\Lambda_{Q}$ decays}

Let us now consider the contributions of the various
$\Lambda_{Q}\rightarrow\Lambda_{Q^{\prime}}^{*}$ transitions to the
Bjorken sum rule. Again we use the same trick as in the mesonic case in
that we first calculate the longitudinal transition amplitudes. In the
baryonic case one needs the Clebsch-Gordon decomposition
\be
u_{\m_{1}\ldots\m_{j}}(\pm\frac{1}{2})=
\sqrt{\frac{j+1}{2j+1}}\sqrt{\frac{j!}{(2j-1)!!}}u(\pm\frac{1}{2})
\e_{\m_{1}}(0)\cdots\e_{\m_{j}}(0)+\cdots.
\label{bjb1}
\ee
where the ellipsis at the end of (\ref{bjb1}) stand for terms involving
transverse
spin $1$ polarisation vectors. These, however, are not needed in
explicit form since they do not contribute to the longitudinal helicity
projection because of the HQET structure of the
transitions. Upon squaring the longitudinal amplitudes, summing over the
respective degenerate partners and dividing out the $K_{L}$-factor
\cite{bhkt} one obtains
\bea
1&=&\vert f(\omega)\vert^{2}+\sum_{L\geq 1}(\omega^{2}-1)^{L}
\frac{L!}{(2L-1)!!}(\vert f_{1}^{(L)}(\omega)\vert^{2}
+2\vert f_{2}^{(L)}(\omega)\vert^{2})\nonumber\\
&&+\frac{2}{3}(\omega^{2}-1)^{2}(\vert f_{1}^{\prime(2)}(\omega)\vert^{2}
+2\vert
f_{3}^{(2)}(\omega)\vert^{2}+9\vert f_{4}^{(2)}(\omega)\vert^{2})+\ldots\,.
 \eea
Here $f(\omega)$ is the single form factor appearing in the ground state
to ground state heavy $\Lambda_{Q}$ decay treated in \cite{hkkt} and
\cite{hlkkt}, with $f(1)=1$.
In this sum rule we have included all the contributions up to the first
positive parity excitations. However, for the general L case we have
included only one of the many highest weight contributions. The other
possible highest weight states will give exactly the same form of
contributions as in the sum above. We have also not
shown the contributions from the other types of higher excitations i.e.
other than highest weight. However our method of construction is
general enough to allow us to calculate the contribution of the decay to
any arbitrary excitation given its angular momentum structure.

\subsubsection{Bjorken sum rule for $\Sigma_{Q}(\Omega_{Q})$ decays}

Following the same general procedure one can calculate the contribution
of any excitation to the Bjorken sum rule for the decay of $\Sigma_{Q}$
or $\Omega_{Q}$. We write the sum rule as:
\bea
1&=&\vert g_{1}^{(0)}\vert^{2}\frac{(2+\omega^{2})}{3}+
\vert g_{2}^{(0)}\vert^{2}\frac{(\omega^{2}-1)^{2}}{3}-
(g_{1}^{(0)}g_{2}^{(0)*}
+g_{1}^{(0)*}g_{2}^{(0)})\frac{\omega(\omega^{2}-1)}{3}\nonumber\\
&&+\sum_{L\geq 1}\{\frac{1}{3}\vert g_{1}^{(L)}\vert^{2}
\frac{(L+1)!}{(2L-1)!!}L(\omega^{2}-1)^{L}\nonumber\\
&&+\frac{1}{3}\frac{(L+1)!}{(2L+1)!!}(\omega^{2}-1)^{L}[\vert
g_{2}^{(L)}\vert^{2}(\omega^{2}-1)^{2}+\vert g_{3}^{(L)}\vert^{2}(L+1)
((L+1)\omega^{2}+(L+2))\nonumber\\
&&+(g_{2}^{(L)}g_{3}^{(L)*}+g_{3}^{(L)}g_{2}^{(L)*})(L+1)\omega(\omega^{2}-1)]
\nonumber\\
&&+\frac{2}{3}\vert g_{4}^{(L)}\vert^{2}\frac{1}{L}\frac{(L+1)!}{(2L-1)!!}
(\omega^{2}-1)^{L}\nonumber\\
&&+\frac{1}{3}\frac{(L-1)!}{(2L-1)!!}(2L-1)(\omega^{2}-1)^{L-2}[\vert
g_{5}^{(L)}\vert^{2}(\omega^{2}-1)^{2}+\vert g_{6}^{(L)}\vert^{2}
((L-1)^{2}\omega^{2}+L(L-1))\nonumber\\
&&+(g_{5}^{(L)}g_{6}^{(L)*}+g_{6}^{(L)}g_{5}^{(L)*})(L-1)(\omega^{2}-1)\omega]\}
\nonumber\\
&&+\frac{2}{3}\vert g_{7}^{(2)}\vert^{2}(\omega^{2}-1)^{2}+\ldots.
\eea
where the form factors $g_{1}^{(0)}$ and $g_{2}^{(0)}$ parametrize the $\Sigma$
s-wave to s-wave decays, \cite {hkkt}, \cite{hlkkt},
\be
(-g_{1}^{(0)}g^{\m\n}+g_{2}^{(0)}v_{1}^{\m}v_{2}^{\n})
\left( \begin{array}{c}
\bar{u}_{\m}\\ \frac{1}{\sqrt{3}}
\bar{u}\g_{5}\g_{\bot_{2}\m}
 \end{array}\right)
\g_{\lambda}(1-\g_{5})\left( \begin{array}{c}
u_{\n}\\ \frac{1}{\sqrt{3}}\g_{\bot_{1}\n}\g_{5}u \end{array}\right)\,.
\ee
As one can see in subsection 5.1.2, there is no form factor
$g_{6}^{(1)}$, whereas it seems to be present in the sum above. However,
this is misleading as the factor $(L-1)$ always occurring with
$g_{6}^{(L)}$ kills $g_{6}^{(1)}$.

As in the $\Lambda_{Q}$ case, the above sum rule contains all the
contributions up to the first positive parity excitations if one also
includes the contributions form the form factors $g_{i}^{\prime(2)}$,
$i=2\ldots6 $,
mentioned after eq. (\ref{136}). These give the same formal structure as
the $g_{i}{(2)}$, $i=2\ldots 6$. Again, for
the general L case we have included only one of the many highest weight
contributions and have not shown the contributions from the other types
of (non-highest weight) excitations, which can of course be calculated
easily, if so desired. The Bjorken sum rule for $\Sigma_{Q}(\Omega_{Q})$
decays has also been
studied in \cite{x}. However, in that work the contribution of
$g_{1}^{(1)}$ is missing and further the factor in front of the
$g_{4}^{(1)}$ contribution is $\frac{12}{9}$ rather than $\frac{20}{9}$.

\section{Conclusions}

In this paper we have presented a general method to construct
wavefunctions for arbitrary orbital excitations of heavy hadrons. In
doing this we have utilised the representations of the group
${\cal L}\otimes O(3,1)$. We then used these wave functions to calculate
the form factors for the weak transitions of s-wave heavy hadrons to
arbitrarily excited  heavy hadrons. Of course, given these wave
functions we can also study the form factor structure for the
transitions between arbitrary heavy excited states as the need arises. The
contributions of excited states
to the Bjorken sum rule have also been worked out in some detail. One can
also use these same
projection operators for studying transitions from heavy to light
hadrons. This is currently under investigation and the results will be
presented elsewhere.

\section{Acknowledgements}
JGK would like to thank Prof. Abdus Salam and the IAEA for hospitality
at the ICTP during the course of this work. We would like to thank J.
Landgraf for checking most of the formulae in this paper as well as
deriving the $L\geq2$ contributions to the $\Omega$-type Bjorken sum rule.

\appendix
\section{Normalisation}
In this appendix we present a general approach to the normalisation of
the projection operators for mesons and baryons. The polarisation
vectors for a meson of spin $J$ are normalised as
\be
\e^{*\m_{1}\ldots\m_{J}}\e_{\m_{1}\ldots\m_{J}}=(-1)^{J}\,,
\label{a1}
\ee
and the generalised Rarita-Schwinger spinors for baryons with spin
$(J+\frac{1}{2})$ are normalised such that
\be
\bar{u}^{\m_{1}\dots\m_{J}}u_{\m_{1}\dots\m_{J}}=(-1)^{J}2M\,.\label{a2}
\ee
Recall that both the meson and baryon polarisation tensors are symmetric
and traceless in all the indices.

The meson projection operators are relatively easy to normalise
because they are all of the form
\be
N^{\m_{1}\ldots\m_{L}}(k)(\frac{1+\vslash}{2})\bar{\G}_{\m_{1}}
\ldots\m_{L}(\frac{1-\vslash}{2})\,. \label{a3}
\ee
Thus to normalise $\bar{\G}_{\m_{1}\ldots\m_{L}}$ we consider
\be
Tr(\frac{1-\vslash}{2})\bar{\G}^{*}_{\m_{1}\ldots\m_{L}}
(\frac{1+\vslash}{2})\bar{\G}_{\n_{1}\ldots\n_{L}}
G^{\m_{1}\ldots\m_{L};\n_{1}\ldots\n_{L}}=2M\,,\label{a4}
\ee
where
\be
G^{\m_{1}\ldots\m_{L};\n_{1}\ldots\n_{L}}=\int d^{4}k\,d^{4}k^{\prime}
N^{\m_{1}\ldots\m_{L}}(k^{\prime})N^{\n_{1}\ldots\n_{L}}(k)f(k,k^{\prime})
\label{a5}
\ee
and $f(k,k^{\prime})$ is an appropriate weight function. This
$G^{\m_{1}\ldots\m_{L};\n_{1}\ldots\n_{L}}$ is the required ``metric"
tensor to take the trace over the Lorentz indices in eq. (\ref{a4}). This
tensor $G$ must reflect the symmetry and trace properties of
$N^{\m_{1}\ldots\m_{L}}$. Hence it must be separately symmetric and
traceless in the sets of indices $(\m_{1},\ldots,\m_{L})$ and
$(\n_{1},\ldots,\n_{L})$. Also
\be
v_{\m_{1}}G^{\m_{1}\ldots\m_{L};\n_{1}\ldots\n_{L}}=
v_{\n_{1}}G^{\m_{1}\ldots\m_{L};\n_{1}\ldots\n_{L}}=0\,.
\ee
Of course, in general, such a tensor consists of many terms but
fortunately all our polarisation  tensors have at least $(L-1)$ indices
and are traceless so that we need only the first few terms in $G$. These
are
\bea
G^{\m_{1}\ldots\m_{L};\n_{1}\ldots\n_{L}}&=&
\frac{1}{L!}[g_{\bot}^{\m_{1}\ldots\m_{L};\n_{1}\ldots\n_{L}}\nonumber\\
&&-\frac{2}{2L-1}\sum_{i=2}^{L}g_{\bot}^{\m_{1}\m_{i}}
g_{\bot}^{\n_{1}\m_{2}\ldots\m_{i-1}\m_{i}\ldots\m_{L};\n_{2}\ldots\n_{L}}]+\ldots.
\label{a6}
\eea
The dots indicate terms which necessarily contain
$g_{\bot}^{\m_{2}\m_{3}}$ plus other such terms which annihilate on the
$\e$'s. In the above equation we have used the following definitions
\bea
g_{\bot}^{\m_{1}\ldots\m_{L};\n_{1}\ldots\n_{L}}&=&
\sum_{i=1}^{L}g_{\bot}^{\m_{1}\n_{i}}
g_{\bot}^{\m_{2}\ldots\m_{L};\n_{1}\ldots\n_{i-1}\n_{i+1}\ldots\n_{L}}
\nonumber\\
\vdots &&\nonumber\\
g_{\bot}^{\m_{1}\m_{2}\m_{3};\n_{1}\n_{2}\n_{3}}&=&
g_{\bot}^{\m_{1}\n_{1}}(g_{\bot}^{\m_{2}\n_{2}}g_{\bot}^{\m_{3}\n_{3}}+
g_{\bot}^{\m_{2}\n_{3}}g_{\bot}^{\m_{3}\n_{2}})\nonumber\\
&&+g_{\bot}^{\m_{1}\n_{2}}(g_{\bot}^{\m_{2}\n_{1}}g_{\bot}^{\m_{3}\n_{3}}
+g_{\bot}^{\m_{2}\n_{3}}g_{\bot}^{\m_{3}\n_{1}})\nonumber\\
&&+g_{\bot}^{\m_{1}\n_{3}}(g_{\bot}^{\m_{2}\n_{1}}g_{\bot}^{\m_{3}\n_{2}}
+g_{\bot}^{\m_{2}\n_{2}}g_{\bot}^{\m_{3}\n_{1}})\,.
\eea
We have a further simplification as in the normalisation equation
(\ref{a4}) we can drop the subscript $\bot$ from the $g^{\m\n}$'s
appearing in the definition of the tensor $G$ as the velocity vector
$v_{\m}$ will always annihilate on the $\bar{\G}$'s. All our
normalisations of the mesonic wavefunctions have been done with the
``metric" tensor in eq. (\ref{a6}).

For the general baryonic case, the situation is very complicated.
However, in the present paper, we have only considered baryonic
projection operators which are either fully symmetric in the Lorentz
indices or at most have the form
\be
A^{[\m_{1}\m_{2}]\m_{3}\dots\m_{L}}\Psi_{[\m_{1}\m_{2}]\m_{3}\dots\m_{L}}\,,
\ee
where $A^{[\m_{1}\m_{2}]\m_{3}\dots\m_{L}}$ is a tensor antisymmetric
in $\m_{1}\m_{2}$ and symmetric under interchange of the other labels and
traceless
with respect to any pair of indices. For the symmetric $\Psi$'s we use
the same ``metric" as in eq.(\ref{a6}), whereas for the mixed symmetric
$\Psi$'s we need the following:
\bea
G^{[\m\m_{1}]\m_{2}\ldots\m_{L};[\n\n_{1}]\n_{2}\ldots\n_{L}}&=&
\frac{1}{2(L-1)!}[(g_{\bot}^{\m\n}g_{\bot}^{\m_{1}\n_{1}}
-g_{\bot}^{\m\n_{1}}g_{\bot}^{\m_{1}\n})
g_{\bot}^{\m_{2}\ldots\m_{L};\n_{2}\ldots\n_{L}}\nonumber \\
&&+\frac{1}{L}\sum_{i=2}^{L}g_{\bot}^{\m\m_{i}}(g_{\bot}^{\m_{1}\n}
g_{\bot}^{\n_{1}\m_{2}\ldots\m_{i-1}\m_{i+1}\ldots\m_{L};\n_{2}\ldots\n_{L}}
\nonumber \\
&&-g_{\bot}^{\m_{1}\n_{1}}
g_{\bot}^{\n\m_{2}\ldots\m_{i-1}\m_{i+1}\ldots\m_{L};\n_{2}\ldots\n_{L}})
\nonumber\\
&&-\frac{1}{L}\sum_{i=2}^{L}g_{\bot}^{\m_{1}\m_{i}}(g_{\bot}^{\m\n}
g_{\bot}^{\n_{1}\m_{2}\ldots\m_{i-1}\m_{i+1}\ldots\m_{L};\n_{2}\ldots\n_{L}}
\nonumber\\
&&-g_{\bot}^{\m\n_{1}}
g_{\bot}^{\n\m_{2}\ldots\m_{i-1}\m_{i+1}\ldots\m_{L};\n_{2}\ldots\n_{L}})]\nonumber\\
&&\ldots.
\eea
The extra terms not shown necessarily involve $g_{\bot}^{\m_{2}\m_{3}}$
and generally $g_{\bot}^{\m_{i}\m_{j}}$ with $j>i\geq 2$. These extra terms
annihilate on the heavy side spinor functions $\Psi_{\m\m_{1}\ldots\m_{L}}$.

\section{Mixed Symmetry Baryon Projection Operators}

The baryon projection operators, which we have considered in this
paper, contain tensor spinors
$\Psi_{\m_{1}\ldots\m_{L}}$ which are either fully symmetric in the
indices or are of mixed symmetry of the form
$[\m_{1}\m_{2}]\m_{2}\ldots\m_{L}$. These spinors satisfy the Dirac
equation and the transversality condition
\be
v^{\m_{i}}\Psi_{\m_{1}\ldots\m_{i}\ldots\m_{L}}=0\label{b1}\,,
\ee
on each label. To construct these $\Psi$'s we note once again that if $\f$
satisfies
the Dirac equation then $\g_{\bot\m}\g_{5}\f$ and
$\g_{\bot\m_{1}}\g_{\bot\m_{2}}\f$ also satisfy the Dirac equation plus,
obviously, the transversality condition \ref{b1}. Thus the construction
of the fully symmetric and  fully antisymmetric tensors is obvious.

We now demonstrate how we have constructed the mixed tensors e.g. eqs.
(\ref{dw8})-(\ref{dw10}), (\ref{dw16}) and the generalisation to eq.
(\ref{81}). To construct the projection operator (\ref{dw8}) we need to
construct a traceless, mixed symmetry tensor from $\chi^{1,\m}_{\bot}$
and $N^{\{\m_{1}\m_{2}\}}$ to obtain total angular momentum $2$. We decompose
\be
\begin{Young}
\cr
\end{Young}\;
\bigotimes\;
\begin{Young}
& \cr
\end{Young}\;
=\;
\begin{Young}
&& \cr
\end{Young}\;
\bigoplus\;
\begin{Young}
& \cr
\cr
\end{Young}\,.
\ee

Thus to construct the mixed symmetric tensor we simply subtract the
fully symmetric tensor from the direct product.
The fully symmetric tensor is
\be
\f^{\{\m\m_{1}\m_{2}\}}=\frac{1}{3}[\chi_{\bot}^{1,\m}N^{\{\m_{1}\m_{2}\}}
+\chi_{\bot}^{1,\m_{1}}N^{\{\m\m_{2}\}}+\chi_{\bot}^{1,\m_{2}}N^{\{\m\m_{1}\}}]
\label{b3}
\ee
and hence the mixed symmetry tensor is
\be
\chi_{\bot}^{1,\m}N^{\{\m_{1}\m_{2}\}}-\f^{\{\m\m_{1}\m_{2}\}}\,.
\ee
We make this traceless and write the mixed symmetric, traceless
tensor as
\bea
&\frac{1}{3}[2\chi_{\bot}^{1,\m}N^{\{\m_{1}\m_{2}\}}
-\chi_{\bot}^{1,\m_{1}}N^{\{\m\m_{2}\}}
-\chi_{\bot}^{1,\m_{2}}N^{\{\m\m_{1}\}}&\nonumber\\
&+g_{\bot}^{\m_{1}\m_{2}}\chi_{\bot\n}^{1}N^{\{\n\m\}}
-\frac{1}{2}g_{\bot}^{\m\m_{1}}\chi_{\bot\n}^{1}N^{\{\n\m_{2}\}}
-\frac{1}{2}g_{\bot}^{\m\m_{2}}\chi_{\bot\n}^{1}N^{\{\n\m_{1}\}}]&\,.
\label{b4}
\eea
Now for the $\frac{5}{2}^{+}$ case, eq. (\ref{dw9}), we need to multiply
the above tensor by a similar mixed symmetry, traceless tensor
constructed with $\g_{\bot\m}\g_{5}u_{\m_{1}\m_{2}}$ i.e.
\be
\frac{2}{3}\g_{\bot\m}\g_{5}u_{\m_{1}\m_{2}}
-\frac{1}{3}\g_{\bot\m_{1}}\g_{5}u_{\m\m_{2}}
-\frac{1}{3}\g_{\bot\m_{2}}\g_{5}u_{\m\m_{1}}\,.
\label{b5}
\ee
Hence we obtain
\bea
&\frac{1}{3}[2\chi_{\bot}^{1,\m}N^{\{\m_{1}\m_{2}\}}
-\chi_{\bot}^{1,\m_{1}}N^{\{\m\m_{2}\}}
-\chi_{\bot}^{1,\m_{2}}N^{\{\m\m_{1}\}}&\nonumber\\
&+g_{\bot}^{\m_{1}\m_{2}}\chi_{\bot\n}^{1}N^{\{\n\m\}}
-\frac{1}{2}g_{\bot}^{\m\m_{1}}\chi_{\bot\n}^{1}N^{\{\n\m_{2}\}}
-\frac{1}{2}g_{\bot}^{\m\m_{2}}\chi_{\bot\n}^{1}N^{\{\n\m_{1}\}}]&\nonumber\\
\times&\frac{1}{3}[2\g_{\bot\m}\g_{5}u_{\m_{1}\m_{2}}
-\g_{\bot\m_{1}}\g_{5}u_{\m\m_{2}}-\g_{\bot\m_{2}}\g_{5}u_{\m\m_{1}}]&\,.
\label{b6}
\eea
We simplify this by noting that both the factors in the product are
symmetric under $\m_{1}\leftrightarrow\m_{2}$, so that we can write the
product as
\bea
&\frac{1}{3}[2\chi_{\bot}^{1,\m}N^{\{\m_{1}\m_{2}\}}
-\chi_{\bot}^{1,\m_{1}}N^{\{\m\m_{2}\}}
-\chi_{\bot}^{1,\m_{2}}N^{\{\m\m_{1}\}}&\nonumber\\
&+g_{\bot}^{\m_{1}\m_{2}}\chi_{\bot\n}^{1}N^{\{\n\m\}}
-\frac{1}{2}g_{\bot}^{\m\m_{1}}\chi_{\bot\n}^{1}N^{\{\n\m_{2}\}}
-\frac{1}{2}g_{\bot}^{\m\m_{2}}\chi_{\bot\n}^{1}N^{\{\n\m_{1}\}}]&\nonumber\\
\times&\frac{2}{3}[\g_{\bot\m}\g_{5}u_{\m_{1}\m_{2}}
-\g_{\bot\m_{1}}\g_{5}u_{\m\m_{2}}]&\,.
\label{b7}
\eea
But now the second factor is antisymmetric under $\m\leftrightarrow\m_{1}$.
Therefore it projects out the antisymmetric piece from the first term
leading to
\bea
&\frac{1}{3}[\chi_{\bot}^{1,\m}N^{\{\m_{1}\m_{2}\}}
-\chi_{\bot}^{1,\m_{1}}N^{\{\m\m_{2}\}}
-\frac{1}{2}g_{\bot}^{\m\m_{2}}\chi_{\bot\n}^{1}N^{\{\n\m_{1}\}}
+\frac{1}{2}g_{\bot}^{\m_{1}\m_{2}}\chi_{\bot\n}^{1}N^{\{\n\m\}}]&\nonumber\\
\times
&(\g_{\bot\m}\g_{5}u_{\m_{1}\m_{2}}-\g_{\bot\m_{1}}\g_{5}u_{\m\m_{2}})\,.&
\label{b8}
\eea
This is precisely eqn. (\ref{dw8}) combined with eq. (\ref{dw9}) upto a
normalisation factor.

To construct the other $(\frac{3}{2}^{+})$ , eq. (\ref{dw10}), member
of the degenerate multiplet  $(\frac{5}{2}^{+},\frac{3}{2}^{+})$ we
proceed as follows. Here we need to construct a mixed symmetry tensor
from $\g_{\bot\m},\g_{\bot\m_{1}}$ and $u_{\m_{2}}$. We can take either
\be
\begin{Young}\cr\end{Young}\;\bigotimes\;\begin{Young}\cr\end{Young}\;=\;
\begin{Young}&\cr\end{Young}\;\bigoplus\;\begin{Young}\cr\cr\end{Young}\,
\ee
and then
\be
\begin{Young}&\cr\end{Young}\;\bigotimes\;\begin{Young}\cr\end{Young}\;=\;
\begin{Young}&&\cr\end{Young}\;\bigoplus\;\begin{Young}&\cr\cr\end{Young}_{1}\,
\ee

or

\be
\begin{Young}\cr\cr\end{Young}\;\bigotimes\;\begin{Young}\cr\end{Young}\;=\;
\begin{Young}\cr\cr\cr\end{Young}\;\bigoplus\;\begin{Young}&\cr\cr\end{Young}_{2}\,.
\ee

To make contact with the other member of the multiplet and also for
later generalisation to the $L>2$ case we follow the first route. We
first construct the symmetric tensor
\be
F_{\m_{1}\m_{2}} =\frac{1}{2}(\g_{\bot\m_{1}}u_{\m_{2}}
+\g_{\bot\m_{2}}u_{\m_{1}})\,.
\label{b9}
\ee
Hence the traceless, mixed symmetry tensor is
\be
\frac{2}{3}\g_{\bot\m}F_{\m_{1}\m_{2}}
-\frac{1}{3}\g_{\bot\m_{1}}F_{\m\m_{2}}-\frac{1}{3}\g_{\bot\m_{2}}F_{\m\m_{1}}\,.
\label{b10}
\ee
Following the same procedure as in the $\frac{5}{2}^{+}$ case above the
product
\bea
&\frac{1}{3}[2\chi_{\bot}^{1,\m}N^{\{\m_{1}\m_{2}\}}
-\chi_{\bot}^{1,\m_{1}}N^{\{\m\m_{2}\}}
-\chi_{\bot}^{1,\m_{2}}N^{\{\m\m_{1}\}}&\nonumber\\
&+g_{\bot}^{\m_{1}\m_{2}}\chi_{\bot\n}^{1}N^{\{\n\m\}}
-\frac{1}{2}g_{\bot}^{\m\m_{1}}\chi_{\bot\n}^{1}N^{\{\n\m_{2}\}}
-\frac{1}{2}g_{\bot}^{\m\m_{2}}\chi_{\bot\n}^{1}N^{\{\n\m_{1}\}}]&\nonumber\\
\times&\frac{1}{3}[2\g_{\bot\m}F_{\m_{1}\m_{2}}
-\g_{\bot\m_{1}}F_{\m\m_{2}}-\g_{\bot\m_{2}}F_{\m\m_{1}}]&
\label{b11}
\eea
reduces to
\bea
&\frac{1}{3}[\chi_{\bot}^{1,\m}N^{\{\m_{1}\m_{2}\}}
-\chi_{\bot}^{1,\m_{1}}N^{\{\m\m_{2}\}}
-\frac{1}{2}g_{\bot}^{\m\m_{2}}\chi_{\bot\n}^{1}N^{\{\n\m_{1}\}}
+\frac{1}{2}g_{\bot}^{\m_{1}\m_{2}}\chi_{\bot\n}^{1}N^{\{\n\m\}}]&\nonumber\\
\times &\frac{1}{2}\{[\g_{\bot\m},\g_{\bot\m_{1}}]u_{\m_{2}}
+(\g_{\bot\m}\g_{\bot\m_{2}}u_{\m_{1}}
-\g_{\bot\m_{1}}\g_{\bot\m_{2}}u_{\m})\}&
\label{b12}
\eea
which is the same as eq. (\ref{dw10}) upto normalisation. The mixed
symmetry projection operators, eqs. (\ref{sdw4})-(\ref{sdw6}), for the
$\Sigma$-type first positive parity excitations are constructed in the
same way.

We now come to the construction of the projection operators
(\ref{dw14})-(\ref{dw16}). Here the $J_{q_{1}q_{2}}=2$ is represented by
the mixed symmetry tensor arising from the product of the antisymmetric tensor,
$N^{[\m_{1}\m_{2}]}$, defined in eq. (\ref{dw2}) and $\chi_{\bot}^{1,\m}$.
The decomposition is
\be
\begin{Young}\cr\cr\end{Young}\;\bigotimes\;\begin{Young}\cr\end{Young}\;=\;
\begin{Young}&\cr\cr\end{Young}\;\bigoplus\;\begin{Young}\cr\cr\cr\end{Young}\,.
\ee
The fully antisymmetric tensor is
\be
T^{[\m\m_{1}\m_{2}]}=\frac{1}{3}[\chi_{\bot}^{1,\m}N^{[\m_{1}\m_{2}]}
+\chi_{\bot}^{1,\m_{1}}N^{[\m_{2}\m]}+\chi_{\bot}^{1,\m_{2}}N^{[\m\m_{1}]}]\,.
\label{b13}
\ee
The mixed symmetry tensor is then
\be
\chi_{\bot}^{1,\m}N{[\m_{1}\m_{2}]}-T^{[\m\m_{1}\m_{2}]}=
\frac{1}{3}(2\chi_{\bot}^{1,\m}N^{[\m_{1}\m_{2}]}-
\chi_{\bot}^{1,\m_{1}}N^{[\m_{2}\m]}-\chi_{\bot}^{1,\m_{2}}N^{[\m\m_{1}]})\,.
\label{b14}
\ee
Note that this is antisymmetric under $\m_{1}\leftrightarrow\m_{2}$ and
annihilates on the mixed symmetry tensor constructed earlier for the
$\frac{5}{2}^{+}$, eq. (\ref{b5}),
which was symmetric under $\m_{1}\leftrightarrow\m_{2}$. However,
instead of eq. (\ref{b5}) we are free to choose another mixed symmetry
tensor, (change of labels only!),
\be
\frac{2}{3}\g_{\bot\m_{1}}\g_{5}u_{\m_{2}\m}-
\frac{1}{3}\g_{\bot\m_{2}}\g_{5}u_{\m\m_{1}}-
\frac{1}{3}\g_{\bot\m}\g_{5}u_{\m_{1}\m_{2}}\,.
\label{b15}
\ee
This is symmetric under $\m_{2}\leftrightarrow\m$ and does not
annihilate on the mixed symmetry tensor eq. (\ref{b14}). Actually to proceed
further we have to make eq. (\ref{b14}) traceless. Doing that we can write
the projection operator for the $\frac{5}{2}^{+}$ as
\bea
&(\frac{2}{3}\chi_{\bot}^{1,\m}N^{[\m_{1}\m_{2}]}
-\frac{1}{3}\chi_{\bot}^{1,\m_{1}}N^{[\m_{2}\m]}
-\frac{1}{3}\chi_{\bot}^{1,\m_{2}}N^{[\m\m_{1}]}
-\frac{1}{2}g_{\bot}^{\m\m_{1}}\chi_{\bot\n}^{1}N^{[\n\m_{2}]}
+\frac{1}{2}g_{\bot}^{\m\m_{2}}\chi_{\bot\n}^{1}N^{[\n\m_{1}]})&\nonumber\\
&\times\frac{2}{3}(2\g_{\bot\m_{1}}\g_{5}u_{\m_{2}\m}
-\g_{\bot\m_{2}}\g_{5}u_{\m\m_{1}}-\g_{\bot\m}\g_{5}u_{\m_{1}\m_{2}})\,.&
\label{b16}
\eea
Then using the fact that the second factor is symmetric under
$\m\leftrightarrow\m_{2}$ we can write the above as
\bea
&(\chi_{\bot}^{1,\m}N^{[\m_{1}\m_{2}]}
-\frac{1}{2}g_{\bot}^{\m\m_{1}}\chi_{\bot\n}^{1}N^{[\n\m_{2}]}
+\frac{1}{2}g_{\bot}^{\m\m_{2}}\chi_{\bot\n}^{1}N^{[\n\m_{1}]})&\nonumber\\
&\times\frac{2}{3}(2\g_{\bot\m_{1}}\g_{5}u_{\m_{2}\m}
-\g_{\bot\m_{2}}\g_{5}u_{\m\m_{1}}-\g_{\bot\m}\g_{5}u_{\m_{1}\m_{2}})\,.&
\label{b17}
\eea
Now the first factor is antisymmetric under $\m_{1}\leftrightarrow\m_{2}$
and will thus pick out the corresponding antisymmetric piece from the
second bracket finally leading to
\bea
&(\chi_{\bot}^{1,\m}N^{[\m_{1}\m_{2}]}
-\frac{1}{2}g_{\bot}^{\m\m_{1}}\chi_{\bot\n}^{1}N^{[\n\m_{2}]}
+\frac{1}{2}g_{\bot}^{\m\m_{2}}\chi_{\bot\n}^{1}N^{[\n\m_{1}]})&\nonumber\\
&\times\frac{1}{2}(\g_{\bot\m_{1}}\g_{5}u_{\m_{2}\m}
-\g_{\bot\m_{2}}\g_{5}u_{\m\m_{1}})\,,&
\label{b18}
\eea
which is precisely eqs. (\ref{dw14}) and (\ref{dw15}) upto a normalisation
factor.

For the $\frac{3}{2}^{+}$ partner, eq. (\ref{dw16}), of this
$\frac{5}{2}^{+}$ we need to construct, as earlier, a mixed symmetry tensor
from
two $\g_{\bot}$'s and from the Rarita-Schwinger spinor, $u_{\m}$.
Following by now familiar steps such a mixed symmetry tensor is
\be
\frac{2}{3}\g_{\bot\m}G_{\m_{1}\m_{2}}-\frac{1}{3}\g_{\bot\m_{1}}G_{\m_{2}\m}
-\frac{1}{3}\g_{\bot\m_{2}}G_{\m\m_{1}}\,,
\label{b19}
\ee
where
\be
G_{\m_{1}\m_{2}}=\frac{1}{2}(\g_{\bot\m_{1}}u_{\m_{2}}
-\g_{\bot\m_{2}}u_{\m_{1}})\,.
\label{b20}
\ee
Multiplying the tensor in eq. (\ref{b19}) into the mixed symmetry tensor
in eq. (\ref{b14}),
appropriately made traceless, and following by now a familiar procedure
we arrive at the projection operator eq. (\ref{dw16}), for the
$\frac{3}{2}^{+}$, upto a normalisation factor.

We now come to the mixed symmetric tensors arising in the highest weight L-wave
states, i.e. we want to construct eq. (\ref{78c}) combined with eq. (\ref{81}).
The fully symmetric tensor constructed from $\chi_{\bot}^{1,\m}$ and the
fully symmetric tensor $N^{\m_{1}\ldots\m_{L}}$ is
\be
T^{\{\m\m_{1}\ldots\m_{L}\}}=
\frac{1}{L+1}(\chi_{\bot}^{1,\m}N^{\m_{1}\ldots\m_{L}}
+\sum_{i=1}^{L}\chi_{\bot}^{1,\m_{i}}
N^{\m_{1}\ldots\m_{i-1}\m\m_{i+1}\ldots\m_{L}})\,.
\label{b21}
\ee
Hence the mixed symmetry tensor is just
\bea
T_{ms}^{\m\m_{1}\ldots\m_{L}}&=&\chi_{\bot}^{1,\m}N^{\m_{1}\ldots\m_{L}}
-T^{\{\m\m_{1}\ldots\m_{L}\}}\nonumber\\
&&=\frac{L}{L+1}\chi_{\bot}^{1,\m}N^{\m_{1}\ldots\m_{L}}
-\frac{1}{L+1}\sum_{i=1}^{L}\chi_{\bot}^{1,\m_{i}}
N^{\m_{1}\ldots\m_{i-1}\m\m_{i+1}\ldots\m_{L}}\,.\label{b22}
\eea
For the $L+\frac{1}{2}$ spinor part of the projector, we follow a similar
construction of
the mixed symmetry tensor from $\g_{\bot\m}$ and the fully symmetric
tensor $\g_{5}u_{\m_{1}\ldots\m_{L}}$ to obtain
\be
\frac{L}{L+1}\g_{\bot\m}\g_{5}u_{\m_{1}\ldots\m_{L}}
-\frac{1}{L+1}\sum_{i=1}^{L}\g_{\bot\m_{i}}\g_{5}
u_{\m_{1}\dots\m_{i-1}\m\m_{i+1}\dots\m_{L}}\,.\label{b23}
\ee
Now making the tensor in eq. (\ref{b22}) traceless and multiplying into
expression (\ref{b23}) we can write
the projection operator for $L+\frac{1}{2}$ as
\bea
&\frac{1}{L+1}[L\chi_{\bot}^{1,\m}N^{\m_{1}\ldots\m_{L}}
-\sum_{i=1}^{L}\chi_{\bot}^{1,\m_{i}}
N^{\m_{1}\ldots\m_{i-1}\m\m_{i+1}\ldots\m_{L}}&\nonumber\\
&-\frac{L-1}{L}\sum_{i=1}^{L}g_{\bot}^{\m\m_{i}}\chi_{\bot\n}^{1}
N^{\m_{1}\ldots\m_{i-1}\n\m_{i+1}\ldots\m_{L}}
+\frac{2}{L}\sum_{i<j}g_{\bot}^{\m_{i}\m_{j}}\chi_{\bot\n}^{1}
N^{\m_{1}\ldots\m_{i-1}\n\m_{i+1}\ldots\m_{j-1}\m\m_{j+1}\ldots\m_{L}}]
\,.&\nonumber\\
&\times\frac{1}{L+1}(L\g_{\bot\m}\g_{5}u_{\m_{1}\ldots\m_{l}}
-\sum_{i=1}^{L}\g_{\bot\m_{i}}\g_{5}
u_{\m_{1}\ldots\m_{i-1}\m\m_{i+1}\ldots\m_{L}})\,.&\label{b24}
\eea
Both these factors are symmetric under any interchange
$\m_{i}\leftrightarrow\m_{j}$ and thus using the fact that these are all
dummy indices we can write
\bea
&\frac{1}{L+1}[L\chi_{\bot}^{1,\m}N^{\m_{1}\ldots\m_{L}}
-\sum_{i=1}^{L}\chi_{\bot}^{1,\m_{i}}
N^{\m_{1}\ldots\m_{i-1}\m\m_{i+1}\ldots\m_{L}}&\nonumber\\
&-\frac{L-1}{L}\sum_{i=1}^{L}g_{\bot}^{\m\m_{i}}\chi_{\bot\n}^{1}
N^{\m_{1}\ldots\m_{i-1}\n\m_{i+1}\ldots\m_{L}}
+\frac{2}{L}\sum_{i<j}g_{\bot}^{\m_{i}\m_{j}}\chi_{\bot\n}^{1}
N^{\m_{1}\ldots\m_{i-1}\n\m_{i+1}\ldots\m_{j-1}\m\m_{j+1}\ldots\m_{L}}]
\,.&\nonumber\\
&\times\frac{L}{L+1}(\g_{\bot\m}\g_{5}u_{\m_{1}\ldots\m_{l}}
-\g_{\bot\m_{1}}\g_{5}
u_{\m\m_{2}\ldots\m_{L}})\,.&\label{b25}
\eea
But now the second factor is antisymmetric under $\m\leftarrow\m_{1}$ and
hence projects out the corresponding antisymmetric part from the first
factor leading to
\bea
&\frac{L}{2(L+1)}(\chi_{\bot}^{1,\m}N^{\m_{1}\ldots}
-\chi_{\bot}^{1,\m_{1}}
N^{\m\m_{2}\ldots\m_{L}}&\nonumber\\
&-\frac{2}{L}\sum_{i=2}^{L}g_{\bot}^{\m\m_{i}}\chi_{\bot\n}^{1}
N^{\m_{1}\ldots\m_{i-1}\n\m_{i+1}\ldots\m_{L}}
+\frac{2}{L}\sum_{i=2}^{L}g_{\bot}^{\m_{1}\m_{i}}\chi_{\bot\n}^{1}
N^{\m\m_{2}\ldots\m_{i-1}\n\m_{i+1}\ldots\m_{L}})\,.&\nonumber\\
&\times(\g_{\bot\m}\g_{5}u_{\m_{1}\ldots\m_{l}}
-\g_{\bot\m_{1}}\g_{5}
u_{\m\m_{2}\ldots\m_{L}})\,.&\label{b26}
\eea
Upto a normalisation factor this is just the first of the eqs. (\ref{81}).
For the second $L-\frac{1}{2}$ member of the degenerate pair we proceed
by first constructing the fully symmetric $L$ index tensor
\be
F_{\m_{1}\ldots\m_{L}}=\frac{1}{L}\sum_{i=1}^{L}\g_{\bot\m_{i}}
u_{\m_{1}\ldots\m_{i-1}\m_{i+1}\ldots\m_{L}}\,.\label{b27}
\ee
Now make the fully symmetric $(L+1)$ tensor from combining the above with
$\g_{\bot\m}$.
\be
\frac{1}{L+1}(\g_{\bot\m}F_{\m_{1}\ldots\m_{L}}
+\sum_{i=1}^{L}\g_{\bot\m_{i}}
F_{\m_{1}\ldots\m_{i-1}\m\m_{i+1}\ldots\m_{L}})\,. \label{b28}
\ee
Thus the mixed symmetric $(L+1)$ index tensor is
\bea
U^{ms}_{\m\m_{1}\ldots\m_{L}}&=&\g_{\bot\m}F_{\m_{1}\ldots\m_{L}}
-\frac{1}{L+1}(\g_{\bot\m}F_{\m_{1}\ldots\m_{L}}
+\sum_{i=1}^{L}\g_{\bot\m_{i}}
F_{\m_{1}\ldots\m_{i-1}\m\m_{i+1}\ldots\m_{L}})\nonumber\\
&=&\frac{1}{L+1}(L\g_{\bot\m}F_{\m_{1}\ldots\m_{L}}
-\sum_{i=1}^{L}\g_{\bot\m_{i}}
F_{\m_{1}\ldots\m_{i-1}\m\m_{i+1}\ldots\m_{L}})\,.\label{b29}
\eea
Now we multiply this into the mixed symmetry tensor of eq. (\ref{b22})
and using the familiar arguments about symmetry and dummy
indices we finally arrive at
\bea
&\frac{L}{2(L+1)}(\chi_{\bot}^{1,\m}N^{\m_{1}\ldots}
-\chi_{\bot}^{1,\m_{1}}
N^{\m\m_{2}\ldots\m_{L}}&\nonumber\\
&-\frac{2}{L}\sum_{i=2}^{L}g_{\bot}^{\m\m_{i}}\chi_{\bot\n}^{1}
N^{\m_{1}\ldots\m_{i-1}\n\m_{i+1}\ldots\m_{L}}
+\frac{2}{L}\sum_{i=2}^{L}g_{\bot}^{\m_{1}\m_{i}}\chi_{\bot\n}^{1}
N^{\m\m_{2}\ldots\m_{i-1}\n\m_{i+1}\ldots\m_{L}})\,.&\nonumber\\
&\times\frac{1}{L}\{[\g_{\bot\m},\g_{\bot\m_{1}}]u_{\m_{2}\ldots\m_{L}}+
(L-1)(\g_{\bot\m}\g_{\bot\m_{2}}u_{\m_{1}\m_{3}\ldots\m_{L}}-
\g_{\bot\m_{1}}\g_{\bot\m_{2}}u_{\m\m_{3}\ldots\m_{L}})\}\,.&\nonumber\\
&&
\label{b30}
\eea
Upto a normalisation factor this is the same as the second of the
eqs. (\ref{81}).

\end{document}